# Physics of Lyα Radiative Transfer

Mark Dijkstra (Institute of Theoretical Astrophysics, University of Oslo)

(Dated: April 7, 2017)

**Contents**

## 1. PREFACE

These lectures notes cover my 8 lectures on the 'Physics of Ly$\alpha$ Radiative Transfer', which I gave at the 46$^{\rm th}$ Saas-Fee winter school that was held in Les Diablerets, Switzerland on March 13-19 2016. These lectures aimed at offering basic insights into Ly$\alpha$ radiative processes including emission processes and Ly$\alpha$ radiative transfer, and highlighting some of the exciting physics associated with these processes. The notes discuss some derivation in greater detail than what was discussed during the lectures. Feel free to contact me with any questions/comment on these notes. I plan to update & expand these notes in time.

The Ly$\alpha$ transfer problem is an exciting problem to learn about and work on. Ly$\alpha$ transfer is deeply rooted in quantum physics, it requires knowledge of statistics, statistical physics/thermodynamics, computational astrophysics, and has applications in a wide range of astrophysical contexts including galaxies, the interstellar medium, the circum-galactic medium, the intergalactic medium, reionization, 21-cm cosmology and astrophysics. In these lectures I will describe the basics of Ly$\alpha$ radiative processes and transfer. These lectures are aimed to be self-contained, and are (hopefully) suitable for anyone with an undergraduate degree in astronomy/physics.

Throughout these notes, I denote symbols that represent vectors in **bold print**. I use CGS units, as is common in the literature. Table I provides an overview of (some of the) symbols that appear throughout these notes.

## 2. INTRODUCTION

Half a century ago, Partridge & Peebles (1967) predicted that the Ly$\alpha$ line should be a good tracer of star forming galaxies at large cosmological distances. This statement was based on the assumption that ionizing photons that are emitted by young, newly formed stars are efficiently reprocessed into recombination lines, of which Ly$\alpha$ contains the largest flux. In the past two decades the Ly$\alpha$ line has indeed proven to provide us with a way to both find and identify galaxies out to the highest redshifts (currently as high as $z = 8.7$, see Zitrin et al., 2015). In addition, we do not only expect Ly$\alpha$ emission from (star forming) galaxies, but from structure formation in general (e.g. Furlanetto et al. 2005). Galaxies are surrounded by vast reservoirs of gas that are capable of both emitting and absorbing Ly$\alpha$ radiation. Observed spatially extended Ly$\alpha$ nebulae (or 'blobs') indeed provide insight into the formation & evolution of galaxies, in ways that complement direct observations of galaxies.

Many new instruments & telescopes[1] are either about to be, or have just been, commissioned that are ideal for targeting the redshifted Ly$\alpha$ line. The sheer number of observed Ly$\alpha$ emitting sources is expected to increase by more than two orders of magnitude at all redshifts $z \sim 2 - 7$. For comparison, this boost is similar to that in the number of known exoplanets as a result of the launch of the Kepler satellite. In addition, sensitive integral field unit spectrographs will allow us to ($i$) detect sources that are more than an order of magnitude fainter than what has been possible so far, ($ii$) take spectra of faint sources, ($iii$) take spatially resolved spectra of the more extended sources, and ($iv$) detect phenomena at surface brightness levels at which diffuse Ly$\alpha$ emission from the environment of galaxies is visible.

In order to optimally benefit form this rapidly growing body of data, we must understand the radiative transfer of Ly$\alpha$ photons. Ly$\alpha$ transfer depends sensitively on the distribution and kinematics of neutral gas, which complicates interpretations of Ly$\alpha$ observations. On the other hand, the close interaction of the Ly$\alpha$ radiation field and gaseous flows in and around galaxies implies that the Ly$\alpha$ line contains information on the scattering medium, and may thus present an opportunity to learn more about atomic hydrogen in gaseous flows in and around galaxies[2].

---

[1] New instruments/telescopes that will revolutionize our ability to target Ly$\alpha$ emission: the Hobby-Eberly Telescope Dark Energy Experiment (HETDEX, http://hetdex.org/ ) will increase the sample of Ly$\alpha$ emitting galaxies by orders of magnitude at $z \sim 2 - 4$; Subaru's Hyper Suprime-Cam (http://www.naoj.org/Projects/HSC/) will provide a similar boost out to $z \sim 7$. Integral Field Unit Spectrographs such as MUSE (https://www.eso.org/sci/facilities/develop/instruments/muse.html, see Bacon et al. 2010) and the Keck Cosmic Web Imager (http://www.srl.caltech.edu/sal/keck-cosmic-web-imager.html, and Martin et al. 2014 for observations carried out with the *Palomar* Cosmic Web Imager) will allow us to map out spatially extended Ly$\alpha$ emission down to $\sim 10$ times lower surface brightness levels, and take spatially resolved spectra. In the (near) future, telescopes such as the James Webb Space Telescope (JWST, http://www.jwst.nasa.gov/ ) and ground based facilities such as the Giant Magellan Telescope ( http://www.gmto.org/) and ESO's E-ELT (http://www.eso.org/public/usa/teles-instr/e-elt/, http://www.tmt.org/ (TMT).

[2] To underline this point: recent observations of Ly$\alpha$ sources (see e.g. CR7, Sobral et al. 2015) have triggered discussion on the formation of direct collapse black holes (e.g. Pallottini et al. 2015), Population III stars (e.g. Visbal et al. 2016), and on the structure of multiphase gases in and around galaxies (McCourt et al. 2017). These recent developments are not discussed in these lecture notes, and reflect that only a handful of new observations of the Ly$\alpha$ have already started highlighting that exciting science can be done with the line.

TABLE I Symbol Dictionary

| Symbol | Definition |
|---|---|
| $k_B$ | Boltzmann constant: $k_B = 1.38 \times 10^{-16}$ erg K$^{-1}$ |
| $h_P$ | Planck constant: $h_P = 6.67 \times 10^{-27}$ erg s |
| $\hbar$ | reduced Planck constant: $\hbar = \frac{h_P}{2\pi}$ |
| $m_p$ | proton mass: $m_p = 1.66 \times 10^{-24}$ |
| $m_e$ | electron mass: $m_e = 9.1 \times 10^{-28}$ g |
| $q$ | electron charge: $q = 4.8 \times 10^{-10}$ esu |
| $c$ | speed of light: $c = 2.9979 \times 10^{10}$ cm s$^{-1}$ |
| $\Delta E_{ul}$ | Energy difference between upper level 'u' and lower level 'l' (in ergs) |
| $\nu_{ul}$ | photon frequency associated with the transition $u \to l$ (in Hz) |
| $f_{ul}$ | the oscillator strength associated with the transition $u \to l$ (dimensionless) |
| $A_{ul}$ | Einstein A-coefficient of the transition $u \to l$ (in s$^{-1}$) |
| $B_{ul}$ | Einstein B-coefficient of the transition $u \to l$: $B_{ul} = \frac{2h_P \nu_{ul}^3}{c^2} A_{ul}$ (in erg cm$^{-2}$ s$^{-1}$) |
| $B_{lu}$ | Einstein B-coefficient of the transition $l \to u$: $B_{lu} = \frac{g_u}{g_l} B_{ul}$ |
| $\alpha_{A/B}$ | case A /B recombination coefficient (in cm$^3$ s$^{-1}$) |
| $\alpha_{nl}$ | recombination coefficient into state $(n,l)$ (in cm$^3$ s$^{-1}$) |
| $g_{u/l}$ | statistical weight of upper/lower level of a radiative transition (dimensionless) |
| $\nu_\alpha$ | photon frequency associated with the Ly$\alpha$ transition: $\nu_\alpha = 2.47 \times 10^{15}$ Hz |
| $\omega_\alpha$ | angular frequency associated with the Ly$\alpha$ transition: $\omega_\alpha = 2\pi\nu_\alpha$ |
| $\lambda_\alpha$ | wavelength associated with the Ly$\alpha$ transition: $\lambda_\alpha = 1215.67$ Å |
| $A_\alpha$ | Einstein A-coefficient of the Ly$\alpha$ transition: $A_\alpha = 6.25 \times 10^8$ s$^{-1}$ |
| $T$ | gas temperature (in K) |
| $v_{\rm th}$ | velocity dispersion (times $\sqrt{2}$): $v_{\rm th} = \sqrt{\frac{2k_B T}{m_p}}$ |
| $v_{\rm turb}$ | turbulent velocity dispersion |
| $b$ | Doppler broadening parameter : $b = \sqrt{v_{\rm th}^2 + v_{\rm turb}^2}$ |
| $\Delta\nu_\alpha$ | Doppler induced photon frequency dispersion:$\Delta\nu_\alpha = \nu_\alpha \frac{b}{c}$ (in Hz) |
| $x$ | 'normalized' photon frequency: $x = (\nu - \nu_\alpha)/\Delta\nu_\alpha$ (dimensionless) |
| $\sigma_\alpha(x)$ | Ly$\alpha$ absorption cross-section at frequency $x$ (in cm$^2$), $\sigma_\alpha(x) = \sigma_{\alpha,0}\phi(x)$ |
| $\sigma_{\alpha,0}$ | Ly$\alpha$ absorption cross-section at line center, $\sigma_{\alpha,0} = 5.9 \times 10^{-14} \left(\frac{T}{10^4 \text{ K}}\right)^{-1/2}$ cm$^2$ |
| $\phi(x)$ | Voigt profile (dimensionless) |
| $a_v$ | Voigt parameter: $a_v = A_\alpha/[4\pi\Delta\nu_\alpha] = 4.7 \times 10^{-4}(T/10^4 \text{ K})^{-1/2}$ |
| $I_\nu$ | specific intensity (in erg s$^{-1}$ Hz$^{-1}$ cm$^{-2}$ sr$^{-1}$) |
| $J_\nu$ | angle averaged specific intensity (in erg s$^{-1}$ Hz$^{-1}$ cm$^{-2}$ sr$^{-1}$) |

## 3. THE HYDROGEN ATOM AND INTRODUCTION TO LY$\alpha$ EMISSION MECHANISMS

### 3.1. Hydrogen in our Universe

It has been known from almost a century that hydrogen is the most abundant element in our Universe. In 1925 Cecilia Payne demonstrated in her PhD dissertation that the Sun was composed primarily of hydrogen and helium. While this conclusion was controversial[3] at the time, it is currently well established that hydrogen accounts for the majority of baryonic mass in our Universe: the fluctuations in the Microwave Background as measured by the Planck satellite (Planck Collaboration et al., 2016) imply that baryons account for 4.6% of the Universal energy density, and that hydrogen accounts for 76% of the total baryonic mass. The remaining 24% is in the form of helium (see Fig 1). The leading constraints on these mass ratios come from Big Bang Nucleosynthesis, which predicts a hydrogen abundance of $\sim 75\%$ by mass for the inferred Universal baryon density ($\Omega_b h^2 = 0.022$, Planck Collaboration et al., 2016). Additional constraints come from observations of hydrogen and emission lines of extragalactic, metal poor HII regions (Izotov et al. 2014, Aver et al. 2015).

---

[3] See `https://en.wikipedia.org/wiki/Cecilia_Payne-Gaposchkin`.

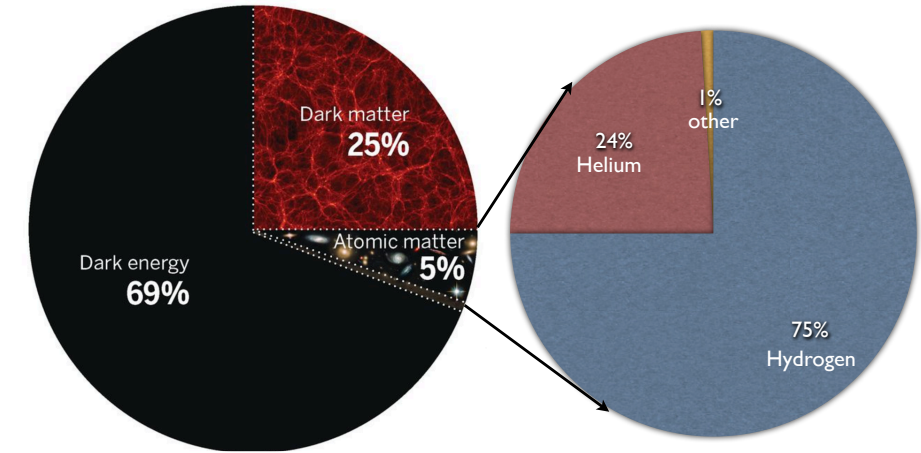

FIG. 1 Relative contributions to Universal energy/mass density. Most ($\sim 70\%$) of the Universal energy content is in the form of dark energy, which is an unknown hypothesized form of energy which permeates all of space, and which is responsible for the inferred acceleration of the expansion of the Universe. In addition to this, $\sim 25\%$ is in the form of dark matter, which is a pressureless fluid which acts only gravitationally with ordinary matter. Only $\sim 5\%$ of the Universal energy content is in the form of ordinary matter like baryons, leptons etc. Of this component, $\sim 75\%$ of all baryonic matter is in Hydrogen, while the remaining 25% is almost entirely Helium. Observing lines associated with atomic hydrogen is therefore an obvious way to go about studying the Universe.

Because of its prevalence throughout the Universe, lines associated with atomic hydrogen have provided us with a powerful window on our Universe. The 21-cm hyperfine transition was observed from our own Milky Way by Ewen & Purcell in 1951 (Ewen & Purcell, 1951), shortly after it was predicted to exist by Jan Oort in 1944. Observations of the 21-cm line have allowed us to perform precise measurements of the distribution and kinematics of neutral gas in external galaxies, which provided evidence for dark matter on galactic scales (e.g. Bosma, 1978). Detecting the redshifted 21-cm emission from galaxies at $z > 0.5$, and from atomic hydrogen in the diffuse (neutral) intergalactic medium represent the main science drivers for many low frequency radio arrays that are currently being developed, including the Murchinson Wide Field Array[4], the Low Frequency Array[5], The Hydrogen Epoch of Reionization Array (HERA)[6], the Precision Array for Probing the Epoch of Reionization (PAPER)[7], and the Square Kilometer Array[8].

Similarly, the Ly$\alpha$ transition has also revolutionized observational cosmology: observations of the Ly$\alpha$ forest in quasar spectra has allowed us to measure the matter distribution throughout the Universe with unprecedented accuracy. The Ly$\alpha$ forest still provides an extremely useful probe of cosmology on scales that are not accessible with galaxy surveys, and/or the Cosmic microwave background. The Ly$\alpha$ forest will be covered extensively in the lectures by J.X. Prochaska. So far, the most important contributions to our understanding of the Universe from Ly$\alpha$ have come from studies of Ly$\alpha$ absorption. However, with the commissioning of many new instruments and telescopes, there is tremendous potential for Ly$\alpha$ in emission. Because Ly$\alpha$ is a resonance line, and because typical astrophysical environments are optically thick to Ly$\alpha$, we need to understand the radiative transfer to be able to fully exploit the observations of Ly$\alpha$ emitting sources.

---

[4] http://www.mwatelescope.org/
[5] http://www.lofar.org/
[6] http://reionization.org/
[7] http://eor.berkeley.edu/
[8] https://www.skatelescope.org/

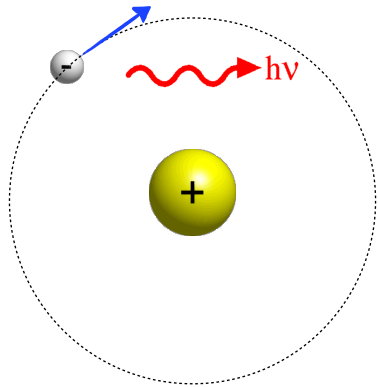

FIG. 2 In the classical picture of the hydrogen atom, an electron orbits a central proton at $v \sim \alpha c$. The acceleration that the electron experiences causes it to emit electromagnetic waves and loose energy. This causes the electron to spiral inwards into the proton on a time-scale of $\sim 10^{-11}$ s. In the classical picture, hydrogen atoms are highly unstable, short-lived objects.

### 3.2. The Hydrogen Atom: The Classical & Quantum Picture

The classical picture of the hydrogen atom is that of an electron orbiting a proton. In this picture, the electrostatic force binds the electron and proton. The equation of motion for the electron is given by

$$\frac{q^2}{r^2} = \frac{m_e v_e^2}{r}, \tag{1}$$

where $q$ denotes the charge of the electron and proton, and the subscript 'e' ('p') indicates quantities related to the electron (proton). The acceleration the electron undergoes thus equals $a_e = \frac{v_e^2}{r} = \frac{q^2}{r^2 m_e}$. When a charged particle accelerates, it radiates away its energy in the form of electromagnetic waves. The total energy that is radiated away by the electron per unit time is given by the Larmor formula, which is given by

$$P = \frac{2}{3} \frac{q^2 a_e^2}{c^3}. \tag{2}$$

The total energy of the electron is given by the sum of its kinetic and potential energy, and equals $E_e = \frac{1}{2} m_e v_e^2 - \frac{q^2}{r} = -\frac{q^2}{2r}$. The total time it takes for the electron to radiative away all of its energy is thus given by

$$t = \frac{E_e}{P} = \frac{3c^3}{4r a_e^2} = \frac{3r^3 m_e^2 c^3}{4q^4} \approx 10^{-11} \text{ s}, \tag{3}$$

where we substituted the Bohr radius for $r$, i.e. $r = a_0 = 5.3 \times 10^{-9}$ cm. In the classical picture, hydrogen atoms would be highly unstable objects, which is clearly problematic and led to the development of quantum mechanics.

In quantum mechanics, electron orbits are *quantized*: In Niels Bohr's model of the atom, electrons can only reside in discrete orbitals. While in such an orbital, the electron does not radiate. It is only when an electron transitions from one orbital to another that it emits a photon. Quantitatively, the total angular momentum of the electron $L \equiv m_e v_e r$ can only taken on discrete values $L = n\hbar$, where $n = 1, 2, ...,$ and $\hbar$ denotes the reduced Planck constant (Table I). The total energy of the electron is then

$$E_e(n) = -\frac{q^4 m_e}{2n^2 \hbar^2} = -\frac{E_0}{n^2}, \tag{4}$$

where $E_0 = 13.6$ eV denotes 1 Rydberg, which corresponds to the binding energy of the electron in its ground state. In quantum mechanics, the total energy of an electron bound to a proton to form a hydrogen atom can only take on a discrete set of values, set by the 'principal quantum number' $n$. The quantum mechanical picture of the hydrogen atom differs from the classical one in additional ways: the electron orbital of a given quantum state (an orbital characterized

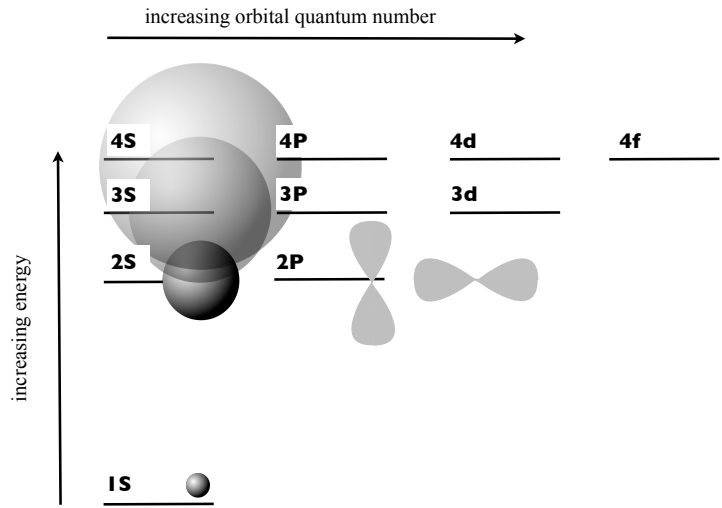

increasing orbital quantum number

increasing energy

| | | | |
|---|---|---|---|
| 4S | 4P | 4d | 4f |
| 3S | 3P | 3d | |
| 2S | 2P | | |

1S

FIG. 3 The total energy of the quantum states of the hydrogen atom, and a simplified representation of the associated quantum mechanical wavefunction describing the electron. The level denoted with '1s' denotes the ground state and has a total energy $E = -13.6$ eV. The wavefunction is spherically symmetric and compact. The extent/size of the wavefunction increases with quantum number $n$. The eccentricity/elongation of the wavefunction increases with quantum number $l$. The orientation of non-spherical wavefunction can be represented by a third quantum number $m$.

by quantum number $n$) does not correspond to the classical orbital described above. Instead, the electron is described by a quantum mechanical wavefunction $\psi(\mathbf{r})$, (the square of) which describes the probability of finding the electron at some location $\mathbf{r}$. The functional form of these wavefunctions are determined by the Schrödinger equation. We will not discuss the Schrödinger equation in these lectures, but will simply use that it implies that the quantum mechanical wavefunction $\psi(\mathbf{r})$ of the electron is characterized fully by *two* quantum numbers: the principal quantum number $n$, and the orbital quantum number $l$. The orbital quantum number $l$ can only take on the values $l = 0, 1, 2, ..., n - 1$. The electron inside the hydrogen atom is fully characterized by these two numbers. The classical analogue of requiring two numbers to characterize the electron wavefunction is that we need two numbers to characterize the classical orbit of the electron around the proton, namely energy $E$ and total angular momentum $L$.

The diagram in Figure 3 shows the total energy of different quantum states in the hydrogen, and a sketch of the associated wavefunctions. This Figure indicates that

- The lowest energy state corresponds to the $n = 1$ state, with an energy of $E = -13.6$ eV. For the state $n = 1$, the orbital quantum number $l$ can only take on the value $l = 0$. This state with $(n, l) = (1, 0)$ is referred to as the '1s'-state. The '1' refers to the value of $n$, while the 's' is a historical way (the 'spectroscopic notation') of labelling the '$l = 0$'-state. This Figure also indicates (schematically) that the wavefunction that describes the 1$s$-state is spherically symmetric. The 'size' or extent of this wavefunction relates to the classical atom size in that the expectation value of the radial position of the electron corresponds to the Bohr radius $a_0$, i.e. $\int dV r |\psi_{1s}(\mathbf{r})|^2 = a_0$.

- The second lowest energy state, $n = 2$, has a total energy $E = E_0/n^2 = -3.4$ eV. For this state there exist two quantum states with $l = 0$ and $l = 1$. The '2s'-state is again characterized by a spherically symmetric wavefunction, but which is more extended. This larger physical extent reflects that in this higher energy state, the electron is more likely to be further away from the proton, completely in line with classical expectations. On the other hand, the wavefunction that describes the '2p'-state ($n = 2$, $l = 1$) is not spherically symmetric, and consists of two 'lobes'. The elongation that is introduced by these lobes can be interpreted as the electron being on an eccentric orbit, which reflects the increase in the electron's orbital angular momentum.

- The third lowest energy state $n = 3$ has a total energy of $E = E_0/n^2 = -1.5$ eV. The size/extent of the orbital/wavefunction increases further, and the complexity of the shape of the orbitals increases with $n$ (see e.g. `https://en.wikipedia.org/wiki/Atomic_orbital` for illustrations). Loosely speaking, the quantum number $n$ denotes the extent/size of the wavefunction, $l$ denotes its eccentricity/elongation. The orientation of non-spherical wavefunction can be represented by a third quantum number $m$.

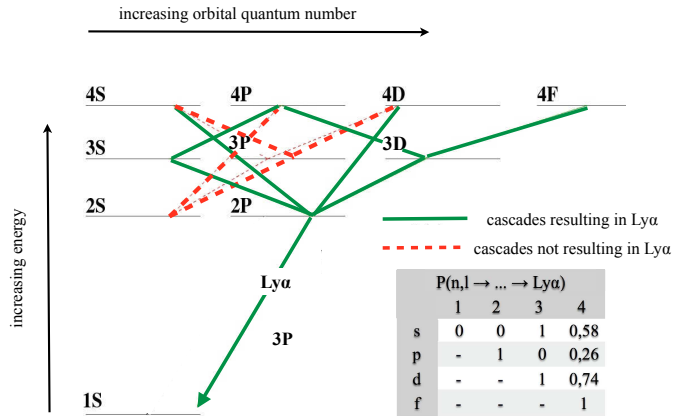

FIG. 4 Atoms in any state with $n > 1$ radiatively cascade back down to the ground (1s) state. The quantum mechanical selection rules only permit transitions where $|\Delta l| = 1$. These transitions are indicated with colored lines connecting the different quantum states. The *green solid lines* indicate radiative cascades that result in the emission of a Ly$\alpha$ photon, while *red dotted lines* indicate transitions that do not. We have omitted all direct radiative transitions $np \rightarrow 1s$: this corresponds to the 'case-B' approximation, which assumes that the recombining gas is optically thick to all Lyman-series photons, and that these photons would be re-absorbed immediately. The table in the lower right corner indicates Ly$\alpha$ production probabilities from various states: e.g. the probability that at atom in the $4s$ state produces a Ly$\alpha$ photon is $\sim 0.58$. *Credit: from Figure 1 of Dijkstra 2014, Lyman Alpha Emitting Galaxies as a Probe of Reionization, PASA, 31, 40D.*

### 3.3. Radiative Transitions in the Hydrogen Atom: Lyman, Balmer, ..., Pfund, .... Series

We discussed how in the classical picture of the hydrogen atom, the electron ends up inside the proton after $\sim 10^{-11}$ s. In quantum mechanics, the electron is only stable in the ground state (1s). The life-time of an atom in any excited state is very short, analogous to the instability of the atom in the classical picture. Transitions between different quantum states have been historically grouped into series, and named after the discoverer of these series. The series include

- **The Lyman series**. A series of radiative transitions in the hydrogen atom which arise when the electron goes from $n \geq 2$ to $n = 1$. The first line in the spectrum of the Lyman series - named Lyman $\alpha$ (hereafter, Ly$\alpha$) - was discovered in 1906 by Theodore Lyman, who was studying the ultraviolet spectrum of electrically excited hydrogen gas. The rest of the lines of the spectrum were discovered by Lyman in subsequent years.

- **The Balmer series**. The series of radiative transitions from $n \geq 3$ to $n = 2$. The series is named after Johann Balmer, who discovered an empirical formula for the wavelengths of the Balmer lines in 1885. The Balmer-$\alpha$ (hereafter H$\alpha$) transition is in the red, and is responsible for the reddish glow that can be seen in the famous Orion nebula.

- Following the Balmer series, we have the **Paschen series** ($n \geq 4 \rightarrow n = 3$), the **Brackett series** ($n \geq 5 \rightarrow 4$), the **Pfund series** ($n \geq 6 \rightarrow 5$), .... Especially Pfund-$\delta$ is potentially an interesting probe (Östlin & Hayes, 2009-2016 private communication).

Quantum mechanics does not allow radiative transitions between just any two quantum states: these radiative transitions must obey the 'selection rules'. The simplest version of the selection rules - which we will use in these lectures - is that only transitions of the form $|\Delta l| = 1$ are allowed. A simple interpretation of this is that photons carry a (spin) angular momentum given by $\hbar$, which is why the angular momentum of the electron orbital must change by $\pm \hbar$ as well. Figure 4 indicates allowed transitions, either as *green solid lines* or as *red dashed lines*. Note that the Lyman-$\beta, \gamma, ...$ transitions ($3p \rightarrow 1s$, $4p \rightarrow 1s$, ...) are not shown on purpose. As we will see later in the lectures, while these transitions are certainly allowed, in realistic astrophysical environments it is better to simply

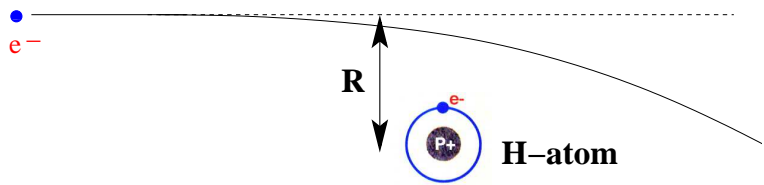

FIG. 5 Cooling radiation at the atomic level: an interaction between an electron and a hydrogen atom can leave the hydrogen atom in an excited state, which can produce a Ly$\alpha$ photon. The energy carried by the Ly$\alpha$ photon comes at the expense of the kinetic energy of the electron. Ly$\alpha$ emission by the hydrogen atom thus cools the gas.

ignore them.

Consider an electron in some arbitrary quantum state $(n, l)$. The electron does not spend much time in this state, and radiatively decays down to a lower energy state $(n', l')$. This lower energy state is again unstable [unless $(n', l') = (1, 0)$], and the electron again radiatively decays to an even lower energy state $(n'', l'')$. Ultimately, all paths lead to the ground state, even those paths that go through the $2s$ state. While the selection rules do not permit transitions of the form $2s \to 1s$, these transitions can occur, if the atom emits *two* photons (rather than one). Because these two-photon transitions are forbidden, the life-time of the electron in the $2s$ state is many orders of magnitude larger than almost all other quantum states (it is $\sim 8$ orders of magnitude longer than that of the $2p$-state), and this quantum state is called 'meta-stable'. The path from an arbitrary quantum state $(n, l)$ to the ground state via a sequence of radiative decays is called a 'radiative cascade'.

The *green solid lines* in Figure 4 show radiative cascades that result in the emission of a Ly$\alpha$ photon. The *red dashed lines* show the other radiative cascades. The table in the *lower right corner* shows the probability that a radiative cascade from quantum state $(n, l)$ produces a Ly$\alpha$ photon. This probability is denoted with $P(n, l \to ... \to \text{Ly}\alpha)$. For example, the probability that an electron in the $2s$ orbital gives rise to Ly$\alpha$ is zero. The probability that an electron in the $3s$ orbital gives Ly$\alpha$ is 1. This is because the only allowed radiative cascade to the ground state from $3s$ is $3s \to 2p \to 1s$. This last transition corresponds to the Ly$\alpha$ transition. For $n \geq 4$ the probabilities become non-trivial, as we have to compute the likelihood of different radiative cascades. We discuss this in more detail in the next section.

### 3.4. Ly$\alpha$ Emission Mechanisms

A hydrogen atom emits Ly$\alpha$ once its electron is in the $2p$ state and decays to the ground state. We mentioned qualitatively how radiative cascades from a higher energy state can give rise to Ly$\alpha$ production. Electrons can end up these higher energy quantum states (any state with $n > 1$) in two different ways:

1. **Collisions.** The 'collision' between an electron and a hydrogen atom can leave the atom in an excited state, at the expense of kinetic energy of the free electron. This process is illustrated in Figure 5. This process converts thermal energy of the electrons, and therefore of the gas as whole, into radiation. This process is also referred to as Ly$\alpha$ production via 'cooling' radiation. We discuss this process in more detail in § 4.1, and in which astrophysical environments it may occur in § 5.

2. **Recombination.** Recombination of a free proton and electron can leave the electron in any quantum state $(n, l)$. Radiative cascades to the ground state can then produce a Ly$\alpha$ photon. As we discussed in § 3.3, we can compute the probability that each quantum state $(n, l)$ produces a Ly$\alpha$ photon during the radiative cascade down to the ground-state. If we sum over all these quantum states, and properly weigh by the probability that the freshly combined electron-proton pair ended up in state $(n, l)$, then we can compute the probability that a recombination event gives us a Ly$\alpha$ photon. We discuss the details of this calculation in § 4.2. Here, we simply discuss the main results.

The *upper panel* of Figure 6 shows the total probability $P(\text{Ly}\alpha)$ that a Ly$\alpha$ photon is emitted per recombination event as a function of gas temperature $T$. This plot contains two lines. The *solid black line* represents 'Case-A', which refers to the most general case where we allow the electron and proton to recombine into *any* state $(n, l)$, and where we allow for all radiative transitions permitted by the selection rules. The *dashed black line* shows 'Case-B', which refers to the case where we do not allow for ($i$) direct recombination into the ground state, which produces an ionizing photon, and (*ii*) radiative transitions of the higher order Lyman series, i.e. Ly$\beta$, Ly$\gamma$, Ly$\delta$,.... Case-B represents that most astrophysical gases efficiently re-absorb higher order Lyman series and ionizing photons, which effectively 'cancels out' these transitions (see § 4.2 for more discussion on this). This Figure shows that for gas at $T = 10^4$ K and case-B recombination, we have $P(\mathrm{Ly}\alpha) = 0.68$. This value '0.68' is often encountered during discussions on Ly$\alpha$ emitting galaxies. It is worth keeping in mind that the probability $P(\mathrm{Ly}\alpha)$ increases with decreasing gas temperature and can be as high as $P(\mathrm{Ly}\alpha) = 0.77$ for $T = 10^3$ K (also see Cantalupo et al. 2008). The *red open circles* represent the following two fitting formulae

$$P_{\mathrm{A}}(\mathrm{Ly}\alpha) = 0.41 - 0.165 \log T_4 - 0.015(T_4)^{-0.44} \tag{5}$$
$$P_{\mathrm{B}}(\mathrm{Ly}\alpha) = 0.686 - 0.106 \log T_4 - 0.009(T_4)^{-0.44},$$

where $T_4 \equiv T/10^4$ K. The fitting formula for case-B is taken from Cantalupo et al. (2008).

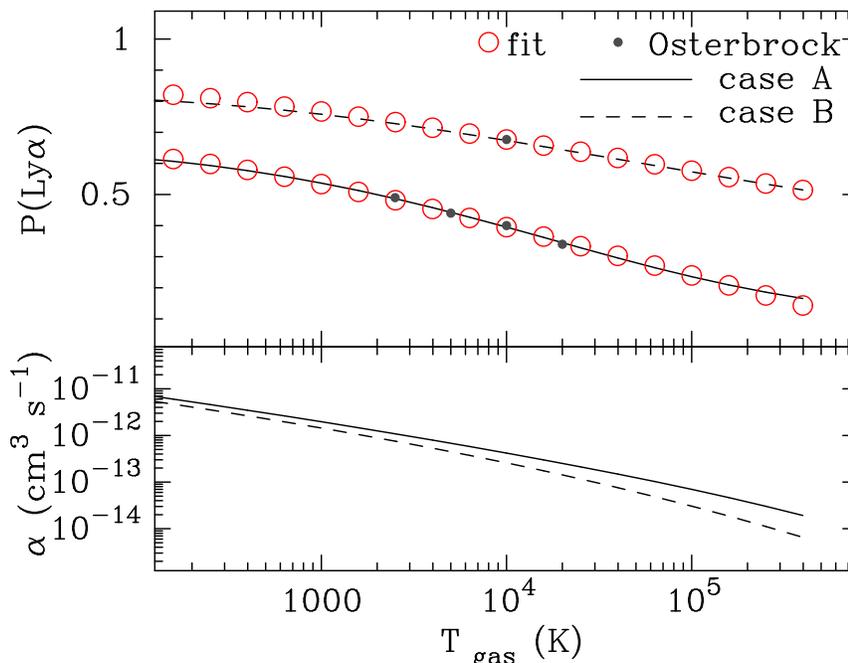

FIG. 6 The *top panel* shows the probability $P(\mathrm{Ly}\alpha)$ that a recombination event leads to the production of a Ly$\alpha$ photon, as a function of gas temperature $T$. The *upper dashed line* (*lower solid line*) corresponds to 'case B' ('case A'). The *lower panel* shows the recombination rate $\alpha(T)$ (in cm$^3$ s$^{-1}$) at which electrons and protons recombine. The *solid line* (*dashed line*) represents case-B (case-A). The *red open circles* represent fitting formulae (Eq 5). *Credit: from Figure 2 of Dijkstra 2014, Lyman Alpha Emitting Galaxies as a Probe of Reionization, PASA, 31, 40D.*

## 4. A CLOSER LOOK AT LY$\alpha$ EMISSION MECHANISMS & SOURCES

The previous section provided a brief description of physical processes that give rise to Ly$\alpha$ emission. Here, we discuss these in more detail, and also link them to astrophysical sources of Ly$\alpha$.

### 4.1. Collisions

Collisions involve an electron and a hydrogen atom. The efficiency of this process depends on the relative velocity of the two particles. The Ly$\alpha$ production rate therefore includes the product of the number density of both species, and the rate coefficient $q_{1s2p}(P[v_e])$ which quantifies the velocity dependence of this process ($P[v_e]$ denotes the velocity distribution of electrons). If we assume that the velocity distribution of electrons is given by a Maxwellian distribution, then $P[v_e]$ is uniquely determined by temperature $T$, and the rate coefficient becomes a function of temperature, $q_{1s2p}(T)$. The total Ly$\alpha$ production rate through collisional excitation is therefore

$$R_{\text{coll}}^{\text{Ly}\alpha} = n_e n_{\text{H}} q_{1s2p} \text{ cm}^{-3} \text{ s}^{-1}. \tag{6}$$

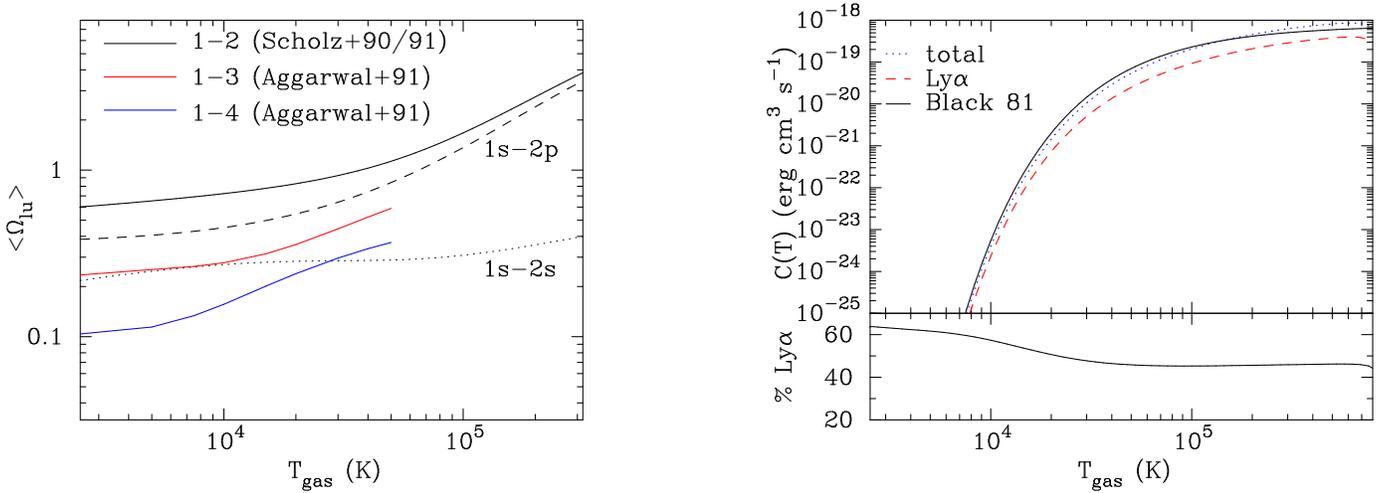

FIG. 7 *Left:* Velocity averaged collision strengths $\langle\Omega_{lu}\rangle$ are plotted as a function of temperature for the transitions $1s \to 2s$ (*dotted line*), $1s \to 2p$ (*dashed line*), and for their sum $1s \to 2$ (*black solid line*) as given by Scholz et al. (1990); Scholz & Walters (1991). Also shown are velocity averaged collision strengths for the $1s \to 3$ (*red solid line*, obtained by summing over all transitions $3s, 3p$ and $3d$), and $1s \to 4$ (*blue solid line*, obtained by summing over all transitions $4s, 4p, 4d$ and $4f$) as given by Aggarwal et al. (1991). Evaluating the collision strengths becomes increasingly difficult towards higher $n$ (see text). *Right:* The *blue-dotted line* in the *top panel* shows the total cooling rate per H-nucleus that one obtaines by collisionally exciting H atoms into all states $n \leq 4$. For comparison, the *red dashed* line shows the total cooling rate as a result of collisional excitation of the 2p state, which is followed by a downward transition through emission of a Ly$\alpha$ photon. All cooling rates increase rapidly around $T \sim 10^4$ K. The *lower panel* shows the ratio (in%) of these two cooling rates. This plot shows that $\sim 60\%$ of the total gas cooling rate is in the form of Ly$\alpha$ photons at $T \sim 10^4$ K, and that this ratio decreases to $\sim 45-50\%$ towards higher gas temperatures. At the gas temperatures at which cooling via line excitation is important, $T \lesssim 10^5$ K (see text), $\sim 45-60\%$ of this cooling emerges as Ly$\alpha$ photons. Also shown for comparison as the *black solid line* is the often used fitting formula of Black (1981), and modified following Cen (1992).

In general, the rate coefficient $q_{lu}$ is expressed[9] in terms of a 'velocity averaged collision strength' $\langle\Omega_{lu}\rangle$ as

$$q_{lu} = \frac{h_P^2}{(2\pi m_e)^{3/2}(k_B T)^{1/2}} \frac{\langle\Omega_{lu}\rangle}{g_l} \exp\left(-\frac{\Delta E_{lu}}{k_B T}\right) = 8.63 \times 10^{-6} T^{-1/2} \frac{\langle\Omega_{lu}\rangle}{g_l} \exp\left(-\frac{\Delta E_{lu}}{k_B T}\right) \text{ cm}^3 \text{ s}^{-1}. \tag{7}$$

Calculating the collision strength is a very complex problem, because for the free-electron energies of interest, the free electron spends a relatively long time near the target atom, which causes distortions in the bound electron's wavefunctions. Complex quantum mechanical interactions may occur, and especially for collisional excitation into higher-n states, multiple scattering events become important (see Bely & van Regemorter, 1970, and references therein). The most reliable collision strengths in the literature are for the $1s \to nL$, with $n < 4$ and $L < d$ (Aggarwal 1983, Scholz et al. 1990, Osterbrock & Ferland 2006). The *left panel* of Figure 7 shows the velocity averaged collision strength for several transitions. There exist some differences in the calculations between these different groups. Collisional excitation rates still appear uncertain at the 10-20% level.

---

[9] The subscripts 'l' and 'u' refer to the 'lower' and 'upper' energy states, respectively

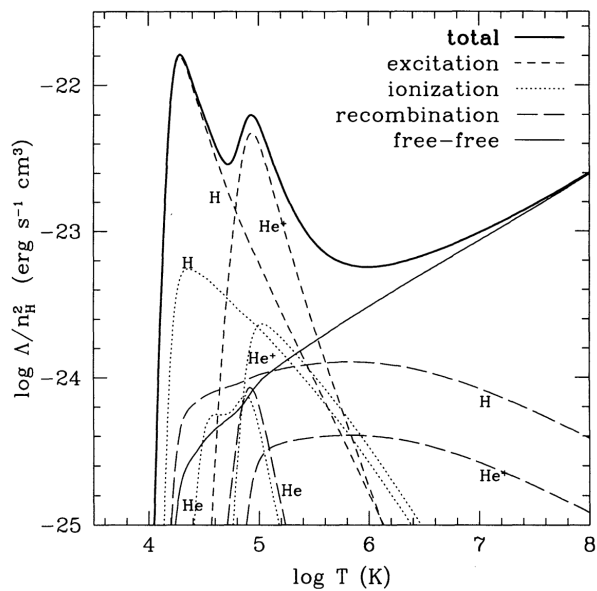

FIG. 8 This figure shows the temperature dependence of the cooling rate of primordial gas, under the assumption of collisional ionization equilibrium (i.e. the ionization state of the gas is set entirely by its temperature). Gas cooling due to collisional excitation of atomic hydrogen becomes important at $\log T/K \sim 4$, and reaches a maximum at $\log T/K \sim 4.2$, beyond which collisions can ionize away atomic hydrogen. At $\log T/K \gtrsim 4.6$ cooling is dominated by collisional excitation of singly ionized helium. Figure taken from Thoul & Weinberg (1995).

As we mentioned above, radiation is produced in collisions at the expense of the gas' thermal energy. The total rate at which the gas looses thermal energy, i.e. cools, per unit volume is

$$\frac{dE_{\text{th}}}{dV dt} = n_e n_{\text{H}} C(T), \tag{8}$$

where

$$C(T) = \sum_u q_{1s \to u} \Delta E_{1s \to u} \text{ erg cm}^3 \text{ s}^{-1}. \tag{9}$$

Here, the sum is over all excited states 'u'. The *blue-dotted line* in the *top right panel* of Figure 7 shows $C(T)$ including collisional excitation into all states $n \leq 4$. The cooling rate rises by orders of magnitude around $T \sim 10^4$ K, and reflects the strong temperature-dependence of the number density of electrons that are moving fast enough to excite the hydrogen atom. For comparison, the *red dashed* line shows the contribution to $C(T)$ from only collisional excitation into the 2p state, which is followed by a downward transition through emission of a Ly$\alpha$ photon. The *lower panel* shows the ratio (in%) of these two rates. This plot shows that $\sim 60\%$ of the total gas cooling rate is in the form of Ly$\alpha$ photons at $T \sim 10^4$ K, and that this ratio decreases to $\sim 45 - 50\%$ towards higher gas temperatures. The *black solid line* is an often-used analytic fitting formula by Black (1981)

$$C(T) = 7.5 \times 10^{-19} \frac{\exp\left(-\frac{118348}{T}\right)}{(1 + T_5^{1/2})} \text{ erg cm}^3 \text{ s}^{-1}. \tag{10}$$

It is good to keep in mind that over the past few decades, the hydrogen collision strengths have changed quite substantially, which can explain the difference between these curves.

The cooling rate per unit volume depends on the product of $C(T)$, $n_e$, and $n_{\text{H}}$, and therefore on the ionization state of the gas. If we also assume that the ionization state of the gas is determined entirely by its temperature (the gas is then said to be in 'collisional ionization equilibrium'), then the total cooling rate per unit volume is a function of temperature only (and *overall* gas density squared). Figure 8 shows that the cooling curve increases dramatically around $T \sim 10^4$ K, which is due to the corresponding increase in $C(T)$ (see Fig 7). The cooling curve reaches a maximum at $\log T \sim 4.2$, which is because at higher $T$ collisional ionization of hydrogen removes neutral hydrogen, which eliminates the collisional excitation cooling channel. For a cosmological mixture of H and He, collisional excitation of singly ionized Helium starts dominating at $\log T \sim 4.6$ (see Fig 8).

### 4.2. Recombination

The capture of an electron by a proton generally results in a hydrogen atom in an excited state $(n, l)$. Once an atom is in a quantum state $(n, l)$ it radiatively cascades to the ground state $n = 1$, $l = 0$ via intermediate states $(n', l')$. The probability that a radiative cascade from the state $(n, l)$ results in a Ly$\alpha$ photon is given by

$$P(n, l \to \text{Ly}\alpha) = \sum_{n', l'} P(n, l \to n', l') P(n', l' \to \text{Ly}\alpha). \tag{11}$$

This may not feel satisfactory, as we still need to compute $P(n', l' \to \text{Ly}\alpha)$, which is the same quantity but for $n' < n$. In practice, we can compute $P(n', l' \to \text{Ly}\alpha)$ by starting at low values for $n'$, and then work towards increasingly high $n$. For example, the probability that a radiative cascade from the $(n, l) = (3, 1)$ state (i.e. the 3p state) produces a Ly$\alpha$ photon is 0, because the selection rules only permit[10] the transitions $(3, 1) \to (2, 0)$ and $(3, 1) \to (1, 0)$. The first transition leaves the H-atom in the 2s state, from which it can only transition to the ground state by emitting two photons (Breit & Teller, 1940). On the other hand, a radiative cascade from the $(n, l) = (3, 2)$ state (i.e. the 3d state) will certainly produce a Ly$\alpha$ photon, since the only permitted cascade is $(3, 2) \to (2, 1) \overset{\text{Ly}\alpha}{\to} (1, 0)$. Similarly, the only permitted cascade from the 3s state is $(3, 0) \to (2, 1) \overset{\text{Ly}\alpha}{\to} (1, 0)$, and $P(3, 0 \to \text{Ly}\alpha) = 1$. For $n > 3$, multiple radiative cascades down to the ground state are generally possible, and $P(n, l \to \text{Ly}\alpha)$ takes on values other than 0 or 1 (see e.g. Spitzer & Greenstein, 1951, for numerical values). Figure 4 also contains a table that shows the probability $P(n', l' \to \text{Ly}\alpha)$ for $n \leq 5$.

When the selection rules permit radiative cascades from a quantum state $(n, l)$ into multiple states $(n', l')$, then this probability is given by the 'branching ratio' which represents the ratio of the decay rate into state $(n', l')$ into all permitted states, i.e.

$$P(n, l \to n', l') = \frac{A_{n, l, n', l'}}{\sum_{n'', l''} A_{n, l, n'', l''}}, \tag{12}$$

in which $A_{n, l, n', l'}$ denotes the Einstein A-coefficient[11] for the $nl \to n'l'$ transition, where the quantum mechanical selection rules only permit transitions for which $|l - l'| = 1$. The probability that a radiative cascade from an arbitrary quantum state $(n, l)$ gives rise to a Ly$\alpha$ photon can be computed once we know the Einstein-coefficients $A_{n, l, n', l'}$.

The probability that an arbitrary recombination event results in a Ly$\alpha$ photon follows naturally, if we know the probability that recombination leaves the atom in state $(n, l)$. That is

$$P(\text{Ly}\alpha) = \sum_{n_{\min}}^{\infty} \sum_{l=0}^{n-1} \frac{\alpha_{nl}(T)}{\alpha_{\text{tot}}(T)} P(n, l \to \text{Ly}\alpha), \tag{15}$$

---

[10] As we referred to in § 3.4, Figure 4 schematically depicts permitted radiative cascades in a four-level H atom. *Green solid lines* depict radiative cascades that result in a Ly$\alpha$ photon, while *red dotted lines* depict radiative cascades that do not yield a Ly$\alpha$ photon.

[11] This coefficient is given by

$$A_{n, l, n', l'} = \frac{64\pi^4 \nu_{ul}^3}{3h_P c^3} \frac{\max(l', l)}{2l + 1} e^2 a_0^2 [M(n, l, n', l')]^2, \tag{13}$$

where fundamental quantities $e$, $c$, $h_P$, and $a_0$ are given in Table I, $h_P \nu_{ul}$ denotes the energy difference between the upper (n,l) and lower (n',l') state. The matrix $M(n, l, n', l')$ involves an overlap integral that involves the radial wavefunctions of the states $(n, l)$ and $(n', l')$:

$$M(n, l, n', l') = \int_0^\infty P_{n,l}(r) r^3 P_{n', l'}(r) dr. \tag{14}$$

Analytic expressions for the matrix $M(n, l, n', l')$ that contain hypergeometric functions were derived by Gordon (1929). For the Ly$\alpha$ transition $M(n, l, n', l') = M(2, 1, 1, 0) = \sqrt{6}(128/243)$ (Hoang-Binh, 1990).

where the first term denotes the fraction of recombination events into the $(n, l)$ state, in which $\alpha_{\rm tot}$ denotes the total recombination coefficient $\alpha_{\rm tot}(T) = \sum_{n_{\rm min}}^{\infty} \sum_{l=0}^{n-1} \alpha_{nl}(T)$. The temperature-dependent state specific recombination coefficients $\alpha_{nl}(T)$ can be found in for example Burgess (1965) and Rubiño-Martín et al. (2006). The value of $n_{\rm min}$ depends on the physical conditions of the medium in which recombination takes place, and two cases bracket the range of scenarios commonly encountered in astrophysical plasmas:

- '*case-A*' recombination: recombination takes place in a medium that is optically thin at all photon frequencies. In this case, direct recombination to the ground state is allowed and $n_{\rm min} = 1$.

- '*case-B*' recombination: recombination takes place in a medium that is opaque to all Lyman series[12] photons (i.e. Ly$\alpha$, Ly$\beta$, Ly$\gamma$, ...), and to ionizing photons that were emitted following direct recombination into the ground state. In the so-called 'on the spot approximation', direct recombination to the ground state produces an ionizing photon that is immediately absorbed by a nearby neutral H atom. Similarly, any Lyman series photon is immediately absorbed by a neighbouring H atom. This case is quantitatively described by setting $n_{\rm min} = 2$, and by setting the Einstein coefficient for all Lyman series transitions to zero, i.e. $A_{np,1s} = 0$.

The probability $P({\rm Ly}\alpha)$ that we obtain from Eq 15 was plotted in Figure 6 assuming case-A (*solid line*) and case-B (*dashed line*) recombination. The temperature dependence comes in entirely through the temperature dependence of the state-specific recombination coefficients $\alpha_{nl}(T)$. As we mentioned earlier, for case-B recombination, we have $P({\rm Ly}\alpha) = 0.68$ at $T = 10^4$ K. It is worth keeping in mind that our calculations technically only apply in a low density medium. For 'high' densities, collisions can 'mix' different $l$-levels at a fixed $n$. In the limit of infinitely large densities, collisional mixing should cause different $l-$levels to be populated following their statistical weigths [i.e $n_{nl} \propto (2l-1)$]. Collisions can be important in realistic astrophysical conditions, as we discuss in more detail in § 5.

## 5. ASTROPHYSICAL LY$\alpha$ SOURCES

Now that we have specified different physical mechanisms that give rise to the production of a Ly$\alpha$ photon, we discuss various astrophysical sites of Ly$\alpha$ production.

### 5.1. Interstellar HII Regions

Interstellar HII regions are the most prominent sources of Ly$\alpha$ emission in the Universe. Hot, (mostly) massive and young stars produce ionizing photons in their atmospheres which are efficiently absorbed in the interstellar medium, and thus create ionized HII regions. Recombining protons and electrons give rise to Ly$\alpha$, H$\alpha$, etc lines. These lines are called 'nebular' lines. One of the most famous nebulae is the Orion nebula, which is visible with the naked eye in the constellation of Orion. The reddish glow is due to the H$\alpha$ line, which at $\lambda = 6536$ Å falls in the middle of the red part of the visual spectrum, that is produced as recombination emission. We showed previously that there is a $P({\rm Ly}\alpha) = 0.68$ probability that a Ly$\alpha$ photon is produced per case-B recombination event at $T = 10^4$ K. A similar analysis can yield the probability that an H$\alpha$ photon is produced is: $P({\rm H}\alpha) \sim 0.45$. The total ratio of the Ly$\alpha$ to H$\alpha$ flux is therefore $\sim 8$. It is interesting to realize that the total Ly$\alpha$ luminosity that is produced in the Orion nebula is almost an order of magnitude larger than the flux contained in the H$\alpha$ line, which is prominently visible in Figure 9.

Recombinations in HII regions in the ISM balance photoionization by ionizing photons produced by the hot stars. The total recombination rate in an equilibrium[13] HII region therefore equals the total photoionization rate in the nebulae, i.e. the total rate at which ionizing photons are absorbed in the HII region. If a fraction $f_{\rm esc}^{\rm ion}$ of ionizing photons is *not* absorbed in the HII region (and hence escapes), then the total Ly$\alpha$ production rate in recombinations is

$$\dot{N}_{{\rm Ly}\alpha}^{\rm rec} = P({\rm Ly}\alpha)(1 - f_{\rm esc}^{\rm ion})\dot{N}_{\rm ion} \approx 0.68(1 - f_{\rm esc}^{\rm ion})\dot{N}_{\rm ion}, \quad \text{case - B, T} = 10^4 \text{ K}, \tag{16}$$

---

[12] At gas densities that are relevant in most astrophysical plasmas, hydrogen atoms predominantly populate their electronic ground state ($n = 1$), and the opacity in the Balmer lines is generally negligible. In theory one can introduce *case-C/D/E/...* recombination to describe recombination in a medium that is optically thick to Balmer/Paschen/Bracket/... series photons.

[13] The condition of equilibrium is generally satisfied in ordinary interstellar HII regions. In expanding HII regions, e.g. those that exist in the intergalactic medium during cosmic reionization (which is discussed later), the total recombination rate is less than the total rate at which ionising photons are absorbed

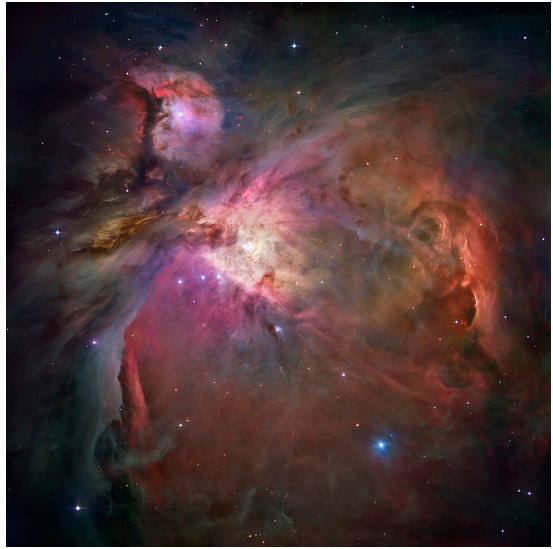

FIG. 9 The Orion nebula is one of the closest nearby nebulae, and visible with the naked eye. The reddish glow is due to Hα that was produced as recombination radiation. These same recombination events produce a Lyα flux that is about 8× larger. Recombination line emission is often referred to as 'nebular' emission. *Credit: ESA/Hubble.*

where $\dot{N}_{\rm ion}$ ($\dot{N}_{\rm Ly\alpha}^{\rm rec}$) denotes the rate at which ionizing (Lyα recombination) photons are emitted. The production rate of ionizing photons, $\dot{N}_{\rm ion}$, relates to the abundance of short-lived massive stars, and therefore closely tracks the star formation rate. For a Salpeter (1955) initial mass function (IMF) with mass limits $M_{\rm low} = 0.1 M_\odot$ and $M_{\rm high} = 100 M_\odot$ we have (Kennicutt, 1998)

$$\dot{N}_{\rm ion} = 9.3 \times 10^{52} \times {\rm SFR}({\rm M}_\odot/{\rm yr}) \ {\rm s}^{-1} \ \Rightarrow L_\alpha = 1.0 \times 10^{42} \times {\rm SFR}({\rm M}_\odot/{\rm yr}) \ {\rm erg \ s}^{-1} \quad ({\rm Salpeter}, \ Z = Z_\odot). \tag{17}$$

Equation 17 strictly applies to galaxies with continuous star formation over timescales of $10^8$ years or longer. This equation is commonly adopted in the literature. A Kroupa IMF gives us a slightly higher Lyα luminosity for a given SFR:

$$L_\alpha = 1.7 \times 10^{42} \times {\rm SFR}({\rm M}_\odot/{\rm yr}) \ {\rm erg \ s}^{-1} \quad ({\rm Kroupa}, \ Z = Z_\odot). \tag{18}$$

Finally, for a fixed IMF the Lyα production rate increases towards lower metallicities: stellar evolution models combined with stellar atmosphere models show that the effective temperature of stars of fixed mass become hotter with decreasing gas metallicity (Tumlinson & Shull 2000, Schaerer 2002). The increased effective temperature of stars causes a larger fraction of their bolometric luminosity to be emitted as ionizing radiation. We therefore expect galaxies that formed stars from metal poor (or even metal free) gas, to be strong sources of nebular emission. Schaerer (2003) provides the following fitting formula for $\dot{N}_{\rm ion}$ as a function of *absolute* gas metallicity[14] $Z_{\rm gas}$

$$\log \dot{N}_{\rm ion} = -0.0029 \times (\log Z_{\rm gas} + 9.0)^{2.5} + 53.81 + \log {\rm SFR}({\rm M}_\odot/{\rm yr}) \quad ({\rm Salpeter}). \tag{19}$$

**Warning**: note that this fitting formula is valid for a Salpeter IMF in the mass range $M = 1 - 100 M_\odot$. If we substitute $Z = Z_\odot = 0.02$ we get log $\dot{N}_{\rm ion} = 53.39$, which is a factor of $\sim 2.6$ times larger than that given by Eq 17. This difference is due to the different lower-mass cut-off of the IMF.

The previous discussion always assumed case-B recombination when converting the production rate of ionizing photon into a Lyα luminosity. However, at $Z \lesssim 0.03 Z_\odot$ departures from case-B *increases* the Lyα luminosity relative to case-B (e.g. Raiter et al., 2010). This increase of the Lyα luminosity towards lower metallicities is due to two effects: (*i*) the increased temperature of the HII region as a result of a suppressed radiative cooling efficiency of metal-poor gas.

---

[14] It is useful to recall that solar metallicity $Z_\odot = 0.02$.

The enhanced temperature in turn increases the importance of collisional processes, which enhances the rate at which collisions excite the $n = 2$ level, and which can transfer atoms from the $2s$ into the $2p$ states; (ii) harder ionizing spectra emitted by metal poor(er) stars. When higher energy photons (say $E_\gamma \gtrsim 50$ eV) photoionize a hydrogen atom, then it releases an electron with a kinetic energy that is $E = E_\gamma - 13.6$ eV. This energetic electron can heat the gas, collisionally excite hydrogen atoms, and collisionally ionize other hydrogen atoms (see e.g. Shull & van Steenberg 1985). Ionizations triggered by energetic electrons created after photoionization by high energy photons are called 'secondary' ionizations. The efficiency with which high energy photons induce secondary ionizations depends on the energy of the photon, and strongly on the ionization state of the gas (see Fig 3 in Shull & van Steenberg 1985): in a highly ionized gas, energetic electrons rapidly loose their energy through Coulomb interactions with other charged particles. A useful figure to remember is that a 1 keV photon can ionize $\sim 25$ hydrogen atoms in a fully neutral medium. Raiter et al. (2010) provide a simple analytic formula which captures these effects:

$$\dot{N}^{\rm rec}_{\rm Ly\alpha} = f_{\rm coll} P (1 - f^{\rm ion}_{\rm esc}) \dot{N}_{\rm ion} \quad (\text{non} - \text{case B}), \tag{20}$$

where $P \equiv \langle E_{\gamma,\rm ion} \rangle / 13.6$ eV, in which $\langle E_{\gamma,\rm ion} \rangle$ denotes the mean energy of ionising photons[15]. Furthermore, $f_{\rm coll} \equiv \frac{1 + a n_{\rm HII}}{b + c n_{\rm HII}}$, in which $a = 1.62 \times 10^{-3}$, $b = 1.56$, $c = 1.78 \times 10^{-3}$, and $n_{\rm HI}$ denotes the number density of hydrogen nuclei. Eq 20 resembles the 'standard' equation, but replaces the factor 0.68 with $P f_{\rm coll}$, which can exceed unity. Eq 20 implies that for a fixed IMF, the Ly$\alpha$ luminosity may be boosted by a factor of a few. Incredibly, for certain IMFs the Ly$\alpha$ line may contain 40% of the total bolometric luminosity[16] of a galaxy.

## 5.2. The Circumgalactic/Intergalactic Medium (CGM/IGM)

Not only nebulae are sources of Ly$\alpha$ radiation. Most of our Universe is in fact a giant Ly$\alpha$ source. Observations of spectra of distant quasars reveal a large collection of Ly$\alpha$ absorption lines. This so-called 'Ly$\alpha$ forest' is discussed in more detail in the lecture notes by X. Prochaska. Observations of the Ly$\alpha$ forest imply that the intergalactic medium is highly ionized, and that the temperature of intergalactic gas is $T \sim 10^4$ K. Observations of the Ly$\alpha$ forest can be reproduced very well if we assume that gas is photoionized by the Universal "ionizing background" that permeates the entire Universe, and that is generated by adding the contribution from all ionizing sources[17]. The *residual* neutral fraction of hydrogen atoms in the IGM is $x_{\rm HI} \equiv \frac{n_{\rm HI}}{n_{\rm HI} + n_{\rm HII}} = \frac{n_e \alpha_{\rm B}(T)}{\Gamma_{\rm ion}}$, where $\Gamma_{\rm ion}$ denotes the photoionization rate by the ionizing background (with units s$^{-1}$).

From § 4.2 we know that each recombination event in the gas produces $\sim 0.68$ Ly$\alpha$ photons. First, we note that the *recombination time* of a proton and electron in the intergalactic medium is

$$t_{\rm rec} \equiv \frac{1}{\alpha_{\rm B}(T) n_e} = 9.3 \times 10^9 \left( \frac{1+z}{4} \right)^{-3} \left( \frac{1+\delta}{1} \right)^{-1} T_4^{0.7} \text{ yr}, \tag{22}$$

where we used that fully ionized gas at mean density of the Universe, $\delta = 0$, has $n_e = \bar{n} = \frac{\Omega_{\rm m} \rho_{\rm crit,0}}{\mu m_{\rm p}} (1+z)^3 \sim 2 \times 10^{-7} (1+z)^3$ cm$^{-3}$, in which $\rho_{\rm crit,0} = 1.88 \times 10^{-29} h^2$ ($H_0 \equiv 100h$ km s$^{-1}$ Mpc$^{-1}$) denotes the critical density of the Universe today. We also approximated the case-B recombination coefficient as $\alpha_{\rm B}(T) = 2.6 \times 10^{-13} T_4^{-0.7}$ cm$^3$ s$^{-1}$, in which we have adopted the notation $T_4 \equiv (T/10^4$ K$)$. We can compare this number to the Hubble time which is $t_{\rm Hub} \equiv \frac{1}{H(z)} \sim 3([1+z]/4)^{-3/2}$ Gyr (where the last approximation is valid strictly for $z \gg 1$). The recombination time for gas, at mean density ($\delta = 0$) thus exceeds the Hubble time for most of the existence of the Universe. Only at

---

[15] That is, $\langle E_{\gamma,\rm ion} \rangle \equiv h_{\rm P} \frac{\int_{13.6 \text{ eV}}^{\infty} d\nu f(\nu)}{\int_{13.6 \text{ eV}}^{\infty} d\nu f(\nu)/\nu}$, where $f(\nu)$ denotes the flux density.

[16] Another useful measure for the 'strength' of the Ly$\alpha$ line is the *equivalent width* (EW, which was discussed in much more detail in the lectures by J.X. Prochaska) of the line:

$$\text{EW} \equiv \int d\lambda \ (F(\lambda) - F_0)/F_0, \tag{21}$$

which measures the total line flux compared to the continuum flux density just redward (as the blue side can be affected by intergalactic scattering, see § 10.2) of the Ly$\alpha$ line, $F_0$. For a Salpeter IMF in the range $0.1 - 100 M_\odot$, $Z = Z_\odot$, the UV-continuum luminosity density, $L_\nu^{\rm UV}$, relates to SFR as $L_\nu^{\rm UV} = 8 \times 10^{27} \times {\rm SFR}(M_\odot/{\rm yr})$ erg s$^{-1}$ Hz$^{-1}$. The corresponding equivalent width of the Ly$\alpha$ line would be EW$\sim 70$ Å (Dijkstra & Westra 2010). The equivalent width can reach a few thousand Å for Population III stars/galaxies forming stars with a top-heavy IMF (see Raiter et al. 2010).

[17] All sources within a radius equal to the mean free path of ionizing photons, $\lambda_{\rm ion}$. For more distant sources ($r > \lambda_{\rm ion}$), the ionizing flux is reduced by an additional factor $\exp(-r/\lambda_{\rm ion})$.

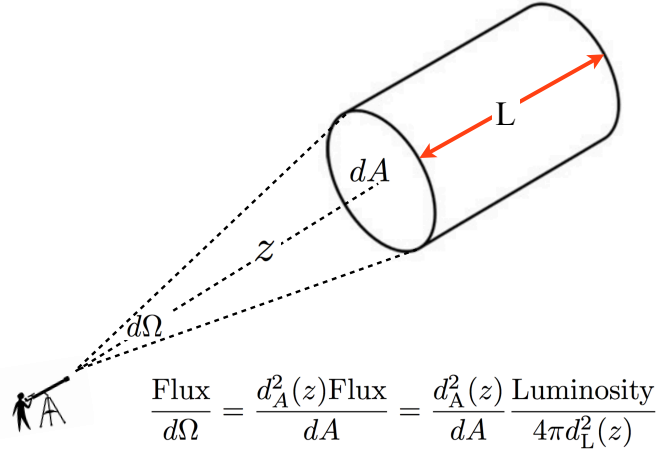

$$\frac{\text{Flux}}{d\Omega} = \frac{d_A^2(z)\text{Flux}}{dA} = \frac{d_A^2(z)}{dA}\frac{\text{Luminosity}}{4\pi d_L^2(z)}$$

FIG. 10 The adopted geometry for calculating the Lyα surface brightness of recombining gas in the IGM at redshift $z$.

$z \gtrsim 8$ does the recombination time become shorter than the Hubble time. However, at lower redshifts recombination (of course) still happens, and we expect the Universe as a whole to emit Lyα recombination radiation (see e.g. Martin et al. 2014).

Lyα emission from the CGM/IGM differs from interstellar (nebular) Lyα emission in two ways: (*i*) Lyα emission from the CGM/IGM occurs over a spatially extended region. Spatially extended Lyα emission is better characterized by its *surface brightness* (flux per unit area on the sky) than by its overall flux. We present a general formalism for computing the Lyα surface brightness below; and (*ii*) Lyα emission is powered by *external* sources. For example, recombination radiation from the CGM/IGM balances photoionization by either the ionizing background (generate by a large numbers of star forming galaxies and AGN), or a nearby source. This conversion of externally - i.e. *not* within the same galaxy (or even cloud) - generated ionizing radiation into Lyα is known as '*fluorescence*' (e.g. Hogan & Weymann 1987, Gould & Weinberg 1996). Fluorescence generally corresponds to emission of radiation by some material following absorption by radiation at some other wavelength. Certain minerals emit radiation in the optical when irradiated by UV radiation. Fluorescent materials cease to glow immediately when the irradiating source is removed. In the case of Lyα, fluorescently produced Lyα is a product of recombinaton cascade, just as the case of nebular Lyα emission.

Here, we present the general formalism for computing the Lyα surface brightness level. The total flux 'Flux' from some redshift $z$ that we receive per unit solid angle $d\Omega$ equals (see Fig 10 for a visual illustration of the adopted geometry)

$$\frac{\text{Flux}}{d\Omega} = \frac{d_A^2(z)\text{Flux}}{dA} = \frac{d_A^2(z)}{dA}\frac{\text{Luminosity}}{4\pi d_L^2(z)} \tag{23}$$

where we used the definition of solid angle $d\Omega \equiv dA/d_A^2(z)$, in which $d_A(z)$ denotes the angular diameter distance to redshift $z$. We also used that Flux = Luminosity$/[4\pi d_L^2(z)]$, in which $d_L(z)$ denotes the luminosity distance to redshift $z$. We know that $d_L(z) = d_A(z)(1+z)^2$. Furthermore, the total luminosity that we receive from $dA$ depends on the length of the cylinder $L$, as Luminosity $= \epsilon_{\text{Ly}\alpha} \times dA \times L$. Here, $\epsilon_{\text{Ly}\alpha}$ denotes the Lyα emission per unit volume. Plugging all of this into Eq (23) we get

$$\frac{\text{Flux}}{d\Omega} = \frac{\epsilon_{\text{Ly}\alpha}L}{4\pi(1+z)^4} \equiv S, \tag{24}$$

where we defined $S$ to represent surface brightness. We compute the surface brightness for a number of scenarios next:

1. **Recombination in the diffuse, low density IGM.** First, we compute the surface brightness of Lyα from the diffuse, low density, IGM, denoted with $S_{\text{IGM}}$. The expansion of the Universe causes photons that are emitted over

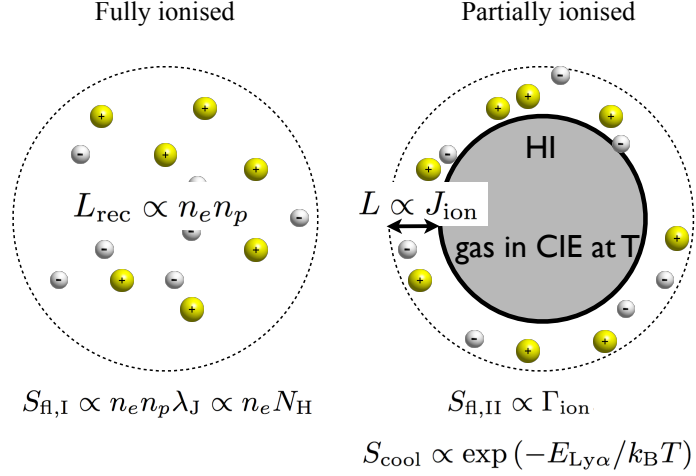

FIG. 11 A schematic representation of the dominant processes that give rise to extended Lyα emission in the CGM/IGM. *Left:* the surface brightness from fluorescence from recombination in fully ionized gas scales as $S_{\mathrm{fl,I}} \propto n_e n_p \lambda_{\mathrm{J}}$, where $\lambda_{\mathrm{J}}$ denotes the Jeans length (see Eq 28). This surface brightness can also be written as $S_{\mathrm{fl,I}} \propto n_e N_{\mathrm{H}}$ where $N_{\mathrm{H}}$ denotes the total column of ionized gas (see Eq 27). *Right:* sufficiently dense clouds can self-shield and form a neutral core of gas, surrounded by an ionized 'skin'. The thickness of this skin is set by the mean free path of ionizing photons, $\lambda_{\mathrm{mfp}}$. The surface brightness of recombination emission that occurs inside this skin is $S_{\mathrm{fl,II}} \propto \Gamma_{\mathrm{ion}}$, where $\Gamma_{\mathrm{ion}}$ denotes the photoionization rate (see Eq 31). The neutral core can also produce Lyα radiation following collisional excitation. The surface brightness of this emission is $S_{\mathrm{cool}} \propto \exp\left(-E_{\mathrm{Ly\alpha}}/k_{\mathrm{B}}T\right)$.

a line-of-sight length $L$ to be spread out in frequency by an amount

$$\frac{d\nu_\alpha}{\nu_\alpha} = \frac{dv}{c} = \frac{H(z)L}{c} \Rightarrow L = \frac{cd\nu_\alpha}{\nu_\alpha H(z)}. \tag{25}$$

When we substitute this into Eq 24 we get

$$S_{\mathrm{IGM}} = \frac{c\epsilon_{\mathrm{Ly\alpha}}}{4\pi(1+z)^4 H(z)} \frac{d\nu_\alpha}{\nu_\alpha} \approx 10^{-21}\left(\frac{1+z}{4}\right)^{0.5}\left(\frac{1+\delta}{1}\right)^2\left(\frac{d\nu_\alpha/\nu_\alpha}{0.1}\right)T_4^{-0.7} \; \mathrm{erg\ s^{-1}\ cm^{-2}\ arcsec^{-2}} \quad \mathrm{for}\ z \gg 1, \tag{26}$$

where we obtained numerical values by substituting that for recombining gas we have $\epsilon_{\mathrm{Ly\alpha}} = 0.68 h_p \nu_\alpha n_e n_p \alpha_{\mathrm{B}}(T)$ (see § 4.2, and that $n_e = n_p \propto (1+z)^3$). We adopted $d\nu \sim 0.1\nu$, which represents the width of narrow-band surveys, which have been adopted in searches for distant Lyα emitters. For comparison, stacking analyses and deep exposure with e.g. MUSE go down to $SB \sim 10^{-19}\ \mathrm{erg\ s^{-1}\ cm^{-2}\ arcsec^{-2}}$ (e.g. Rauch et al. 2008, Steidel et a. 2010, Matsuda et al. 2012, Momose et al. 2014, Wisotzksi et al. 2016, Xue et al. 2017). Directly observing recombination radiation from a representative patch of IGM is still well beyond our capabilities, but would be fantastic as it would allow us to map out the distribution of baryons throughout the Universe.

2. **Fluorescence from recombination in fully ionized dense gas (Fluorescence case I).** Eq 26 shows that the surface brightness increases as $(1+\delta)^2$, and the prospects for detection improve dramatically for overdensities of $\delta \gg 1$. The largest overdensities of intergalactic gas are found in close proximity to galaxies. We therefore expect recombining gas in close proximity to galaxies to be potentially visible in Lyα. Note however that this denser gas in close proximity to galaxies is not comoving with the Hubble flow, and Eq 25 cannot be applied. Instead, we need to specify the line-of-sight size of the cloud $L$. Schaye (2001) has shown that the characteristic size of overdense, growing perturbations is the local Jeans length $\lambda_{\mathrm{J}} \equiv \frac{c_s}{\sqrt{G\rho}} \approx 145([1+z]/4)^{-3/2}(1+\delta)^{-1/2}T_4^{1/2}$ kpc. If we adopt that $L = \lambda_J$, then we obtain[18]

$$S_{\text{fl,I}} = \frac{\epsilon_{\text{Ly}\alpha} \lambda_J}{4\pi(1+z)^4} \approx 1.5 \times 10^{-21} \left(\frac{1+z}{4}\right)^{1/2} \left(\frac{1+\delta}{100}\right)^{3/2} T_4^{-0.2} \text{ erg s}^{-1} \text{ cm}^{-2} \text{ arcsec}^{-2}. \tag{28}$$

For higher densities the enhanced recombination efficiency of the gas gives rise to an enhanced equilibrium neutral fraction, i.e. $x_{\text{HI}} \propto n_e$. With an enhanced neutral fraction, the gas more rapidly 'builds up' a neutral column density $N_{\text{HI}} \gtrsim 10^{-17} \text{ cm}^{-1}$, above which the cloud starts self-shielding against the ionizing background. Quantitatively, if gas is photoionized at a rate $\Gamma_{\text{ion}}$, then the total column density of neutral hydrogen through the cloud is

$$N_{\text{HI}} = \lambda_J(\delta) \times n_H \frac{\alpha_B(T)n_e}{\Gamma_{\text{ion}}} \approx 6 \times 10^{17} \left(\frac{1+\delta}{10^3}\right)^{3/2} \left(\frac{\Gamma_{\text{ion}}}{10^{-12} \text{ s}^{-1}}\right)^{-1} \left(\frac{1+z}{4}\right)^{9/2} T_4^{-0.7} \text{ cm}^{-2}, \tag{29}$$

where we used that $n_{\text{HI}} = x_{\text{HI}} n_H = \frac{\alpha_B(T)n_e}{\Gamma_{\text{ion}}}$. The gas becomes self-shielding when $N_{\text{HI}} \gtrsim \sigma_{\text{ion}}^{-1} \sim 10^{17} \text{ cm}^{-2}$, which translates to $\delta \gtrsim 500(\Gamma_{\text{ion}}/10^{-12} \text{ s}^{-1})^{2/3}([1+z]/4)^{-3} T_4^{0.47}$ (Schaye 2001, also see Rahmati et al. 2013 for a much more extended discussion on self-shielding gas). Once gas become denser than this, it can self-shield against an ionizing background, and form a neutral core surrounded by an ionized 'skin' (see Fig 11).

3. **Fluorescence from recombination from the skin of dense clouds (Fluorescence case II).** For dense clouds that are capable of self-shielding, only the ionized 'skin' emits recombination radiation. The total surface brightness recombination radiation from this skin depends on its density and thickness, the latter depending directly on the amplitude of the ionizing background. This can be most clearly seen from Eq 24, and replacing $L = \lambda_{\text{mfp}}$, where $\lambda_{\text{mfp}}$ denotes the mean free path of ionizing photons into the cloud. This mean free path is given by

$$\lambda_{\text{mfp}} = \frac{1}{n_{\text{HI}}\sigma_{\text{ion}}} \Rightarrow \lambda_{\text{mfp}} = \frac{1}{\sigma_{\text{ion}} x_{\text{HI}} n_H} = \frac{\Gamma_{\text{ion}}}{\sigma_{\text{ion}} \alpha_B(T) n_e^2}, \tag{30}$$

where we used that $x_{\text{HI}} = \alpha_B(T)n_e/\Gamma_{\text{ion}}$. Substituting $\lambda_{\text{mfp}}$ for $L$ in Eq 24 gives for the surface brightness of the skin:

$$S_{\text{fl,II}} = \frac{0.68 h_p \nu_\alpha \Gamma_{\text{ion}}}{4\pi(1+z)^4 \sigma_{\text{ion}}} \approx 1.3 \times 10^{-20} \left(\frac{\Gamma_{\text{ion}}}{10^{-12} \text{ s}^{-1}}\right) \left(\frac{1+z}{4}\right)^{-4} \text{ erg s}^{-1} \text{ cm}^{-2} \text{ arcsec}^{-2}, \tag{31}$$

where the $T$-dependence has cancelled out. A more precise calculation of the surface brightness of fluorescent Ly$\alpha$ emission, which takes into account the spectral shape of the ionizing background as well as the frequency dependence of $\lambda_{\text{mfp}}$, is presented by Cantalupo et al. (2005, note that this calculation introduces only a minor change to the calculated surface brightness).

4. **'Cooling' by dense, neutral gas.** The neutral core of the cloud - the part of the cloud which is truly self-shielded - produces Ly$\alpha$ radiation through collisional excitation. As we mentioned earlier, the rate at which Ly$\alpha$ is produced in collisions depends sensitively on temperature ($\propto \exp(-E_{\text{Ly}\alpha}/k_b T)$, see Eq 7). Note however, that this is a cooling process, and thus must balance some heating mechanism. Once we know the heating rate of the gas, we can almost immediately compute the Ly$\alpha$ production rate.

The direct environment of galaxies, also known as the 'circum galactic medium' (CGM), represents a complex mixture of hot and cold gas, of metal poor gas that is being accreted from the intergalactic medium and metal enriched gas that is driven out of either the central, massive galaxy or from the surrounding lower mass satellite galaxies. Figure 12 shows a snap-shot from a cosmological hydrodynamical simulation (Agertz et al., 2009) which nicely illustrates this complexity. A disk galaxy (total baryonic mass $\sim 2 \times 10^{10} M_\odot$) sits in the center of the snap-shot, taken

---

[18] Another way to express $S_{\text{fl,I}}$ is by replacing $Ln_p = N_H$, where $N_H$ denotes the total column density of hydrogen ions (i.e. protons), which yields (see Hennawi et al. 2015)

$$S_{\text{fl,I}} = \frac{0.68 h_p \nu_\alpha n_e \alpha_B(T) N_H}{4\pi(1+z)^4} \approx 2.0 \times 10^{-21} \left(\frac{1+z}{4}\right)^{-4} \left(\frac{n_e}{10^{-3} \text{ cm}^{-3}}\right) \left(\frac{N_H}{10^{20} \text{ cm}^{-2}}\right) T_4^{-0.7} \text{ erg s}^{-1} \text{ cm}^{-2} \text{ arcsec}^{-2}. \tag{27}$$

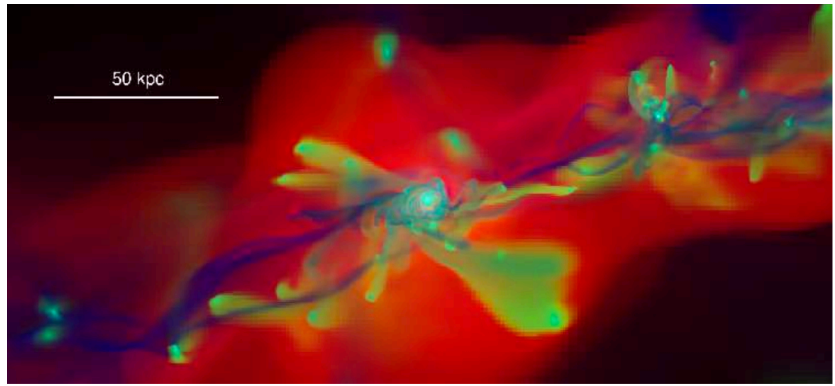

FIG. 12 A snap-shot from a cosmological hydrodynamical simulation by Agertz et al. (2009) which illustrates the complexity of the circumgalactic gas distribution. A disk galaxy sits in the center of the snap-shot, taken at $z = 3$. The *blue filaments* show dense gas that is being accreted. The *red gas* has been shock heated to the virial temperature of the dark matter halo hosting this galaxy ($T_{vir} \sim 10^6$ K). The green clouds show metal rich gas that was stripped from smaller galaxies. This complex mixture of circumgalactic gas produces Ly$\alpha$ radiation through recombination, cooling, and fluorescence. *Credit: from Figure 1 of Agertz et al. 2009, Disc formation and the origin of clumpy galaxies at high redshift, MNRAS, 397L, 64A.*

at $z = 3$. The blue filaments show dense gas that is being accreted. This gas is capable of self-shielding. The red gas has been shock heated to the virial temperature ($T_{vir} \sim 10^6$ K) of the dark matter halo hosting this galaxy. The green clouds show metal rich gas that was driven out of smaller galaxies. This complex mixture of gas produces Ly$\alpha$ via all channels described above: there exists fully ionized gas that is emitting recombination radiation with a surface brightness given by Eq 28, the densest gas is capable of self-shielding and will emit both recombination and cooling radiation.

We currently have observations of Ly$\alpha$ emission from the circum-galactic medium at a range of redshifts, covering a range of surface brightness levels:

1. Ly$\alpha$ emission extends further the UV continuum in nearby star forming galaxies. The *left panel* of Figure 13 shows an example of a false-color image of galaxy # 1 from the **L**yman **A**lpha **R**eference **S**ample ( Östlin et al. 2014, Guaita et al. 2015). In this image, *red* indicates H$\alpha$, *green* traces the far-UV continuum, while *blue* traces the Ly$\alpha$.

2. Stacking analyses have revealed the presence of spatially extended Ly$\alpha$ emission around Lyman Break Galaxies (Hayashino et al., 2004; Steidel et al., 2010) and Ly$\alpha$ emitters at surface brightness levels in the range SB$\sim 10^{-19} - 10^{-18}$ erg s$^{-1}$ cm$^{-2}$ arcsec$^{-2}$ (Matsuda et al., 2012; Momose et al., 2014).

3. Deep imaging with MUSE has now revealed emission at this level around individual star forming galaxies, which further confirms that this emission is present ubiquitously (Wisotzki et al., 2016).

4. These previously mentioned faint halos are reminiscent of Ly$\alpha$ 'blobs', which are spatially extended Ly$\alpha$ sources *not* associated with radio galaxies (more on these next, Francis et al. 1996, Steidel et al. 2000, Matsuda et al. 2004, Dey et al. 2005, Matsuda et al. 2011, Prescott et al. 2012). A famous example of "blob # 1" is shown in the *right panel* of Figure 13 (from Matsuda et al. 2011). This image shows a 'pseudo-color' image of a Ly$\alpha$ blob. The *red* and *blue* really trace radiation in the red and blue filters, while the *green* traces the Ly$\alpha$. The *upper right panel* shows how large the Andromeda galaxy would look on the sky if placed at $z = 3$, to put the size of the blob in perspective. The brightest Ly$\alpha$ blobs have line luminosities of $L_\alpha \sim 10^{44}$ erg s$^{-1}$ (e.g. Dey et al. 2005), though recently several monstrous blobs have been discovered, that are much brighter than this, including the 'Slug' nebula with a Ly$\alpha$ luminosity of $L_\alpha \sim 10^{45}$ erg s$^{-1}$ (Cantalupo et al. 2014, also see Cai et al. 2017 for a similar monster), and the 'Jackpot' nebula, which has a luminosity of $L_\alpha \sim 2 \times 10^{44}$ erg s$^{-1}$ (and contains a quadruple-quasar system, Hennawi et al. 2015).

5. The most luminous Ly$\alpha$ nebulae have traditionally been associated (typically) with High-redshift Radio Galaxies (HzRGs, e.g. McCarthy 1993, Reuland et al. 2003, Van Breugel et al. 2006) with luminosities in excess of $L \sim 10^{45}$ erg s$^{-1}$.

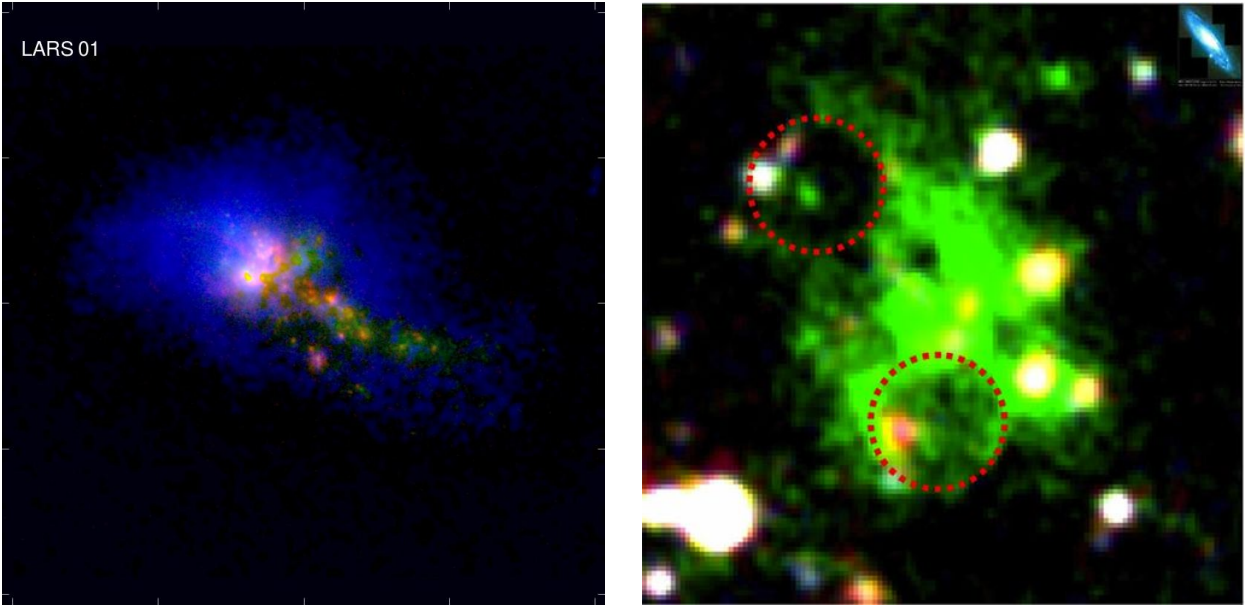

FIG. 13 *Left:* A false color image of 'LARS1' (galaxy #1 from the Lyα Reference Sample, *Credit: from Figure 1 of Hayes et al. 2013 ©AAS. Reproduced with permission.*). *Red* indicates Hα emission, while *green* traces far-UV continuum. The *blue light* traces the Lyα which extends much further than other radiation. *Right:* A pseudo color image of Lyα blob 1 (LAB1). Here, *red* and *blue* light traces emission from the *V* and *B* bands, respectively. The *green light* traces Lyα emission (*Credit: from Figure 2 of Matsuda et al. 2011, The Subaru Lyα blob survey: a sample of 100-kpc Lyα blobs at z= 3, MNRAS, 410L, 13M*). For comparison, the *upper right* shows the Andromeda galaxy to get a sense for the scale of 'giant' Lyα blobs.

The origin of extended Lyα emission is generally unclear. To reach surface brightness levels of $\sim 10^{-19} - 10^{-17}$ erg s$^{-1}$ cm$^{-2}$ arcsec$^{-2}$ we need a density exceeding (see Eq 28)

$$\delta \gtrsim \left(1.5 \times 10^3 - 3.5 \times 10^4\right) \times \left(\frac{1+z}{4}\right)^{-1/3} T_4^{-0.13} \tag{32}$$

To keep this gas photoionized requires a large $\Gamma_{\rm ion}$:

$$\Gamma_{\rm ion} \gtrsim (5 - 600) \times 10^{-12} \left(\frac{1+z}{4}\right)^{4.5} T_4^{-0.7} \ {\rm s}^{-1}, \tag{33}$$

which is $\sim 10 - 10^3$ times larger than the values inferred from observations of the Lyα forest at this redshift. In case the gas starts to self-shield, then Eq 31 shows that in order to reach $S \sim 10^{-19} - 10^{-17}$ erg s$^{-1}$ cm$^{-2}$ arcsec$^{-2}$ we need an almost identical boost in $\Gamma_{\rm ion} \sim 10^{-11} - 10^{-9}$ s$^{-1}$. This enhanced intensity of the ionizing radiation field is expected in close proximity to ionizing sources (e.g. Mas-Ribas & Dijkstra, 2016): the photoionization rate at a distance $r$ from a source that emits $\dot{N}$ion ionizing photons per second is

$$\Gamma = \dot{N}_{\rm ion} \frac{\sigma_{\rm ion} f_{\rm esc}^{\rm ion}}{4\pi r^2} = 5 \times 10^{-10} \left(\frac{f_{\rm esc}^{\rm ion}}{1.0}\right) \left(\frac{\dot{N}_{\rm ion}}{10^{54} {\rm s}^{-1}}\right) \left(\frac{r}{10 \ {\rm kpc}}\right)^{-2}, \tag{34}$$

where $f_{\rm esc}^{\rm ion}$ denotes the fraction of ionizing photons that escapes from the central source into the environment. The production rate of ionizing photons can be linked to the star formation rate via Eq 19. Alternatively, ionizing radiation may be powered by an accretion disk surrounding a black hole of mass $M_{\rm BH}$. Assuming Eddington accretion onto the black hole, and adopting a template spectrum of a radio-quiet quasar, we have

$$\dot{N}_{\rm ion} = 6.5 \times 10^{53} \left(\frac{M_{\rm BH}}{10^6 \ M_\odot}\right) \ {\rm s}^{-1} \tag{35}$$

which assumes a broken power-law spectrum of the form $f_\nu \propto \nu^{-0.5}$ for 1050 Å$< \lambda < 1450$ Å, and $f_\nu \propto \nu^{-1.5}$ for $\lambda < 1050$ Å (Bolton et al., 2011). It therefore seems that fluorescence can explain the observed values of the surface brightness in close proximity to ionizing sources. An impressive recent example of this process is described by Borisova et al. (2016), who found luminous, spatially extended Ly$\alpha$ halos around *each* of the 17 brightest radio-quiet quasars with MUSE (also see North et al. 2012 for earlier hints of the presence of extended Ly$\alpha$ halos around a high fraction of radio-quiet quasars, and see Haiman & Rees 2001 for an early theoretical prediction).

Finally, Ly$\alpha$ cooling radiation gives rise to spatially extended Ly$\alpha$ radiation (Haiman et al. 2000, Fardal et al. 2001), and provides a possible explanation for Ly$\alpha$ 'blobs' (Dijkstra & Loeb 2009, Goerdt et al. 2010, Faucher-Giguère et al. 2010, Rosdahl & Blaizot 2012, Martin et al. 2015). In these models, the Ly$\alpha$ cooling balances 'gravitational heating' in which gravitational binding energy is converted into thermal energy in the gas. Precisely how gravitational heating works is poorly understood. Haiman et al. (2000) propose that the gas releases its binding energy in a series of 'weak' shocks as the gas navigates down the gravitational potential well. These weak shocks convert binding energy into thermal energy over a spatially extended region[19], which is then reradiated primarily as Ly$\alpha$. We must therefore accurately know and compute all the heating rates in the ISM (Faucher-Giguère et al. 2010, Cantalupo et al. 2012, Rosdahl & Blaizot 2012) to make a robust prediction for the Ly$\alpha$ cooling rate. These heating rates include for example photoionization heating, which requires coupled radiation-hydrodynamical simulations (as Rosdahl & Blaizot, 2012), or shock heating by supernova ejecta (e.g. Shull & McKee, 1979).

The previous discussion illustrates that it is possible to produce spatially extended Ly$\alpha$ emission from the CGM at levels consistent with observations, via all mechanisms described in this section[20]. This is one of the main reasons why we have not solved the question of the origin of spatially extended Ly$\alpha$ halos yet. In later lectures, we will discuss how Ly$\alpha$ spectral line profiles (and polarization measurements) contain physical information on the scattering/emitting gas, which can help distinguish between different scenarios.

## 6. STEP 1 TOWARDS UNDERSTANDING LY$\alpha$ RADIATIVE TRANSFER: LY$\alpha$ SCATTERING CROSS-SECTION

The goal of this section is to present a classical derivation of the Ly$\alpha$ absorption cross-section. This classical derivation gives us the proper functional form of the real cross-section, but that differs from the real expression by a factor of order unity, due to a quantum mechanical correction. Once we have evaluated the magnitude of the cross-section, it is apparent that most astrophysical sources of Ly$\alpha$ emission are optically thick to this radiation, and that we must model the proper Ly$\alpha$ radiative transfer.

The outline of this section is as follows: we first describe the interaction of a free electron with an electromagnetic wave (i.e. radiation) in the classical picture (see § 6.1). This discussion provides us with an opportunity to introduce the important concepts of the cross-section and phase function. This discussion also sets us up for discussion of the same interaction in § 6.2 for an electron that is *bound* to a proton. The classical picture of this interaction allows us to derive the Ly$\alpha$ absorption cross-section up to a numerical factor of order unity (see § 6.3). We introduce the velocity averaged Voigt-profile for the Ly$\alpha$ cross-section in § 6.4.

### 6.1. Interaction of a Free Electron with Radiation: Thomson Scattering

Figure 14 shows the classical view of the interaction of a free electron with an incoming electro-magnetic wave. The electromagnetic wave consists of an electric field (represented by the *red arrows*) and a magnetic field (represented by the *blue arrows*). The amplitude of the electric field at time $t$ varies as $E(t) = E_0 \sin \omega t$, where $\omega$ denotes the angular frequency of the wave. The electron is accelerated by the electric field by an amount $|a_e|(t) = \frac{q|E|(t)}{m_e}$. The total power radiated by this electron is given by the Larmor formula, i.e. $P_{\rm out}(t) = \frac{2q^2 |a_e(t)|^2}{3c^3}$. The *time average* of this radiated power equals $\langle P_{\rm out} \rangle = \frac{2q^4 E_0^2}{3c^3 m_e^2} \langle \sin^2 \omega t \rangle = \frac{q^4 E_0^2}{3c^3 m_e^2}$. The total power *per unit area* transported by an electromagnetic wave (i.e. the flux) is $F_{\rm in} = \frac{cE_0^2}{8\pi}$. We define the *cross-section* as the ratio of the total radiated power to the total incident

---

[19] It is possible that a significant fraction of the gravitational binding energy is released very close to the galaxy (e.g. when gas free-falls down into the gravitational potential well, until it is shock heated when it 'hits' the galaxy: Birnboim & Dekel, 2003). It has been argued that some compact Ly$\alpha$ emitting sources may be powered by cooling radiation (as in Birnboim & Dekel 2003, Dijkstra 2009, Dayal et al. 2010).

[20] After these lectures, Mas-Ribas et al. (2017) showed that extended Ly$\alpha$ emission can also be produced by faint satellite galaxies which are too faint to be detected individually (also see Lake et al. 2015).

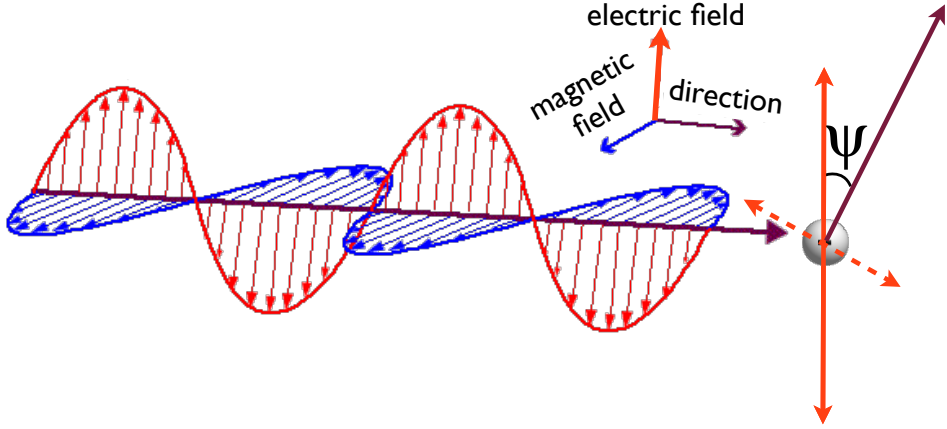

FIG. 14 Classical picture of the interaction of radiation with a free electron. The electric field of the incoming wave accelerates the free-electron. The accelerated electron radiates, and effectively scatters the incoming electromagnetic wave. The cross-section for this process is given by the Thomson cross-section. The angle $\Psi$ denotes the angle between the direction of the outgoing electro-magnetic wave, and the oscillation direction of the electron (which corresponds to the electric vector of the incoming electro-magnetic wave).

flux, i.e.

$$\sigma_{\mathrm{T}} = \frac{P_{\mathrm{out}}}{F_{\mathrm{in}}} = \frac{\frac{q^2 E_0^2}{3 m_e^2 c^3}}{\frac{c E_0^2}{8\pi}} = \frac{8\pi}{3} r_e^2 \approx 6.66 \times 10^{-25} \text{ cm}^2, \tag{36}$$

where $r_e = \frac{q^2}{m_e c^2} = 2.8 \times 10^{-13}$ cm denotes the classical electron radius.

The power of re-emitted radiation is not distributed isotropically across the sky. A useful way to see this is by considering what we see if we observe the oscillating electron along direction $\mathbf{k}_{\mathrm{out}}$. The *apparent* acceleration that the electron undergoes is reduced to $\hat{a}(t) \equiv a(t) \sin \Psi$, where $\Psi$ denotes the angle between $\mathbf{k}_{\mathrm{out}}$ and the oscillation direction, i.e $\cos \Psi \equiv \mathbf{k}_{\mathrm{out}} \cdot \mathbf{e}_{\mathrm{E}}$. Here, the vector $\mathbf{e}_{\mathrm{E}}$ denotes a unit vector pointing in the direction of the E-field. The reduced apparent acceleration translates to a reduced power in this direction, i.e $P_{\mathrm{out}}(\mathbf{k}_{\mathrm{out}}) \propto \hat{a}^2(t) \propto \sin^2 \Psi$.

The outgoing radiation field therefore has a strong directional dependence with respect to $\mathbf{e}_{\mathrm{E}}$. Note however, that for radiation coming in along some direction $\mathbf{k}_{\mathrm{in}}$, the electric vector can point in an arbitrary direction within the plane normal to $\mathbf{k}_{\mathrm{in}}$. For unpolarized incoming radiation, $\mathbf{e}_{\mathrm{E}}$ is distributed uniformly throughout this plane. To compute the angular dependence of the outgoing radiation field with respect to $\mathbf{k}_{\mathrm{in}}$ we need to integrate over $\mathbf{e}_{\mathrm{E}}$. That is[21]

$$P_{\mathrm{out}}(\mathbf{k}_{\mathrm{out}}|\mathbf{k}_{\mathrm{in}}) \propto \int P_{\mathrm{out}}(\mathbf{k}_{\mathrm{out}}|\mathbf{e}_{\mathrm{E}}) P(\mathbf{e}_{\mathrm{E}}|\mathbf{k}_{\mathrm{in}}) d\mathbf{e}_{\mathrm{E}}. \tag{37}$$

We can solve this integral by switching to a coordinate system in which the $x-$axis lies along $\mathbf{k}_{\mathrm{in}}$, and in which $\mathbf{k}_{\mathrm{out}}$ lies in the $x-y$ plane. In this coordinate system $\mathbf{e}_{\mathrm{E}}$ must lie in the $y-z$ plane. We introduce the angles $\phi$ and $\theta$ and write

$$\mathbf{k}_{\mathrm{in}} = (1, 0, 0), \ \mathbf{k}_{\mathrm{out}} = (\cos \theta, 0, \sin \theta), \ \mathbf{e}_{\mathrm{E}} = (0, \cos \phi, \sin \phi). \tag{38}$$

---

[21] We have adopted the notation of probability theory. In this notation, the function $p(y|b)$ denotes the conditional probability density function (PDF) of $y$ given $b$. The PDF for $y$ is then given by $p(y) = \int p(y|b)p(b)db$, where $p(b)$ denotes the PDF for $b$. Furthermore, the joint PDF of $y$ and $b$ is given by $p(y, b) = p(y|b)p(b)$.

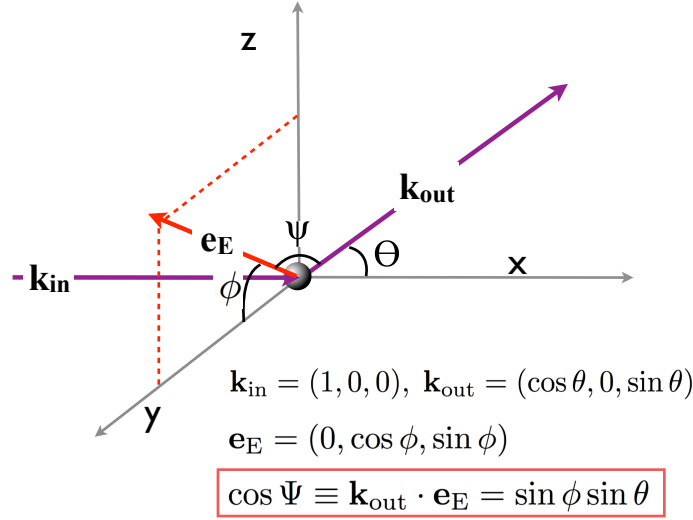

$\mathbf{k}_{\rm in} = (1, 0, 0), \ \mathbf{k}_{\rm out} = (\cos\theta, 0, \sin\theta)$

$\mathbf{e}_{\rm E} = (0, \cos\phi, \sin\phi)$

$\boxed{\cos\Psi \equiv \mathbf{k}_{\rm out} \cdot \mathbf{e}_{\rm E} = \sin\phi \sin\theta}$

FIG. 15 The geometry adopted for calculating the phase-function associated with Thomson scattering.

We can then also get that $\cos\Psi \equiv \mathbf{k}_{\rm out} \cdot \mathbf{e}_{\rm E} = \sin\phi\sin\theta$. This coordinate system is shown in Figure 15, which shows that $\theta$ denotes the angle between $\mathbf{k}_{\rm in}$ and $\mathbf{k}_{\rm out}$. For this choice of coordinates Eq 37 becomes

$$P_{\rm out}(\mathbf{k}_{\rm out}|\mathbf{k}_{\rm in}) \propto \int_0^{2\pi} d\phi \sin^2\Psi \tag{39}$$

where we have used that $P_{\rm out}(\mathbf{k}_{\rm out}|\mathbf{e}_{\rm E}) = P_{\rm out}\sin^2\Psi$, and that $P(\mathbf{e}_{\rm E}|\mathbf{k}_{\rm in}) = $ Constant (i.e. $\mathbf{e}_{\rm E}$ is distributed uniformly in the $y - z$ plane). We have omitted all numerical constants, because we are interested purely in the angular dependence of $P_{\rm out}$. We will determine the precise constants that should preceed the integral below. If we substitute $\sin^2\Psi = 1 - \cos^2\Psi = 1 - \sin^2\phi\sin^2\theta \equiv 1 - A\sin^2\phi$ ($A \equiv \sin^2\theta$), we get

$$P_{\rm out}(\mathbf{k}_{\rm out}|\mathbf{k}_{\rm in}) \propto \int_0^{2\pi} d\phi[1 - A\sin^2\phi] \propto \int_0^{2\pi} d\phi[1 + A - A - A\sin^2\phi] \propto ... \propto (1 + \cos^2\theta). \tag{40}$$

Figure 15 shows that our problem of interest is cylindrically symmetric around the $x$-axis, we therefore have that $P_{\rm out}(\mathbf{k}_{\rm out}|\mathbf{k}_{\rm in})$ only depends on $\theta$, and we will write $P_{\rm out}(\mathbf{k}_{\rm out}|\mathbf{k}_{\rm in}) = P_{\rm out}(\theta)$ for simplicity.

The angular dependence of the re-emitted radiation is quantified by the so-called *Phase-function* (or the *angular redistribution function*), which is denoted with $P(\theta)$.

$$\frac{P_{\rm out}(\theta)}{F_{\rm in}} \equiv \frac{\sigma_{\rm T} P(\theta)}{4\pi}, \quad \text{with} \quad \int_{-1}^{1} d\mu P(\mu) = 2 \quad \left(\text{i.e. } \int d\Omega P(\mu) = 4\pi\right), \tag{41}$$

where $\mu \equiv \cos\theta$. Note that the phase-function relates to the *differential cross-section* simply as $\frac{d\sigma}{d\Omega} \equiv \frac{\sigma P(\theta)}{4\pi}$.

There are two important examples of the phase-function we encounter for Ly$\alpha$ transfer. The first is the one we derived above, and describes 'dipole' or 'Rayleigh' scattering:

$$\boxed{P(\mu) = \frac{3}{4}(1 + \mu^2) \quad \text{dipole.}} \tag{42}$$

The other is for isotropic scattering:

$$\boxed{P(\mu) = 1 \quad \text{isotropic.}} \tag{43}$$

As we will see further on, the phase function associated with Ly$\alpha$ scattering is either described by pure dipole scattering, or by a superposition of dipole and isotropic scattering.

## 6.2. Interaction of a Bound Electron with Radiation: Lorentzian Cross-section

We can understand the expression for the Ly$\alpha$ absorption cross-section via an analysis similar to the one described above. The main difference with the previous analysis is that the electron is not free, but instead orbits the proton at a natural (angular) frequency $\omega_0$. We will treat the electron as a harmonic oscillator[22] with natural frequency $\omega_0$. In the classical picture, the electron radiates as it accelerates and spirals inward. To account for this we will assume that the harmonic oscillator is damped. This damped harmonic oscillator with natural frequency $\omega_0$ is 'forced' by the incoming radiation field that again has angular frequency $\omega$. The equation of motion for this forced, damped harmonic oscillator is

$$\ddot{x} + \Gamma\dot{x} + \omega_0^2 x = \frac{q}{m}E(x,t) = \frac{q}{m}E_0\exp(i\omega t), \tag{44}$$

where $x$ can represent one of the Cartesian coordinates that describe the location of the electron in its orbit. The term $\Gamma\dot{x}$ denotes the friction (or damping) term, which reflects that in the classical picture of a hydrogen atom, the electron spirals inwards over a short timescale (see § 3.2). The term on the RHS simply represents the electric force ($F = qE$) that is exerted by the electric field. We have represented the electro-magnetic wave as $E(t) = E_0\exp(i\omega t)$ (which is a more general way of describing a wave than what we used when considering the free electron). It turns out that this simplifies the analysis. We can find solutions to Eq 44 by substituting $x(t) = x\exp(i\omega t)$. We discuss two solutions here:

- In the absence of the electromagnetic field, this yields a quadratic equation for $\omega$, namely $-\omega^2 + i\Gamma\omega + \omega_0^2 = 0$. This equation has solutions of the form $\omega = \frac{i\Gamma}{2} \pm \sqrt{\omega_0^2 - \Gamma^2/4}$. We assume that $\omega_0 \gg \Gamma$ (which is the case for Ly$\alpha$ as we see below), which can be interpreted as meaning that the electron makes multiple orbits around the nucleus before there is a 'noticeable' change in its position due to radiative energy losses. The solution for $x(t)$ thus looks like $x(t) \propto \exp(-\Gamma t/2)\cos\omega_0 t$. The solution indicates that the electron keeps orbiting the proton with the same natural frequency $\omega_0$, but that it spirals inwards on a characteristic timescale $\Gamma^{-1}$ (also see § 3.2). We will evaluate $\Gamma$ later.

- In the presence of an electromagnetic field, substituting $x(t) = x\exp(i\omega t)$ yields the following solution for the amplitude $x$:

$$x = \frac{qE_0}{m_e}\frac{1}{\omega^2 - \omega_0^2 + i\omega\Gamma} \overset{\omega\approx\omega_0}{\sim} \frac{qE_0}{2m_e\omega_0}\frac{1}{\omega - \omega_0 + i\Gamma/2}, \tag{45}$$

where we used that $(\omega^2 - \omega_0^2) = (\omega - \omega_0)(\omega + \omega_0) \approx 2\omega_0(\omega - \omega_0)$. This last approximation assumes that $\omega \approx \omega_0$. It is highly relevant for Ly$\alpha$ scattering, where $\omega$ and $\omega_0$ are almost always very close together (meaning that $|\omega - \omega_0|/\omega_0 \ll 10^{-2}$).

With the solution for $x(t)$ in place, we can apply the Larmor formula and compute the time averaged power radiated by the accelerated electron:

$$\langle P_{\text{out}}\rangle = \frac{2q^2\langle|\ddot{x}|\rangle^2}{3c^3} = \frac{2q^2\omega^4}{3c^3}\frac{q^2E_0^2}{2 \times 4m_e^2\omega_0^2}\frac{1}{(\omega - \omega_0)^2 + \Gamma^2/4}. \tag{46}$$

Where the highlighted factor of 2 comes in from the time-average: $\langle E^2\rangle = E_0^2/2$. As before, the time average of the total incoming flux $\langle F_{\text{in}}\rangle$ (in erg s$^{-1}$ cm$^{-2}$) of electromagnetic radiation equals $\langle F_{\text{in}}\rangle = \frac{cE_0^2}{8\pi}$. We therefore obtain an expression for the cross-section as:

$$\sigma(\omega) = \frac{\langle P_{\text{out}}\rangle}{\langle F_{\text{in}}\rangle} = \frac{8\pi\langle P_{\text{out}}\rangle}{cE_0^2} = \frac{16\pi q^4\omega^4}{24m_e^2\omega_0^2c^4}\frac{1}{(\omega - \omega_0)^2 + \Gamma^2/4} = \frac{\sigma_T}{4\omega_0^2}\frac{\omega^4}{(\omega - \omega_0)^2 + \Gamma^2/4}, \tag{47}$$

where we substituted the expression for the Thomson cross section $\sigma_T = \frac{8\pi q^4}{3m_e^2c^4}$ (see Eq 36).

---

[22] We can justify this picture as follows: we define the $x-y$ plane to be the plane in which the electron orbits the proton. The $x$−coordinate of the electron varies as $x(t) = x_0\cos\omega_0 t$. The x-component of the electro-static force on the electron varies as $F_x = F_e\frac{x}{r}$, in which $F_e = \frac{q^2}{r^2}$. That is, the equation of motion for the x-coordinate of the electron equals $\ddot{x} = -kx$, where $k = q^2/r^3$.

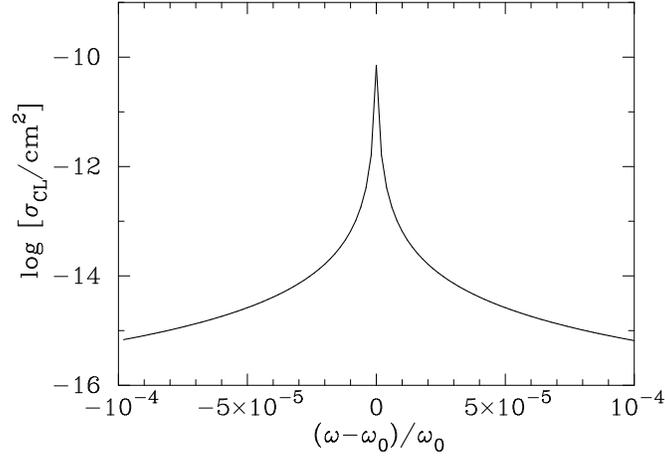

FIG. 16 The Lorentzian profile $\sigma_{\mathrm{CL}}(\omega)$ for the Ly$\alpha$ absorption cross-section. This represents the absorption cross-section of a *single atom*. When averaging over a collection of atoms with a Maxwellian velocity distribution, we obtain the Voigt profile (shown in Fig 17).

The expression for $\sigma(\omega)$ can be recast in a more familiar form when we use the Larmor formula to constrain $\Gamma$. The equation of motion shows that 'friction/damping force' on the electron is $F = -m_e \Gamma \dot{x}$. We can also write this force as $F = \frac{dp_e}{dt} = \frac{m_e}{p_e} \frac{d}{dt} \frac{p_e^2}{2m_e} = \frac{m_e}{p_e} \frac{dE_{\mathrm{kin}}}{dt} = -\frac{m_e}{p_e} P_{\mathrm{out}}$, where $P_{\mathrm{out}}$ denotes the emitted power. Setting the two equal gives us a relation between $\Gamma$ and $P_{\mathrm{out}}$: $-\Gamma m_e \dot{x}^2 = P_{\mathrm{out}}$. Using that $\dot{x} = i\omega_0 x$, $\ddot{x} = -\omega_0^2 x$, and the Larmor formula gives us $\Gamma = \frac{2q^2 \omega_0^2}{3 m_e c^3} = 15 \times 10^8$ s$^{-1}$. With this the 'classical' expression for the Ly$\alpha$ cross section can be recast as

$$\sigma_{\mathrm{CL}}(\omega) = \frac{3\lambda_0^2}{8\pi} \frac{\Gamma^2 (\omega/\omega_0)^4}{(\omega-\omega_0)^2 + \Gamma^2/4} \Rightarrow \sigma_{\mathrm{CL}}(\nu) = \frac{3\lambda_0^2}{8\pi} \frac{(\Gamma/2\pi)^2 (\nu/\nu_0)^4}{(\nu-\nu_0)^2 + \Gamma^2/16\pi^2}, \tag{48}$$

where the subscript 'CL' stresses that we obtained this expression with purely classical physics. If we ignore the $(\omega/\omega_0)^4$ term, then the functional form for $\sigma(\omega)$ is that of a *Lorentzian Profile*. In the second equation we re-expressed the cross-section as a function of frequency $\nu = \omega/(2\pi)$.

Figure 16 shows $\sigma_{\mathrm{CL}}(\omega)$ for a narrow range in $\omega$. There are two things to note:

1. The function is sharply peaked on $\omega_0$, at which $\sigma_0 \equiv \sigma_{\mathrm{CL}}(\omega_0) = \frac{3\lambda_0^2}{8\pi} \sim 7 \times 10^{-11}$ cm$^{-2}$, where $\lambda_0 = 2\pi c/\omega_0$ corresponds to the wavelength of the electromagnetic wave with frequency $\omega_0$. The cross-section falls off by $\gtrsim 5$ orders of magnitude with $|\omega - \omega_0|/\omega_0 \sim 10^{-4}$. Note that the cross-section is many, many orders of magnitude larger than the Thomson cross-section. This enhanced cross-section represents a 'resonance', an '*unusually strong response of a system to an external trigger*'.

2. The $\sigma \propto \omega^4$-dependence implies that the atom is slightly more efficient at scattering more energetic radiation. This correspond to the famous Rayleigh scattering regime, which refers to elastic scattering of light or other electromagnetic radiation by particles much smaller than the wavelength of the radiation (see `https://en.wikipedia.org/wiki/Rayleigh_scattering`), and which explains why the sky is blue[23], and the setting/rising sun red.

### 6.3. Interaction of a Bound Electron with Radiation: Relation to Ly$\alpha$ Cross-Section

The derivation from the previous section was based purely on classical physics. A full quantum-mechanical treatment of the Ly$\alpha$ absorption cross-section is beyond the scope of these lectures, and - interestingly - still subject of ongoing research (e.g. Lee 2003, Bach & Lee 2014, 2015, Alipour et al. 2015). A clear recent discussion on this can be found in Mortlock (2016). There are two main things that change:

---

[23] This is not always the case in the Netherlands or Norway.

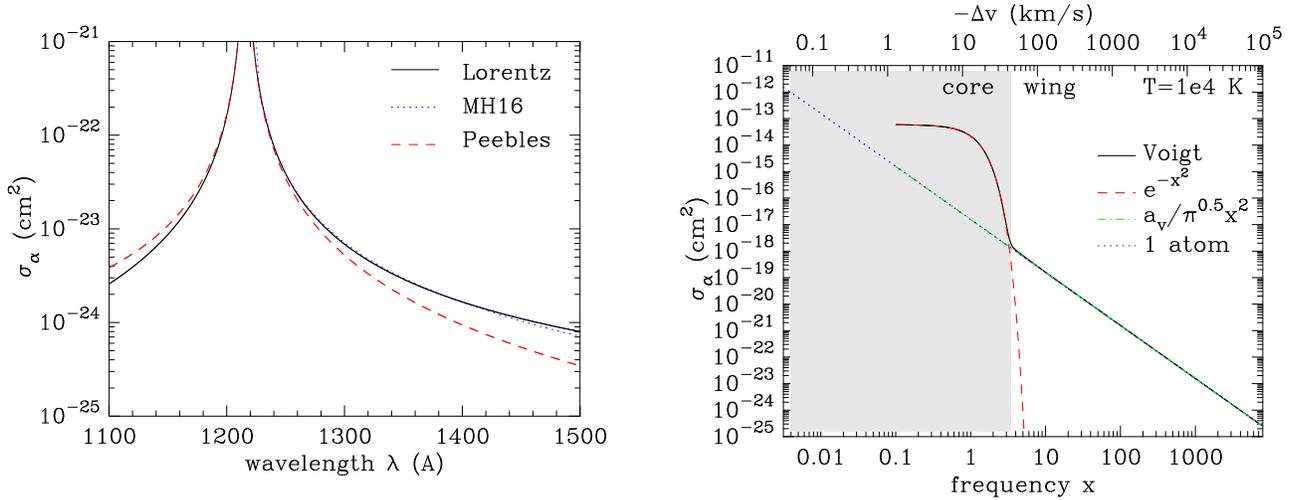

FIG. 17 In the *left panel* we compare the different atom frame Lyα cross sections. The *black solid line* shows the Lorentzian cross-section. The *red dashed line* shows the cross-section given in Peebles (1993), and the *blue dotted line* shows the quantum mechanical cross-section from Mortlock & Hirata (in prep, also see Mortlock 2016). In the *right panel* the *solid line* shows the gas frame (velocity averaged absorption) cross section as given by Eq 55 as a function of the dimensionless frequency $x$ (see text). The *red dashed line* (*green dot-dashed line*) represent the cross section where we approximated the Voigt function as $\exp(-x^2)$ ($a_v/[\sqrt{\pi}x^2]$). Clearly, these approximations work very well in their appropriate regimes. Also shown for comparison as the *blue dotted line* is the symmetric single atom Lorentzian cross section (which was shown in the *left panel* as the *black solid line*). Close to resonance, this single atom cross section provides a poor description of the real cross section, but it does very well in the wings of the line (see text).

1. The parameter $\Gamma$ reduces by a factor of $f_\alpha = 0.4162$, which is known as the 'oscillator' strength, i.e. $\Gamma \to f_\alpha \Gamma \equiv A_\alpha$. Here, $A_\alpha$ denotes the Einstein A-coefficient for the Lyα transition.

2. In detail, a simple functional form (Lorentzian, Eq 48, or see Eq 49 below) for the Lyα cross-section does not exist. The main reason for this is that if we want to evaluate the Lyα cross-section far from resonance, we have to take into account the contributions to the cross-section from the higher-order Lyman-series transitions, and even photoionization. When these contributions are included, the expression for the cross-section involves squaring the sum of all these contributions (e.g. Bach & Lee, 2014, 2015; Mortlock, 2016). Accurate approximations to this expression are possible close to the resonance(s) - when $|\omega - \omega_0|/\omega_0 \ll 1$, which is the generally the case for practical purposes - and these approximations are in excellent agreement with our derived cross-section (see Eq 14 in Mortlock, 2016, and use that $\omega \approx \omega_0$, and that $\Lambda_{12} = \Gamma/2\pi$).

The *left panel* of Figure 17 compares the different Lyα cross sections. The *black solid line* shows the Lorentzian cross-section. The *red dashed line* shows $\sigma_P(\omega)$, which is the cross section that is given in Peebles (1993):

$$\sigma(\omega) = \frac{3\lambda_\alpha^2}{8\pi} \frac{A_\alpha^2(\omega/\omega_\alpha)^4}{(\omega - \omega_\alpha)^2 + A_\alpha^2(\omega/\omega_\alpha)^6/4}. \tag{49}$$

If we ignore the $(\omega/\omega_\alpha)^6$ term in the denominator, then this corresponds exactly to the cross-section that we derived in our classical analysis (see Eq 48, with $\Gamma \to A_\alpha$). This cross-section is obtained from a quantum mechanical calculation, and under the assumption that the hydrogen atom has only two quantum levels (the 1s and 2p levels). The *blue dotted line* shows the cross-section obtained from the full quantum mechanical calculation (see Mortlock 2016 for a discussion, which is based on Mortlock & Hirata in prep) This Figure shows that these cross-sections differ only in the far wings of the line profile. Close to resonance, the line profiles are practically indistinguishable.

### 6.4. Voigt Profile of Lyα Cross-Section

In the previous section we discussed the Lyα absorption cross section in the frame of a single atom. Because each atom has its own velocity, a photon of a fixed frequency $\nu$ will appear Doppler boosted to a slightly different frequency for each atom in the gas. To compute the Lyα absorption cross section for a collection of moving atoms, we must convolve the single-atom cross section with the atom's velocity distribution. That is,

$$\sigma_\alpha(\nu, T) = \int d^3 \mathbf{v} \sigma_\alpha(\nu|\mathbf{v}) f(\mathbf{v}), \tag{50}$$

where $\sigma_\alpha(\nu|\mathbf{v})f$ denotes the Ly$\alpha$ absorption cross-section at frequency $\nu$ for an atom moving at 3D velocity $\mathbf{v}$ (the precise velocity $\mathbf{v}$ changes the frequeny in the frame of the atom). Suppose that the photon is propagating in a direction $\mathbf{n}$. We decompose the atoms three-dimensional velocity vectors into directions parallel ($v_{||}$) and orthogonal ($\mathbf{v}_\perp$) to $\mathbf{n}$. These components are independent and $f(\mathbf{v})d^3\mathbf{v} = f(v_{||})g(\mathbf{v}_\perp)dv_{||}d^2\mathbf{v}_\perp$. The absorption cross-section does not depend on $\mathbf{v}_\perp$ because the frequency of the photon that the atoms 'sees' does not depend on $\mathbf{v}_\perp$. We can therefore write

$$\sigma_\alpha(\nu, T) = \underbrace{\int d^2\mathbf{v}_\perp g(\mathbf{v}_\perp)}_{=1} \int_{-\infty}^{\infty} dv_{||}\sigma_\alpha(\nu|v_{||})f(v_{||}). \tag{51}$$

For a Maxwell-Boltzmann distribution $f(v_{||})dv_{||} = \left(\frac{m_p}{2\pi k_B T}\right)^{1/2} \exp\left(-\frac{m_p v_{||}^2}{2kT}\right)$. If we insert this into Eq 51 and substitute Eq 48 for $\sigma_\alpha(\nu|v_{||}) = \sigma(\nu')$, where $\nu' = \nu\left(1 - \frac{v_{||}}{c}\right)$ we find

$$\sigma_\alpha(\nu, T) = \frac{3(\lambda_0 A_\alpha)^2}{32\pi^3}\left(\frac{m_p}{2\pi k_B T}\right)^{1/2} \int_{-\infty}^{\infty} dv_{||} \frac{\exp\left(-\frac{m_p v_{||}^2}{2kT}\right)}{\left[\nu\left(1 - \frac{v_{||}}{c}\right) - \nu_\alpha\right]^2 + \frac{A_\alpha^2}{16\pi^2}}, \tag{52}$$

where we dropped the term $(\nu/\nu_0)^4$, which is accurate for $\nu \approx \nu_0$ (which is practically always the case). We substitute $y = \sqrt{\frac{m_p v_{||}^2}{2k_B T}} \equiv v_{||}/v_{\rm th}$, and define the *Voigt parameter* $a_v \equiv \frac{A_\alpha}{4\pi\Delta\nu_\alpha} = 4.7\times10^{-4}(T/10^4 \text{ K})^{-1/2}$ in which $\Delta\nu_\alpha = \nu_\alpha v_{\rm th}/c$. We can then recast this expression as

$$\sigma_\alpha(\nu, T) = \frac{3(\lambda_0 A_\alpha)^2}{32\pi^3}\left(\frac{m_p}{2\pi k_B T}\right)^{1/2} \frac{v_{\rm th}}{\Delta\nu_\alpha^2} \int_{-\infty}^{\infty} dy \frac{\exp(-y^2)}{\left[\frac{\nu}{\Delta\nu_\alpha}\left(1 - \frac{yv_{\rm th}}{c}\right) - \frac{\nu_\alpha}{\Delta\nu_\alpha}\right]^2 + a_v^2}. \tag{53}$$

We finally introduce the dimensionless frequency variable $x \equiv (\nu - \nu_\alpha)/\Delta\nu_\alpha$, which we use to express $\nu = \nu_\alpha(1 + xv_{\rm th}/c)$. If we substitute this back into Eq 53 and drop the second order term with $(v_{\rm th}/c)^2$, then we get

$$\sigma_\alpha(\nu, T) = \frac{3\lambda_0^2 A_\alpha^2}{32\pi^3\sqrt{\pi}\Delta\nu_\alpha^2} \int_{-\infty}^{\infty} dy \frac{\exp(-y^2)}{(x-y)^2 + a_v^2} \equiv \frac{3\lambda_0^2 a_v}{2\sqrt{\pi}} H(a_v, x) \equiv \sigma_{\alpha,0}(T)\phi(x) = \tag{54}$$

$$= 5.9\times10^{-14}\left(\frac{T}{10^4 \text{ K}}\right)^{-1/2}\phi(x) \text{ cm}^2$$

and the Voigt function as (e.g. Chluba & Sunyaev, 2009)

$$\phi(x) \equiv H(a_v, x) = \frac{a_v}{\pi} \int_{-\infty}^{\infty} \frac{e^{-y^2}dy}{(y-x)^2 + a_v^2} = \begin{cases} \sim e^{-x^2}[\exp(a_v^2)\text{erfc}(a_v)] \sim e^{-x^2} & \text{core}; \\ \sim \frac{a_v}{\sqrt{\pi}x^2} & \text{wing}. \end{cases} \tag{55}$$

Note that throughout this review we will use both $\phi(x)$ and $H(a_v, x)$ to denote the shape of the Ly$\alpha$ line profile[24]. The transition between core and wing occurs approximately when $\exp(-x^2) = a_v/(\sqrt{\pi}x^2)$.

The *solid line* in the *right panel* of Figure 17 shows the LAB frame cross section - this is also known as the Voigt-profile - as given by Eq 55 as a function of the dimensionless frequency $x$. The *red dashed line* (*green dot-dashed line*) represent the cross section where we approximated the Voigt function as $\exp(-x^2)$ $(a_v/[\sqrt{\pi}x^2])$. Clearly, these approximations work very well in the relevant regimes. Note that this Figure shows that a decent approximation to

---

[24] One has to be a bit careful because in the literature occasionally $\phi(x) = H(a_v, x)/\sqrt{\pi}$, because in this convention the line profile is normalized to 1, i.e. $\int \phi(x)dx = 1$. In our convention $\phi(x = 0) = 1$, while the normalization is $\int \phi(x)dx = \sqrt{\pi}$.

Voigt function at all frequencies is given simply by the sum of these two terms, i.e. $\phi(x) \approx \exp(-x^2) + a_v/(\sqrt{\pi}x^2)$ (Krolik, 1989, provided that $|x| \gtrsim \sqrt{a_v}$). This approximation fails in a very narrow frequency regime where the transition from core to wing occurs. A useful fitting function that is accurate at all $x$ is given in Tasitsiomi (2006). Also shown for comparison as the *blue dotted line* is the symmetric single atom cross section (which was shown in the *left panel* as the *solid line*). Figure 17 shows that close to resonance, this single atom cross section provides a poor description of the real cross section. This is because Doppler motions 'smear out' the sharply peaked cross-section. Far in the wing however, the single atom cross-section provides an excellent fit to the velocity averaged Voigt profile.

One of the key results from this section is that the Ly$\alpha$ cross-section, evaluated at line center and averaged over the velocity distribution of atoms, is tremendous at $\sigma_{\alpha,0}(T) \sim 5.9 \times 10^{-14}(T/10^4\text{ K})^{-1/2}\text{ cm}^{-2}$, which is $\sim 11$ orders of magnitude than the Thomson cross-section. That is, an electron bound to the proton is $\sim 11$ orders of magnitude more efficient at scattering radiation than a free electron when the frequency of that radiation closely matches the natural frequency of the transition. This further emphasises that the electron 'resonantly scatters' the incoming radiation. To put these numbers in context, it is possible to measure the hydrogen column density, $N_{\text{HI}}$, in nearby galaxies. The observed intensity in the 21-cm line translates to typical HI column densities of order $N_{\text{HI}} \sim 10^{19} - 10^{21}$ cm$^{-2}$ (Bosma 1978, Kaberla et al. 2005, Chung et al. 2009), which translates to line center optical depths of Ly$\alpha$ photons of order $\tau_0 \sim 10^7 - 10^8$. This estimate highlights the importance of understanding the transport of Ly$\alpha$ photon out of galaxies. They generally are not expected to escape without interacting with hydrogen gas.

## 7. STEP 2 TOWARDS UNDERSTANDING LY$\alpha$ RADIATIVE TRANSFER: THE RADIATIVE TRANSFER EQUATION

The specific intensity $I_\nu(\mathbf{r}, \mathbf{n}, t)$ is defined as the rate at which energy crosses a unit area, per solid angle, per unit time, as carried by photons of energy $h_p\nu$ in the direction $\mathbf{n}$, i.e. $I_\nu(\mathbf{r}, \mathbf{n}, t) = \frac{d^3 E_\nu}{d\Omega dt dA}$. The change in the spectral/specific intensity of radiation at a location $\mathbf{r}$ that is propagating in direction $\mathbf{n}$ at time $t$ is (e.g. Meiksin, 2009)

$$\mathbf{n} \cdot \nabla I_\nu(\mathbf{r}, \mathbf{n}, t) + \frac{1}{c}\frac{\partial I(\mathbf{r}, \mathbf{n}, t)}{\partial t} = -\alpha_\nu(\mathbf{r})I(\mathbf{r}, t) + j_\nu(\mathbf{r}, \mathbf{n}, t). \tag{56}$$

Here, $\alpha_\nu(\mathbf{r})I(\mathbf{r}, t)$ in the first term on the RHS denotes the 'attenuation coefficient', which accounts for energy loss (gain) due to absorption (stimulated emission). In static media, we generally have that $\alpha_\nu(\mathbf{r})$ is isotropic[25], which is why we dropped its directional dependence. The emission coefficient $j_\nu(\mathbf{r}, \mathbf{n}, t)$ describes the local specific luminosity per solid angle, per unit volume. The random orientation of atoms/molecules (again) generally causes emission coefficient to be isotropic.

For Ly$\alpha$ radiation, photons are not permanently removed following absorption, but they are generally scattered. To account for this scattering, we must add a third term to the RHS on Eq 56:

$$\mathbf{n} \cdot \nabla I_\nu(\mathbf{r}, \mathbf{n}, t) + \frac{1}{c}\frac{\partial I(\mathbf{r}, \mathbf{n}, t)}{\partial t} = -\alpha_\nu(\mathbf{r})I(\mathbf{r}, t) + j_\nu(\mathbf{r}, \mathbf{n}, t) + \int d\nu' \int d^3\mathbf{n}'\ \alpha_{\nu'}(\mathbf{r})I_{\nu'}(\mathbf{r}, \mathbf{n}, t)R(\nu, \nu', \mathbf{n}, \mathbf{n}'). \tag{57}$$

The third term accounts for energy redistribution as a result of scattering. In this term, the so-called '*redistribution function*' $R(\nu, \nu', \mathbf{n}, \mathbf{n}')$ describes the probability that radiation that was originally propagating at frequency $\nu'$ and in direction $\mathbf{n}'$ is scattered into frequency $\nu$ and direction $\mathbf{n}$. We focus on equilibrium solutions and omit the time dependence in Eq 57. Furthermore, we introduce the coordinate $s$, which measures distance along the direction $\mathbf{n}$. The change of the intensity of radiation that is propagating in direction $\mathbf{n}$ with distance $s$ is then

$$\frac{dI_\nu(s, \mathbf{n})}{ds} = -[\ \underbrace{\alpha_\nu^{\text{HI}}(s)}_{\text{I: absorption}}\ +\ \underbrace{\alpha_\nu^{\text{dest}}(s)}_{\text{IV: 'destruction'}}\ ]I_\nu(s, \mathbf{n})\ +\ \underbrace{j_\nu(s)}_{\text{II: emission}}\ +\ \underbrace{\int d\nu' \int d^3\hat{\mathbf{n}}'\ \alpha_{\nu'}(s)I_{\nu'}(s, \hat{\mathbf{n}}')R(\nu, \nu', \mathbf{n}, \hat{\mathbf{n}}')}_{\text{III: scattering}}. \tag{58}$$

In the following subsections we will discuss each term I-IV on the right hand side in more detail.

---

[25] For radiation at some fixed frequency $\nu$ close to the Ly$\alpha$ resonance, the opacity $\alpha_\nu(\mathbf{r}, \mathbf{n})$ depends on $\mathbf{n}$ for non-static media. This directional dependence is taken into account when performing Monte-Carlo Ly$\alpha$ radiative transfer calculations (to be described in § 9).

### 7.1. I: Absorption Term: Lyα Cross Section

The opacity $\alpha_\nu^{\mathrm{HI}}(s) = \left[n_l(s) - \frac{g_l}{g_u} n_u(s)\right] \sigma(\nu)$. The second term within the square brackets corrects the absorption term for stimulated emission. In most astrophysical conditions all neutral hydrogen atoms are in their electronic ground state, and we can safely ignore the stimulated emission term (see e.g. the Appendix of Dijkstra & Loeb 2008a, and Dijkstra et al. 2016). That is, in practice we can simply state that $\alpha_\nu(s) = n_l(s) \sigma_\alpha(\nu) = n_{\mathrm{HI}}(s) \sigma_\alpha(\nu)$. We have derived expressions for the Lyα absorption cross-section in § 6.

### 7.2. II: Volume Emission Term

The emission term is given by

$$j_\nu(s) = \frac{\phi(\nu) h \nu_\alpha}{4\pi^{3/2} \Delta\nu_\alpha} \Big( n_e n_{\mathrm{HI}} q_{1s2p} + n_e n_{\mathrm{HII}} \alpha(T) f_{\mathrm{Ly}\alpha}(T) \Big), \tag{59}$$

where $\phi(\nu)$ is the Voigt profile (see § 6.4), the factor $4\pi$ accounts for the fact that Lyα photons are emitted isotropically into $4\pi$ steradians, and the factor of $\sqrt{\pi}\Delta\nu_\alpha$ arises when converting[26] $\phi(x)$ to $\phi(\nu)$. The first term within the brackets denotes the Lyα production rate as a result of collisional excitation of H atoms by (thermal) electrons, and is given by Eq 7, in which the collision strength $\Omega_{1s2p}(T)$ can be read off from Figure 7. The second term within the brackets denotes the Lyα production rate following recombination, in which both the recombination coefficient and the Lyα production probability depend on both temperature and the opacity of the medium to ionizing and Lyman series photons (see Fig 6, and § 4.2).

### 7.3. III: Scattering Term

The 'redistribution function' $R(\nu, \nu', \mathbf{n}, \mathbf{n}')$ describes the probability that radiation that was originally propagating at frequency $\nu'$ and in direction $\mathbf{n}'$ is scattered into frequency $\nu$ and direction $\mathbf{n}$ (see e.g. Henyey 1940, Zanstra 1949, Unno 1952, Hummer 1962 for early discussions). In practise, this probability depends only on the angle between $\mathbf{n}$ and $\mathbf{n}'$, i.e. $R(\nu, \nu', \mathbf{n}, \mathbf{n}') = R(x_{\mathrm{out}}, x_{\mathrm{in}}, \mu)$, in which $\mu \equiv \cos\theta = \mathbf{n} \cdot \mathbf{n}'$. We have also switched to standard dimensionless frequency coordinates (first introduced in Eq 53), and denote with $x_{\mathrm{out}}$ ($x_{\mathrm{in}}$) the dimensionless frequency of the photon after (before) scattering. Formally, $R(x_{\mathrm{out}}, x_{\mathrm{in}}, \mu) dx_{\mathrm{out}} d\mu$ denotes the probability that a photon of frequency $x_{\mathrm{in}}$ was scattered by an angle in the range $\mu \pm d\mu/2$ into the frequency range $x_{\mathrm{out}} \pm dx_{\mathrm{out}}/2$. Thus, $R(x_{\mathrm{out}}, x_{\mathrm{in}}, \mu)$ is normalized such that $\int_{-1}^{1} d\mu \int_{-\infty}^{\infty} dx_{\mathrm{out}} R(x_{\mathrm{in}}, x_{\mathrm{out}}, \mu) = 1$.

In the remainder of this section, we will compute the redistribution functions $R(x_{\mathrm{in}}, x_{\mathrm{out}}, \mu)$. Following Lee (1974) we will employ the notation of probability theory. In this notation, the function $p(y|b)$ denotes the conditional probability[27] density function (PDF) of $y$ given $b$. The PDF for $y$ is then given by $p(y) = \int p(y|b) p(b) db$, where $p(b)$ denotes the PDF for $b$. Furthermore, the joint PDF of $y$ and $b$ is given by $p(y, b) = p(y|b) p(b)$. We could just as well have written that $p(y, b) = p(b|y) p(y)$, and by setting $p(y|b) p(b) = p(b|y) p(y)$ we get 'Bayes theorem' which states that

$$p(y|b) = \frac{p(b|y) p(y)}{p(b)}. \tag{60}$$

We will use this theorem on several occasions below. We can write the conditional joint PDF for $x_{\mathrm{out}}$ and $\mu$ given $x_{\mathrm{in}}$ as

$$R(x_{\mathrm{out}}, \mu | x_{\mathrm{in}}) = R(x_{\mathrm{out}} | \mu, x_{\mathrm{in}}) P(\mu | x_{\mathrm{in}}). \tag{61}$$

An additional quantity that is of interest in many radiative transfer problems (see for example § 7.5) is the conditional PDF for $x_{\mathrm{out}}$ given $x_{\mathrm{in}}$:

$$R(x_{\mathrm{out}} | x_{\mathrm{in}}) = \int_{-1}^{1} d\mu R(x_{\mathrm{out}} | \mu, x_{\mathrm{in}}) P(\mu | x_{\mathrm{in}}). \tag{62}$$

---

[26] Recall that $\phi(x)$ was normalized to $\int \phi(x) dx = \sqrt{\pi}$ (Eq 55). Substituting $dx = d\nu/\Delta\nu_\alpha$ (see discussion below Eq 53) gives $\int \phi(\nu) d\nu = \sqrt{\pi}\Delta\nu_\alpha$.

[27] In the lectures I illustrated this with an example in which $y$ denotes my happiness, and in which $b$ denotes the number of snowballs that were thrown in my face in the previous 30 minutes. Clearly, $p(y)$ will be different when $b = 0$ or when $b \gg 1$.

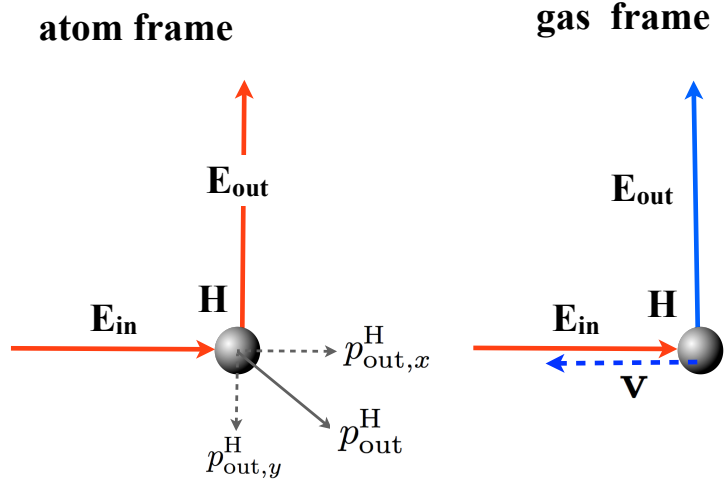

**atom frame**

**gas frame**

FIG. 18 In most astrophysical media, Ly$\alpha$ scattering is *partially coherent*: energy conservation implies that the energy of the photon before and after scattering is the same in the frame of the atom (the 'atom frame' shown on the *left*). In the gas frame however (shown on the *right*) the energy before and after scattering is different. This is because the random thermal motion of the atom induces Doppler shifts to the energy of the photon before and after scattering by an amount that depends on the atoms velocity, and on the scattering angle. Each scattering event induces small changes to the frequency of the Ly$\alpha$ photon. In this case, the atom is moving into the incoming Ly$\alpha$ photon, which causes it to appear at a higher frequency. Scattering by $90^\circ$ does not induce any additional Doppler boost in this example, and the outgoing photon has a higher frequency (which is why we drew it blue, to indicate its newly acquired blue-shift). The picture on the *left* also illustrates the recoil effect: in the atom frame, momentum conservation requires that the hydrogen atom moves after the scattering event. This newly gained kinetic energy of the H-atom comes at the expense of the energy of the Ly$\alpha$ photon. This recoil effect is generally small.

To complete our calculation we have to evaluate the PDFs $R(x_{\mathrm{out}}|\mu, x_{\mathrm{in}})$ and $P(\mu|x_{\mathrm{in}})$. We will do this next.

In most astrophysical conditions, the energy of the Ly$\alpha$ photon before and after scattering is identical in the frame of the absorbing atom. This is because the life-time of the atom in its $2p$ state is only $t = 1/A_\alpha \sim 10^{-9}$ s. In most astrophysical conditions, the hydrogen atom in this state is not 'perturbed' over this short time-interval, and energy conservation forces the energy of the photon to be identical before and after scattering. Because of random thermal motions of the atom, energy conservation in the atom's frame translates to a change in the energy of the incoming and outgoing photon that depends on the velocity of the atom and the scattering direction (see Fig 18 for an illustration). This type of scattering is known as 'partially coherent' scattering[28].

For notational clarity, we denote the propagation direction and dimensionless frequency of the photon before (after) scattering with $\mathbf{k}_{\mathrm{in}}$ and $x_{\mathrm{in}}$ ($\mathbf{k}_{\mathrm{out}}$ and $x_{\mathrm{out}}$). We assume that the scattering event occurs off an atom with velocity vector $\mathbf{v}$. Doppler boosting between the atom and gas frame corresponds to

$$x_{\mathrm{in}}^{\mathrm{atom}} = x_{\mathrm{in}}^{\mathrm{gas}} - \frac{\mathbf{v} \cdot \mathbf{k}_{\mathrm{in}}}{v_{\mathrm{th}}}, \quad \text{gas} \rightarrow \text{atom}; \quad x_{\mathrm{in}}^{\mathrm{gas}} = x_{\mathrm{in}}^{\mathrm{atom}} + \frac{\mathbf{v} \cdot \mathbf{k}_{\mathrm{in}}}{v_{\mathrm{th}}}, \quad \text{atom} \rightarrow \text{gas}. \tag{63}$$

Notice the signs in these equations. When $\mathbf{v}$ and $\mathbf{k}_{\mathrm{in}}$ point in the same direction, the atom is moving away from the incoming photon, which reduces the frequency of the photon in the atom's frame. The expressions are the same for the outgoing photon. Partially coherent scattering dictates that $x_{\mathrm{in}}^{\mathrm{atom}} = x_{\mathrm{out}}^{\mathrm{atom}}$, which allows us to write down the relation between $x_{\mathrm{out}}^{\mathrm{gas}}$ and $x_{\mathrm{in}}^{\mathrm{gas}}$ as

$$x_{\mathrm{out}}^{\mathrm{gas}} = x_{\mathrm{in}}^{\mathrm{gas}} - \frac{\mathbf{v} \cdot \mathbf{k}_{\mathrm{in}}}{v_{\mathrm{th}}} + \frac{\mathbf{v} \cdot \mathbf{k}_{\mathrm{out}}}{v_{\mathrm{th}}}. \tag{64}$$

---

[28] It is possible to repeat the analysis of this section under the assumption that (*i*) the energy of the photon before and after scattering is identical, which is relevant when the gas has zero temperature. This corresponds to 'completely coherent' scattering (*ii*) the energy of the re-emitted photon is completely unrelated to the atom of the incoming photon. This can happen in very dense gas where collisions perturb the atom while in the $2p$ state. This case corresponds to 'completely incoherent' scattering.

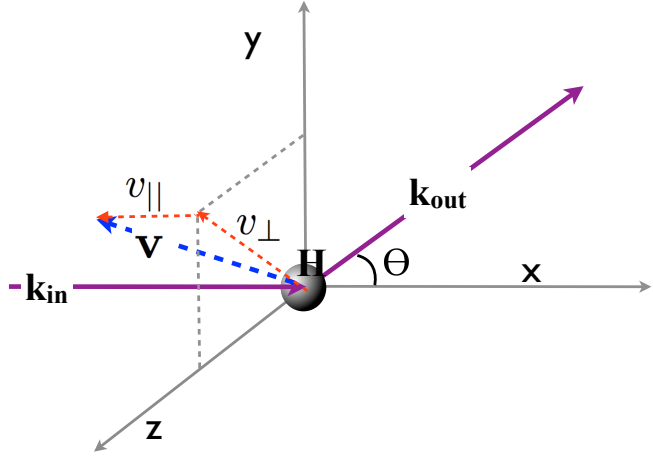

FIG. 19 Schematic depiction of the coordinate system that we used to describe the scattering event. The scattering plane is spanned by the wavevectors of the photon before ($\mathbf{k}_{\mathrm{in}} = [1, 0, 0]$) and after ($\mathbf{k}_{\mathrm{out}} = [\mu, \sqrt{1 - \mu^2}, 0]$) scattering. In other words, the scattering plane corresponds simply to the $x - y$ plane. If we decompose the atom's velocity vector (*the blue dashed vector*) into components parallel ($v_{\parallel}$, *red dotted vector*) and orthogonal ($v_{\perp}$, *red dotted vector*) to the propagation direction of the incoming photon, then we can derive convenient expressions for the conditional probabilities $R(x_{\mathrm{out}}|x_{\mathrm{in}}, \mu)$ (see Fig 20) and $R(x_{\mathrm{out}}|x_{\mathrm{in}})$ (see Fig 21).

We always work in the gas frame and will drop the 'gas' superscript from now on. Eq 64 still misses the effect of *atomic recoil*, which is illustrated in the drawing on the *left* of Figure 18: in the atom frame, the momentum of the photon prior to scattering is $h_{\mathrm{p}}\nu_{\mathrm{in}}/c$ and points to the right. The momentum of the photon after scattering equals $h_{\mathrm{p}}\nu_{\mathrm{out}}/c$ and points up. Momentum would therefore not be conserved in the atom's frame, if it were not for the newly acquired momentum of the hydrogen atom itself. This newly acquired momentum corresponds to newly acquired kinetic energy, which in turn came at the expense of the energy of the Ly$\alpha$ photon. The energy of the Ly$\alpha$ photon is therefore (strictly) not conserved exactly, but reduced by a small amount in each scattering event. This effect is not relevant when $\mathbf{k}_{\mathrm{in}} = \mathbf{k}_{\mathrm{out}}$ (i.e. $\mu = 1$), and maximally important when $\mathbf{k}_{\mathrm{in}} = -\mathbf{k}_{\mathrm{out}}$ (i.e. $\mu = -1$). Demanding momentum conservation in the atom frame gives additional terms of the form[29]

$$x_{\mathrm{out}} = x_{\mathrm{in}} - \frac{\mathbf{v} \cdot \mathbf{k}_{\mathrm{in}}}{v_{\mathrm{th}}} + \frac{\mathbf{v} \cdot \mathbf{k}_{\mathrm{out}}}{v_{\mathrm{th}}} + \underbrace{g(\mu - 1) + \mathcal{O}(v_{\mathrm{th}}^2/c^2)}_{\text{recoil}}, \tag{65}$$

where $g = \frac{h\nu_{\alpha}}{m_p v_{\mathrm{th}} c} = 2.6 \times 10^{-4} (T/10^4 \, \mathrm{K})^{-1/2}$ is the fractional amount of energy that is transferred per scattering event (Field, 1959). Throughout the remainder of this calculation we will ignore recoil. This may be counter-intuitive, as the change in $x$ is of the order $\sim 10^{-4}$ for *each scattering event*, and - as we will see later - Ly$\alpha$ photons can scatter $\gg 10^6$ times. However, (fortunately) this process does not act cumulatively as we will discuss in more detail at the end of this section. Adams (1971) first showed that recoil can generally be safely ignored.

For simplicity, but without loss of generality, we define a coordinate system such that $\mathbf{k}_{\mathrm{in}} = (1, 0, 0)$, and $\mathbf{k}_{\mathrm{out}} = (\mu, \sqrt{1 - \mu^2}, 0)$, i.e. the photon wavevectors lie entirely in the x-y plane (see Fig 19). If we decompose the atom's velocity into components parallel ($v_{\parallel}$) and orthogonal ($v_y$ and $v_z$) to $\mathbf{k}_{\mathrm{in}}$, then $\mathbf{v} = (v_{\parallel}, v_y, v_z)$ and

$$x_{\mathrm{out}} = x_{\mathrm{in}} - \frac{v_{\parallel}}{v_{\mathrm{th}}} + \frac{v_{\parallel}\mu}{v_{\mathrm{th}}} + \frac{v_y\sqrt{1 - \mu^2}}{v_{\mathrm{th}}} \equiv x_{\mathrm{in}} - u + u\mu + w\sqrt{1 - \mu^2}, \tag{66}$$

---

[29] We will not derive this here. The derivation is short. First show that the total momentum of the atom after scattering equals $p_{\mathrm{out}}^{\mathrm{H}} = \frac{h_p\nu_{\mathrm{in}}}{c}\sqrt{2 - 2\mu}$, where $\mu = \mathbf{k}_{\mathrm{in}} \cdot \mathbf{k}_{\mathrm{out}}$. This corresponds to a total kinetic energy $E_e = \frac{|p_{\mathrm{out}}^{\mathrm{H}}|^2}{2m_p}$, which must come at the expense of the Ly$\alpha$ photon. We therefore have $\Delta E = h_p\Delta\nu = \frac{1}{m_p}\left(\frac{h_p\nu_{\mathrm{in}}}{c}\right)^2(1 - \mu)$, which corresponds to $\Delta x = \frac{h\nu_{\alpha}}{m_p v_{\mathrm{th}} c}$ if we approximate that $\nu_{\mathrm{in}} \approx \nu_{\alpha}$.

where we have introduced the dimensionless velocity parameters $u = v_{||}/v_{\rm th}$ and $w = v_y/v_{\rm th}$ (see e.g. Ahn et al., 2000). Note that the value of $v_z$ is irrelevant in this equation, which is because it does not induce any Doppler boost on either the incoming or outgoing photon.

We were interested in calculating $R(x_{\rm out}|\mu, x_{\rm in})$. We can do this by using Eq 66 and applying probability theory. We first write

$$R(x_{\rm out}|\mu, x_{\rm in}) = \mathcal{N} \int_{-\infty}^{\infty} du \int_{-\infty}^{\infty} dw R(x_{\rm out}|\mu, x_{\rm in}, u, w) P(u|\mu, x_{\rm in}) P(w|\mu, x_{\rm in}), \tag{67}$$

where $\mathcal{N}$ is a normalization factor. Eq 66 states that when $x_{\rm out}, x_{\rm in}, \mu$ are fixed, then for a given $u$, solutions only exist when $w \equiv w_u = \frac{x_{\rm out} - x_{\rm in} + u - u\mu}{\sqrt{1-\mu^2}}$. In other words, $R(x_{\rm out}|\mu, x_{\rm in}, u, w)$ is only non-zero when $w = w_u$. We can therefore drop the integral over $w$ and write

$$R(x_{\rm out}|\mu, x_{\rm in}) = \mathcal{N} \int_{-\infty}^{\infty} du P(u|\mu, x_{\rm in}) P(w_u|\mu, x_{\rm in}). \tag{68}$$

The conditional absorption probabilities for both $w$ and $u$ cannot depend on the subsequent emission direction, as the re-emission process is set by quantum mechanics of the wavefunction describing the electron. Therefore we have $P(u|\mu, x_{\rm in}) = P(u|x_{\rm in})$ and $P(w_u|\mu, x_{\rm in}) = P(w_u|x_{\rm in})$. Also note that $w$ denotes the normalized velocity in a direction perpendicular to $\mathbf{k}_{\rm in}$. The incoming photon's frequency, $x_{\rm in}$ - and therefore the absorption probability - cannot depend on $w$. Therefore, $P(w_u|x_{\rm in}) = P(w_u) = \exp(-w_u^2)/\sqrt{\pi}$, where we assumed a Maxwell-Boltzmann distribution of the atoms' velocities.

The expression for $P(u|x_{\rm in})$ is a bit more complicated. From Bayes Theorem (Eq 60) we know that $P(u|x_{\rm in}) = P(x_{\rm in}|u)P(u)/P(x_{\rm in})$, in which $P(x_{\rm in}|u)$ denotes the absorption probability for a single atom that has a speed $u$, and $P(x_{\rm in})$ can be interpreted as a normalization factor. The scattering probability of Ly$\alpha$ photons off atoms with velocity component $u$ must scale with the (single-atom) cross-section, i.e. $P(x_{\rm in}|u) \propto \sigma_\alpha(x_{\rm in}|u) = \frac{3\lambda_\alpha^2}{8\pi} \frac{A_\alpha^2}{[\omega_\alpha(x_{\rm in}-u)v_{\rm th}/c]^2 + A_\alpha^2/4}$ (see Eq 49). If we substitute this into Eq 68 and absorb all factors that can be pulled out of the integral into the normalization constant $\mathcal{N}$, then we get

$$R(x_{\rm out}|\mu, x_{\rm in}) = \mathcal{N} \int_{-\infty}^{\infty} du \frac{\exp(-u^2)}{(x_{\rm in}-u)^2 + a_v^2} \exp\left[-\left(\frac{\Delta x - u(\mu-1)}{\sqrt{1-\mu^2}}\right)^2\right], \tag{69}$$

where we introduced $\Delta x \equiv x_{\rm out} - x_{\rm in}$. It is possible to get an analytic expression[30] for $\mathcal{N}$:

$$R(x_{\rm out}|\mu, x_{\rm in}) = \frac{a_v}{\pi^{3/2}\sqrt{1-\mu^2}\phi(x_{\rm in})} \int_{-\infty}^{\infty} du \frac{\exp(-u^2)}{(x_{\rm in}-u)^2 + a_v^2} \exp\left[-\left(\frac{\Delta x - u(\mu-1)}{\sqrt{1-\mu^2}}\right)^2\right]. \tag{71}$$

Examples of $R(x_{\rm out}|\mu, x_{\rm in})$ as a function of $x_{\rm out}$ are plotted in Figure 20. In the *left panel* we plot $R(x_{\rm out}|\mu, x_{\rm in})$ for $\mu = -0.5$, and $x_{\rm in} = -3.0$ (*red line*), $x_{\rm in} = 0.0$ (*black line*), and $x_{\rm in} = 4.0$ (*blue line*). These frequencies were chosen to represent scattering in the wing (for $x_{\rm in} = 4.0$), in the core (for $x_{\rm in} = 0.0$) and in the transition region ($x_{\rm in} = -3.0$, see Fig 17). In the *right panel* we used $\mu = 0.5$. Figure 20 shows clearly that (*i*) for the vast majority of photons $|x_{\rm in} - x_{\rm out}| \lesssim$ a few. That is, the frequency after scattering is closely related to the frequency before scattering (*ii*) $R(x_{\rm out}|\mu, x_{\rm in})$ can depend quite strongly on $\mu$. This is most clearly seen by comparing the curves for $x_{\rm in} = -3.0$ in the *left* and *right panels*.

---

[30] This can be seen as follows (note that the second line contains colors to clarify how we got from the L.H.S to the R.H.S):

$$\mathcal{N}^{-1}(\mu) = \int_{-\infty}^{\infty} dx_{\rm out} \int_{-\infty}^{\infty} du \frac{\exp(-u^2)}{(x_{\rm in}-u)^2 + a_v^2} \exp\left[-\left(\frac{\Delta x - u(\mu-1)}{\sqrt{1-\mu^2}}\right)^2\right] = \tag{70}$$

$$\color{red}{\int_{-\infty}^{\infty} du \frac{\exp(-u^2)}{(x_{\rm in}-u)^2 + a_v^2}} \color{blue}{\int_{-\infty}^{\infty} d\Delta x \exp\left[-\left(\frac{\Delta x - u(\mu-1)}{\sqrt{1-\mu^2}}\right)^2\right]} = \color{red}{\phi(x_{\rm in})\frac{\pi}{a_v}} \color{blue}{\sqrt{1-\mu^2}\sqrt{\pi}},$$

where we used that $d\Delta x = dx_{\rm out}$. We rewrote the term in red using the definition of $\phi(x)$ (see Eq 55), and that the term in blue is a Gaussian in $\Delta x$.

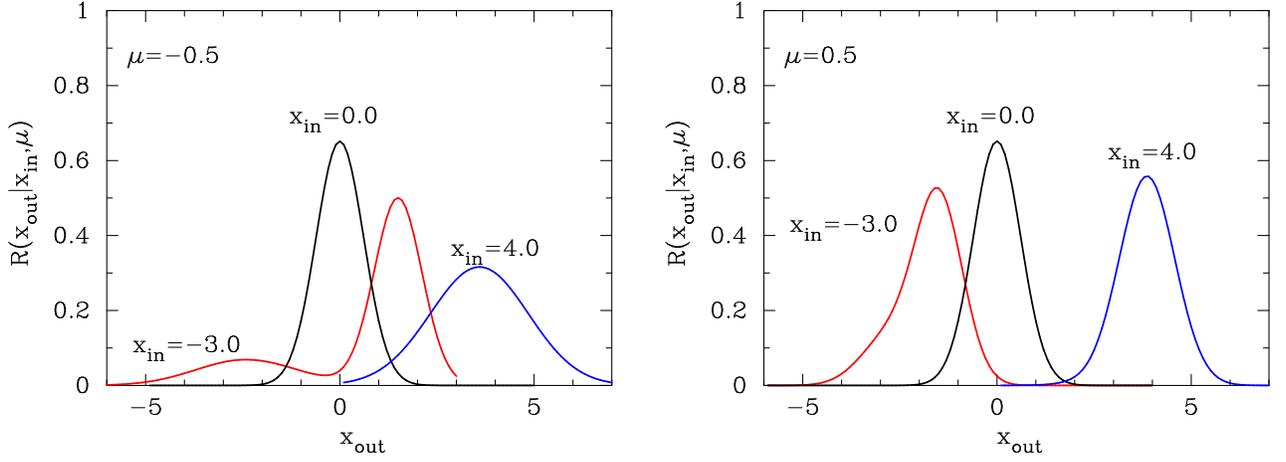

FIG. 20 $R(x_{out}|\mu, x_{in})$ is shown as a function of $x_{out}$. In the *left panel* we plot curves with $\mu = -0.5$, and $x_{in} = -3.0$ (*red line*), $x_{in} = 0.0$ (*black line*), and $x_{in} = 4.0$ (*blue line*). In the *right panel* we changed the sign of $\mu$ to $\mu = 0.5$. This Figure shows that for the vast majority of photons $|x_{in} - x_{out}| \lesssim$ a few. Thus the photon frequencies before and after scattering are closely related. Furthermore, $R(x_{out}|\mu, x_{in})$ can depend quite strongly on $\mu$. This is most clearly seen by comparing the curves for $x_{in} = -3.0$ in the *left* and *right panels*.

In many radiative transfer problem we are mostly interested in the conditional PDF $R(x_{out}|x_{in})$ (see 62) for which one needs to know the 'conditional phase function' $P(\mu|x_{in})$. We introduced the concept of the phase function in § 6.1, and discussed cases of (*i*) isotropic scattering, for which $P(\mu) = 1$, and *dipole scattering* for which $P(\mu) = \frac{3}{4}(1+\mu^2)$. As we will discuss in more detail in § 11.1, Ly$\alpha$ scattering represents a superposition of dipole and isotropic scattering. As we will see, there *is* a weak dependence of the phase function $P(\mu)$ to $x_{in}$, but only over a limited range of frequencies (see § 11). It is generally fine to ignore this effect and state that $P(\mu|x_{in}) = P(\mu)$. The conditional PDF $R(x_{out}|x_{in})$ for isotropic and dipole scattering are given by

$$R_A(x_{out}|x_{in}) = \int_{-1}^{1} d\mu R(x_{out}|\mu, x_{in}) \tag{72}$$

$$R_B(x_{out}|x_{in}) = \frac{3}{4} \int_{-1}^{1} d\mu R(x_{out}|\mu, x_{in})\Big(1+\mu^2\Big),$$

where subscript '$A$' ('$B$') refers to isotropic (dipole) scattering. Figure 21 shows type-A and type-B frequency redistribution functions for $x_{in} = -3.0, 0.0$ and 4.0. Because Ly$\alpha$ scattering is often a superposition of these two, the actual directionally averaged redistribution functions are intermediate between the two cases. As these two cases agree quite closely, they both provide a decent description of actual Ly$\alpha$ scattering.

There are three important properties of these redistribution functions that play an essential role in radiative transfer calculations. We discuss these next.

1. Photons that scatter in the wing of the line are pushed back to the line core by an amount $-\frac{1}{x_{in}}$ (Osterbrock, 1962), i.e.

$$\boxed{\langle \Delta x | x_{in} \rangle = -\frac{1}{x_{in}}}. \tag{73}$$

Demonstrating this requires some calculation. The expectation value for $\Delta x$ per scattering event is given by

$$\langle \Delta x | x_{in} \rangle \equiv \int_{-\infty}^{\infty} \Delta x \; R(x_{out}|x_{in}) dx_{out} = \frac{1}{2} \int_{-\infty}^{\infty} dx_{out} \int_{-1}^{1} d\mu \; \Delta x P(\mu) R(x_{out}|\mu, x_{in}), \tag{74}$$

where the factor of $\frac{1}{2}$ reflects that $\int_{-1}^{1} P(\mu) d\mu = 2$ (see Eq 41). For simplicity we will assume isotropic scattering, for which $P(\mu) = 1$. We previously presented an expression for $R(x_{out}|\mu, x_{in})$ (see Eq 71). Substituting this

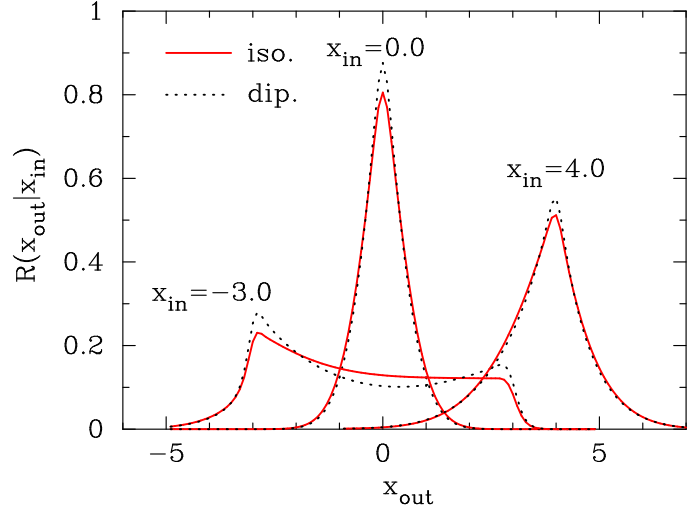

FIG. 21 The conditonal PDF $R(x_{\text{out}}|x_{\text{in}})$, obtained by marginalizing $R(x_{\text{out}}|x_{\text{in}}, \mu)$ (shown in Fig 20) over $\mu$, is shown for isotropic scattering (*red solid lines*, $R_A(x_{\text{out}}|x_{\text{in}})$) and dipole scattering (*black dotted lines*, $R_B(x_{\text{out}}|x_{\text{in}})$) for $T = 10^4$ K. For this gas temperature, $R(x_{\text{out}}|x_{\text{in}})$ is very similar for isotropic and dipole scattering.

expression into the above equation yields

$$\langle \Delta x | x_{\text{in}} \rangle = \frac{a_v}{2\pi^{3/2}\phi(x_{\text{in}})} \int_{-1}^{1} \frac{d\mu}{\sqrt{1-\mu^2}} \int_{-\infty}^{\infty} \frac{du \exp(-u^2)}{(x_{\text{in}} - u)^2 + a_v^2} \int_{-\infty}^{\infty} d\Delta x \ \Delta x \ \exp\left[ -\left( \frac{\Delta x - u(\mu - 1)}{\sqrt{1-\mu^2}} \right)^2 \right], \quad (75)$$

where we replaced the integral over $x_{\text{out}}$ with an integral over $\Delta x$. Note that the term in *blue* also represents a 'standard' integral, which equals [31] $u(\mu - 1)\sqrt{\pi(1-\mu^2)}$. The factor containing $\mu$'s can be taken outside of the integral over $d\Delta x$. The integral over $\mu$ simplifies to $\int_{-1}^{1} d\mu(\mu - 1) = -2$, and we are left with a single integral over $u$. We will further assume that $|x_{\text{in}}| \gg 1$, in which case we get

$$\langle \Delta x | x_{\text{in}} \rangle \sim -\frac{a_v}{\pi\phi(x_{\text{in}})} \int_{-\infty}^{\infty} du \frac{u \exp(-u^2)}{(x_{\text{in}} - u)^2} \sim -\frac{a_v}{\pi\phi(x_{\text{in}})x_{\text{in}}^2} \int_{-\infty}^{\infty} du \ u \exp(-u^2) \Big( \underbrace{1}_{\text{odd } = 0} + \underbrace{2\frac{u}{x_{\text{in}}}}_{\text{even } \neq 0} \Big) =$$

$$= -\frac{2a_v}{\pi\phi(x_{\text{in}})x_{\text{in}}^3} \int_{-\infty}^{\infty} du \ u^2 \exp(-u^2) = -\frac{2a_v}{\pi\phi(x_{\text{in}})x_{\text{in}}^3} \frac{1}{2}\sqrt{\pi} = -\frac{1}{x_{\text{in}}}, \quad (76)$$

where the minus sign appeared after performing the integral over $\mu$. In the last step we used that $\phi(x_{\text{in}}) = a_v/(\sqrt{\pi}x_{\text{in}}^2)$ in the wing of the line profile. This photon is more likely absorbed by atoms that are moving towards the photon, as it would appear closer to resonance for these atoms (i.e. $P(u|x_{\text{in}})$ is larger for those photons with $\mathbf{v} \cdot \mathbf{k}_{\text{in}} < 0$). This photon is likely to experience a Doppler boost to a higher frequency in the frame of the atom. Isotropic (or dipole) re-emission of the photon - on average - conserves this enhancement of the photon's frequency, which pushed it closer to resonance.

This result is very important: it implies that as a Lyα photon is far in the wing at $x_{\text{in}}$, resonant scattering exerts a 'restoring force' which pushes the photon back to line resonance. This restoring force generally overwhelms the energy losses resulting from atomic recoil: Eq 65 indicates that recoil introduces a much smaller average $\Delta x \sim -2.6 \times 10^{-4}(T/10^4 \text{ K})^{-1/2}$, i.e. if a photon finds itself on the red side of line center, then the restoring force pushes the photon back more to line center than recoil pulls it away from it.

2. The r.m.s change in the photon's frequency as it scatters corresponds to 1 Doppler width (Osterbrock 1962).

$$\boxed{\sqrt{\langle \Delta x^2 | x_{\text{in}} \rangle} = 1}. \quad (77)$$

---

[31] Namely that $\int_{-\infty}^{\infty} dx \ x \exp(-a[x - b]^2) = b\sqrt{\pi/a}$ with $a = (1-\mu^2)^{-1}$ and $b = u(\mu - 1)$.

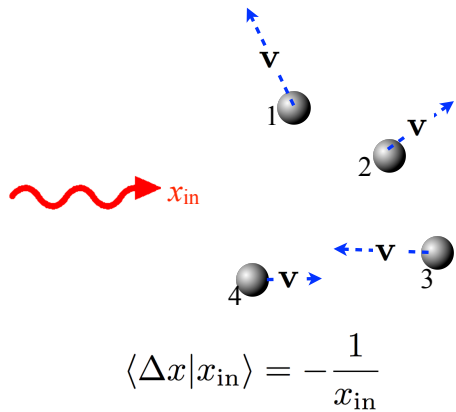

$$\langle \Delta x | x_{\rm in} \rangle = -\frac{1}{x_{\rm in}}$$

FIG. 22 Visual explanation of why resonant scattering tends to push Ly$\alpha$ photons far in the wing of the line profile back to the line core. This represents a Ly$\alpha$ photon far in the red wing of the line profile (i.e. $x_{\rm in} \ll 3$). This photon is more likely to be scattered by atoms that are moving towards the photon (in this case atom number 3), as it appears closer to resonance to them. Scattering by these atoms enhances the photon's frequency, and therefore pushes it back towards the core (this is illustrated in Fig 22). The average change in photon frequency is $\langle \Delta x | x_{\rm in} \rangle = -x_{\rm in}^{-1}$.

This can be derived with a calculation that is very similar to the above calculation.

3. $R(x_{\rm out} | x_{\rm in}) = R(x_{\rm in} | x_{\rm out})$. This can be verified by substituting $u = y - \Delta x$ into Eq 71. After some algebra one obtains an expression in which $y$ replaces $u$, in which $x_{\rm out}$ replaces $x_{\rm in}$, and in which $x_{\rm out}$ replaces $x_{\rm in}$.

These three properties of the redistribution functions offer key insights into the Ly$\alpha$ radiative transfer problem.

### 7.4. IV: 'Destruction' Term

Absorption of a Ly$\alpha$ photon by a hydrogen atom is generally followed by re-emission of the Ly$\alpha$ photon. Moreover, Ly$\alpha$ photons can be absorbed by something different than hydrogen atoms, which also leads to their destruction. We briefly list the most important processes below:

1. **Dust.** Dust grains can absorb Ly$\alpha$ photons. The dust grain can *scatter* the Ly$\alpha$ photon with a probability which is given by its albedo, $A_{\rm d}$. Dust plays an important role in Ly$\alpha$ radiative transfer, and we will return to this later (see § 10.1). Absorption of a Ly$\alpha$ photon by a dust grain increases its temperature, which causes the grain to re-radiate at longer wavelengths, and thus to the destruction of the Ly$\alpha$ photon. This process can be included in the radiative transfer equation by replacing

$$n_{\rm HI} \sigma_{\alpha,0} \phi(x) \to n_{\rm HI}[\sigma_{\alpha,0} \phi(x) + \sigma_{\rm dust}(x)], \tag{78}$$

where $\sigma_{\rm dust}(x)$ denotes the *total* dust cross-section at frequency $x$, i.e. $\sigma_{\rm dust}(x) = \sigma_{\rm dust,a}(x) + \sigma_{\rm dust,s}(x)$, where the subscript 'a' ('s') stands for 'absorption' ('scattering'). Eq 78 indicates that the cross-section $\sigma_{\rm dust}(x)$ is a cross-section *per hydrogen atom*. This definition implies that $\sigma_{\rm dust}(x)$ is not just a property of the dust grain, as it must also depend on the number density of dust grains (if there were no dust grains, then we should not have to add any term). In addition to this, the dust absorption cross section $\sigma_{\rm dust,s}(x)$ (and also the albedo $A_{\rm d}$) must depend on the dust properties. For example, Laursen et al. (2009a) shows that $\sigma_{\rm dust} = 4 \times 10^{-22}(Z_{\rm gas}/0.25 Z_\odot)$ cm$^{-2}$ for SMC type dust (dust with the same properties as found in the Small Magellanic Cloud), and $\sigma_{\rm dust} = 7 \times 10^{-22}(Z_{\rm gas}/0.5 Z_\odot)$ cm$^{-2}$ for LMC (Large Magellanic Cloud) type dust. Here, $Z_{\rm gas}$ denotes the metallicity of the gas. This parametrization of $\sigma_{\rm dust}$ therefore assumes that the number density of dust grains scales linearly with the overall gas metallicity, which is a good approximation for $Z \gtrsim 0.3 Z_\odot$, but for $Z \lesssim 0.3 Z_\odot$ the scatter in dust-to-gas ratio increases (e.g. Draine et al. 2007, Rémy-Ruyer et al. 2014, Schneider et al. 2016). Figure 23 shows $\sigma_{\rm dust}$ for dust properties inferred for the LMC (*solid line*), and SMC (*dashed line*). This Figure shows that the frequency dependence of the dust absorption cross section around the Ly$\alpha$ resonance is weak, and in practise it can be safely ignored. Interestingly, as we will discuss in § 8.3, dust can have a highly frequency dependent impact on the Ly$\alpha$ radiation field, in spite of the weak frequency dependence of the dust absorption cross-section.

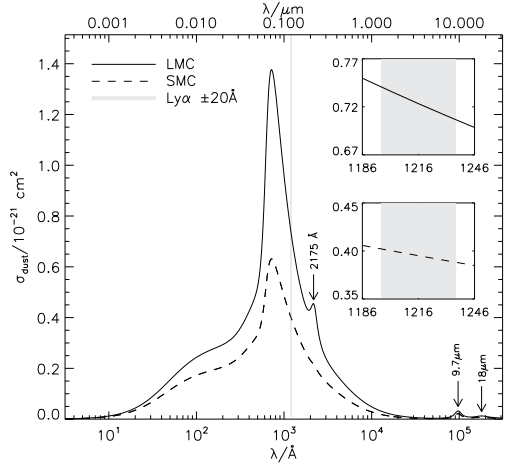

FIG. 23 This Figure shows grain averaged absorption cross section of dust grains *per hydrogen atom* for SMC/LMC type dust (*solid/dashed line*, see text). The *inset* shows the cross section in a narrower frequency range centered on Lyα, where the frequency dependence depends practically linearly on wavelength. However, this dependence is so weak that in practise it can be safely ignored (*Credit: from Figure 1 of Laursen et al. 2009a ©AAS. Reproduced with permission.*). We we discuss in § 8.3, in spite of the weak frequency dependence of the dust absorption cross-section around the Lyα resonance, dust can have a highly frequency dependent impact in Lyα spectra.

2. **Molecular Hydrogen.** Molecular hydrogen has two transitions that lie close to the Lyα resonance: (*a*) the $v = 1 - 2P(5)$ transition, which lies $\Delta v = 99$ km s$^{-1}$ redward of the Lyα resonance, and (*b*) the $1 - 2R(6)$ transition which lies $\Delta v = 15$ km s$^{-1}$ redward of the Lyα resonance. Vibrationally excited $H_2$ may therefore convert Lyα photons into photons in the $H_2$ Lyman bands (Neufeld, 1990, and references therein), and thus effectively destroy Lyα. This process can be included in a way that is very similar to that of dust, and by including

$$n_{\rm HI}\sigma_{\alpha,0}\phi(x) \rightarrow n_{\rm HI}[\sigma_{\alpha,0}\phi(x) + f_{\rm H_2}\sigma_{\rm H_2}(x)], \tag{79}$$

where $f_{\rm H_2} \equiv n_{\rm H_2}/n_{\rm HI}$ denotes the molecular hydrogen fraction. This destruction process is often overlooked, but it is important to realize that Lyα can be destroyed efficiently by molecular hydrogen. Neufeld (1990) provides expressions for fraction of Lyα that is allowed to escape as a function of $f_{\rm H_2}$ and HI column density $N_{\rm HI}$.

3. **Collisional Mixing of the $2s$ and $2p$ Levels.** Lyα absorption puts a hydrogen atom in its $2p$ state, which has a life-time of $t = A_\alpha^{-1} \sim 10^{-9}$ s. During this short time, there is a finite probability that the atom interacts with nearby electrons and/or protons. These interactions can induce transitions of the form $2p \rightarrow 2s$. Once in the $2s$-state, the atom decays back to the ground-state by emitting two photons. This process is known as collisional deexcitation of the $2p$ state. Collisional de-excitation from the $2p$ state becomes more probable at high gas densities, and is predominantly driven by free protons. The probability that this process destroys the Lyα photon, $p_{\rm dest}$, at any scattering event is given by $p_{\rm dest} = \frac{n_p C_{2p2s}}{n_p C_{2p2s} + A_\alpha}$. Here, $n_{\rm p}$ denotes the number density of free protons, and $C_{2p2s} = 1.8 \times 10^{-4}$ cm$^3$ s$^{-1}$ (e.g. Dennison et al., 2005, and references therein) denotes the collisional rate coefficient. This process can be included by rescaling the scattering redistribution function (see § 7.3) to

$$R(x_{\rm out}|x_{\rm in}) \rightarrow R(x_{\rm out}|x_{\rm in}) \times (1 - p_{\rm dest}). \tag{80}$$

This ensures that for each scattering event, there is a finite probability ($p_{\rm dest}$) that a photon is destroyed.

4. **Other.** There are other processes that can destroy Lyα photons, but which are less important. These can trivially be included in Monte-Carlo codes that describe Lyα radiative transfer (see § 9). These processes include: (*i*) Lyα photons can photoionize hydrogen atoms *not* in the ground state. The photoionisation cross-section from the $n = 2$ level by Lyα photons is $\sigma_{\rm ion}^{\rm Ly\alpha} = 5.8 \times 10^{-19}$ cm$^2$ (e.g. Cox 2000, p 108). This requires a non-negligible populations of atoms in the $n = 2$ state which can occur in very dense media (see e.g. Dijkstra et al. 2016) (*ii*) Lyα photons can detach the electron from the H$^-$ ion. The cross-section for this process is $\sigma = 5.9 \times 10^{-18}$ cm$^{-2}$ (e.g. Shapiro & Kang, 1987) for Ly$\alpha$ photons, which is almost an order of magnitude larger than the photoionisation cross-section from the $n = 2$ level at the Ly$\alpha$ frequency. So, unless the H$^-$ number density exceeds $0.1[n_{2p} + n_{2s}]$, where $n_{2s/2p}$ denotes the number density of H-atoms in the 2s/2p state, this process is not important.

### 7.5. Ly$\alpha$ Propagation through HI: Scattering as Double Diffusion Process

Scattering of Ly$\alpha$ photons is often compared to a diffusion process, in which the photons undergo a random walk in space and frequency as they scatter off H atoms. Indeed, known analytic solutions to the radiative transfer equation are possible only under certain idealized scenarios (which are discussed below), for which the radiative transfer equation transforms into a diffusion equation. To demonstrate how this transformation works, and to gain some insight into this diffusion process we rewrite the transfer equation (Eq 58) as a diffusion equation. We first simplify Eq 58 in a number of steps:

1. First, we assume that the Ly$\alpha$ radiation field is isotropic, i.e. scattering completely eliminates any directional dependence of $I_\nu(\mathbf{n})$. Under this assumption we can replace the intensity $I_\nu(\mathbf{n})$ with the angle-averaged intensity $J_\nu \equiv \frac{1}{4\pi} \int d\Omega I_\nu(\mathbf{n})$.

2. Second, we replace frequency $\nu$ with the dimensionless frequency variable $x$, introduced in § 6.4.

3. Third, we ignore destruction processes and rewrite $\alpha_\nu^{\mathrm{HI}}(s) = n_{\mathrm{HI}}(s)\sigma_\alpha(\nu) = n_{\mathrm{HI}}\sigma_{\alpha,0}\phi(x)$.

4. Fourth, we define $d\tau = n_{\mathrm{HI}}(s)\sigma_{\alpha,0}ds$ and obtain

$$\frac{\partial J(x)}{\partial \tau} = -\phi(x)J(x) + S_x(\tau) + \int dx' \phi(x')J(x')R(x|x'), \tag{81}$$

where $S_x(\tau) \equiv j_x(\tau)/(n_{\mathrm{HI}}\sigma_{\alpha,0})$ denotes the 'source' function. In the integral $x$ denotes the frequency of the photon after scattering, and $x'$ denotes the frequency of the photon before scattering. Eq 81 is an integro-differential equation, which is notoriously difficult to solve.

Fortunately, this integro-differential can be transformed into a diffusion equation by Taylor expanding $J(x')\phi(x')$ around $x$ as we demonstrate next. To keep our notation consistent with § 7.3, we denote the photon frequency before (after) scattering with $x_{\mathrm{in}}$ ($x_{\mathrm{out}}$). To shorten the notation, we define $f(x) \equiv J(x)\phi(x)$. We would like to rewrite $f(x_{\mathrm{in}})$ as

$$f(x_{\mathrm{in}}) = f(x_{\mathrm{out}}) + (x_{\mathrm{in}} - x_{\mathrm{out}})\frac{\partial f}{\partial x} + \frac{1}{2}(x_{\mathrm{in}} - x_{\mathrm{out}})^2\frac{\partial^2 f}{\partial x^2} + ... = f(x_{\mathrm{out}}) - \Delta x \frac{\partial f}{\partial x} + \frac{1}{2}\Delta x^2 \frac{\partial^2 f}{\partial x^2} + ..., \tag{82}$$

where the derivatives are evaluated at $x_{\mathrm{out}}$. Dropping terms that contain $\Delta x^3, \Delta x^4, ...$, we can write the scattering term as

$$\int dx_{\mathrm{in}} f(x_{\mathrm{in}}) R(x_{\mathrm{out}}|x_{\mathrm{in}}) \approx \int_{-\infty}^{\infty} dx_{\mathrm{in}} R(x_{\mathrm{out}}|x_{\mathrm{in}})\left(f(x_{\mathrm{out}}) - \Delta x \frac{\partial f}{\partial x} + \frac{1}{2}\Delta x^2 \frac{\partial^2 f}{\partial x^2} + ...\right) =$$

$$\underset{R(x_{\mathrm{in}}|x_{\mathrm{out}})=R(x_{\mathrm{out}}|x_{\mathrm{in}})}{=} f(x_{\mathrm{out}})\underbrace{\int_{-\infty}^{\infty} dx_{\mathrm{in}} R(x_{\mathrm{in}}|x_{\mathrm{out}})}_{=1} - \frac{\partial f}{\partial x}\underbrace{\int_{-\infty}^{\infty} dx_{\mathrm{in}} \Delta x R(x_{\mathrm{in}}|x_{\mathrm{out}})}_{\langle \Delta x | x_{\mathrm{in}} \rangle = -\frac{1}{x_{\mathrm{out}}}, \text{ Eq 76}} + \frac{1}{2}\frac{\partial^2 f}{\partial x^2}\underbrace{\int_{-\infty}^{\infty} dx_{\mathrm{in}} \Delta x^2 R(x_{\mathrm{in}}|x_{\mathrm{out}})}_{\langle \Delta x^2 | x_{\mathrm{in}} \rangle = 1, \text{ Eq 77}} =$$

$$= f(x_{\mathrm{out}}) + \frac{1}{x_{\mathrm{out}}}\frac{\partial f}{\partial x} + \frac{1}{2}\frac{\partial^2 f}{\partial x^2} \tag{83}$$

This term can be further simplified by replacing $f(x_{\mathrm{out}})$ with $\phi(x_{\mathrm{out}})J(x_{\mathrm{out}})$ and taking the derivatives. Note that we derived Eq 77 and Eq 76 assuming that $|x| \gg 1$. This same assumption allows us to further simplify the equation as we see below. For brevity we only write $\phi$ for $\phi(x_{\mathrm{out}})$ etc. The term then becomes

$$J\phi + \frac{\phi}{x_{\mathrm{out}}}\frac{\partial J}{\partial x} + \frac{J}{x_{\mathrm{out}}}\frac{\partial \phi}{\partial x} + \frac{1}{2}\frac{\partial}{\partial x}\left(\phi\frac{\partial J}{\partial x} + J\frac{\partial \phi}{\partial x}\right) = J\phi - \frac{\phi}{x_{\mathrm{out}}}\frac{\partial J}{\partial x} + \frac{J\phi}{x_{\mathrm{out}}^2} + \frac{\phi}{2}\frac{\partial^2 J}{\partial x^2} = \tag{84}$$

$$J\phi\left(1 + \frac{1}{x_{\mathrm{out}}^2}\right) + \frac{1}{2}\frac{\partial}{\partial x}\phi\frac{\partial J}{\partial x} \approx J\phi + \frac{1}{2}\frac{\partial}{\partial x}\phi\frac{\partial J}{\partial x},$$

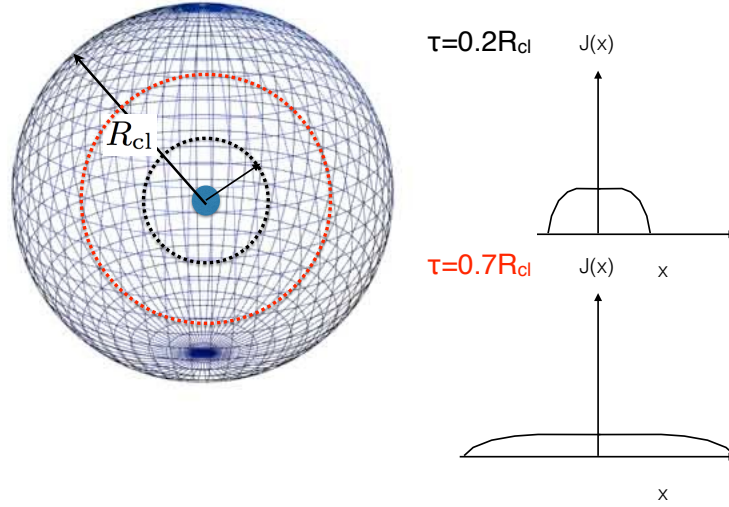

FIG. 24 Lyα scattering can be described as a double diffusion process, in which photons diffuse in frequency space as they scatter. In this case photons are emitted in the center of the sphere, at line center. As photons scatter outwards, the spectrum of Lyα photons broadens. Analytic solutions to the Lyα transfer equation indicate that the spectrum remains remarkably constant over a range of frequencies (this range increases as we move outward from the center of the sphere).

where we used that $\frac{\partial \phi}{\partial x} = -\frac{2\phi}{x}$, $\frac{\partial^2 \phi}{\partial^2 x} = \frac{6\phi}{x^2}$ (see Eq 55), and that $|x_{\rm out}| \gg 1$ in the wing of the line profile. When we substitute this back into Eq 81, and revert to its original notation, we finally find that (see e.g. Rybicki & dell'Antonio, 1994; Rybicki, 2006, and references therein)

$$\boxed{\frac{\partial J}{\partial \tau} = \frac{1}{2}\frac{\partial}{\partial x}\phi(x)\frac{\partial J}{\partial x} + S_x(\tau)}. \tag{85}$$

This equation is known as the 'Fokker-Planck' transfer equation[32], and corresponds to a diffusion equation with a diffusion coefficient $\phi(x)$ (see https://en.wikipedia.org/wiki/Diffusion_equation), where $\tau$ takes the role of the time variable and $x$ denotes the role of space variable. Eq 85 therefore states that as photons propagate outwards (as a result of scattering), they diffuse in frequency direction (but with a slight tendency for the photons to be pushed back to the line core). As photons diffuse further into the wings of the absorption line profile, their mean free path increases, which in turn increases their escape probability. Also note that

## 8. BASIC INSIGHTS AND ANALYTIC SOLUTIONS

In this section, we show that the concept of Lyα transfer as a diffusion process in real and frequency space can offer intuitive insights into some basic aspects of the Lyα transfer process. These aspects include (*i*) how many times a Lyα photon scatters, (*ii*) how long it takes to escape, and (*iii*) the emerging spectrum for a static, uniform medium in § 8.1, for an outflowing/contracting uniform medium in § 8.2, and for a multiphase medium in § 8.3.

### 8.1. Lyα Transfer through Uniform, Static Gas Clouds

We consider a source of Lyα photons in the center of a static, homogeneous sphere, whose line-center optical depth from the center to the edge equals $\tau_0$, where $\tau_0$ is extremely large, say $\tau_0 = 10^7$. This line-centre optical depth corresponds to an HI column density of $N_{\rm HI} = 1.6 \times 10^{20}$ cm$^{-2}$ (see Eq 55). We further assume that the central source emits all Ly$\alpha$ photons at line center (i.e. $x = 0$). As the photons resonantly scatter outwards, they diffuse outward in frequency space. Figure 24 illustrates that as the photons diffuse outwards in real space, the spectral energy distribution of Ly$\alpha$ flux, $J(x)$, broadens. If we were to measure the spectrum of Ly$\alpha$ photons crossing some arbitrary radial shell, then we would find that $J(x)$ is constant up to $\sim \pm x_{\rm max}$ beyond which it drops off fast. For Ly$\alpha$ photons in the core of the line profile, the mean free path is negligible compared to the size of the sphere: the mean free path at frequency $x$ is $[\tau_0 \phi(x)]^{-1}$ *in units of the radius of the sphere*. Because each scattering event changes the frequency of the Ly$\alpha$ photon, the mean free path of each photon changes with each scattering event. From the shape of the redistribution function (see Fig 21) we expect that on rare occasions Ly$\alpha$ photons will be scattered far from resonance into the wing of the line (i.e. $|x| \gtrsim 3$). In the wing, the mean free path of the photon increases by orders of magnitude.

---

[32] There exists various corrections to this Fokker Planck equation. Basko (1981) shows how adding recoil can be incorporated by replacing $\phi(x)\frac{\partial J}{\partial x} \rightarrow \phi(x)\frac{\partial J}{\partial x} + \frac{h_{\rm P}\Delta\nu_\alpha J}{k_{\rm B}T}$ (see Rybicki 2006). Rybicki (2006) introduces further corrections based on requiring that the redistribution satisfies 'detailed balance' (microscopic reversibility of a process: in equilibrium, each elementary process should be equilibrated by its reverse process), which requires that $R(\nu_{\rm in}, \nu_{\rm out})\nu_{\rm out}^2 \exp\left(-\frac{h_{\rm P}\nu_{\rm out}}{k_{\rm B}T}\right) = R(\nu_{\rm out}, \nu_{\rm in})\nu_{\rm in}^2 \exp\left(-\frac{h_{\rm P}\nu_{\rm in}}{k_{\rm B}T}\right)$ instead of our assumed $R(\nu_{\rm in}, \nu_{\rm out}) = R(\nu_{\rm out}, \nu_{\rm in})$.

From the redistribution function we know that - once in the wing - there is a slight tendency to be scattered back into the core of the line profile. Specifically, we showed that $\langle \Delta x | x_{\rm in} \rangle = -\frac{1}{x_{\rm in}}$). We therefore expect photons that find themselves in the wing of the line profile, at frequency $x$, to scatter $N_{\rm scat} \sim x^2$ times before returning to the core. During this 'excursion' back to the core, the photon will diffuse a distance $D \sim \sqrt{N_{\rm scat}} \times \lambda_{\rm mfp}(x) \approx \sqrt{N_{\rm scat}}/[\tau_0 \phi(x)]$ away from the center of the sphere (recall that this is in units of the radius of the sphere). If we now set this displacement equal to the size of the sphere, i.e. $D = \sqrt{N_{\rm scat}}/[\tau_0 \phi(x)] = x/[\tau_0 \phi(x)] = 1$, and solve for $x$ using that $\phi(x) = a_v/[\sqrt{\pi}x^2]$, we find (Adams, 1972; Harrington, 1973; Neufeld 1990)

$$\boxed{x_{\rm p} = \pm[a_v \tau_0/\sqrt{\pi}]^{1/3}}, \tag{86}$$

where $x_{\rm p}$ denotes the frequency at which the emerging spectrum peaks. Photons that are scattered to frequencies[33] $|x| < |x_{\rm p}|$ will return to line center before they escape from the sphere (where they have negligible chance to escape). Photons that are scattered to frequencies $|x| > |x_{\rm p}|$ can escape more easily, but there are fewer of these photons because: (*i*) it is increasingly unlikely that a single scattering event displaces the photon to a larger $|x|$, and (*ii*) photons that wish to reach $|x| \gg |x_{\rm p}|$ through frequency diffusion via a series of scattering events are likely to escape from the sphere before they reach this frequency. We can also express the location of the two spectral peaks at $\pm x_{\rm p}$ in terms of a velocity off-set and an HI column density as

$$\Delta v_{\rm p} = |x_{\rm p}| v_{\rm th} \approx 160 \left(\frac{N_{\rm HI}}{10^{20}~{\rm cm}^{-2}}\right)^{1/3} \left(\frac{T}{10^4~{\rm K}}\right)^{1/6}~{\rm km~s}^{-1}. \tag{87}$$

That is, the full-width at half maximum of the Ly$\alpha$ line can exceed $2\Delta v_{\rm p} \sim 300$ km s$^{-1}$ for a static medium in which the thermal velocity dispersion of the atoms is only $\sim 10$ km s$^{-1}$. Ly$\alpha$ scattering thus broadens spectral lines, which implies that we must exercise caution when interpreting observed Ly$\alpha$ spectra.

We showed above that the spectrum of Ly$\alpha$ photons emerging from the center of an extremely opaque object to have two peaks at $x_{\rm p} \sim \pm[a_v \tau_0/\sqrt{\pi}]^{1/3}$. More generally, $x_{\rm p} = \pm k[a_v \tau_0/\sqrt{\pi}]^{1/3}$, where $k$ is a constant of order unity which depends on geometry (i.e. $k = 1.1$ for a slab [Harrington 1973, Neufeld 1990], and $k = 0.92$ for a sphere [Dijkstra et al. 2006]). This derivation required that photons escaped in a *single excursion*[34]: that is, photons must have been scattered deep enough into the wing (which starts for $|x| > 3$, see Fig 17) to be able to undergo a non-negligible number of wing scattering events before returning to core. So formally our analysis is valid only when $x \gg 3$ when the Ly$\alpha$ photons first start their excursion. Another way of phrasing this requirement is that $x_{\rm p} \gg 3$, or - when expressed in terms of an optical depth $\tau_0$ - when $a_v \tau_0 = \sqrt{\pi}(x_{\rm p}/k)^3 \gtrsim 1600(x_{\rm p}/10)^3$. Indeed, analytic solutions of the full spectrum emerging from static optically thick clouds appear in good agreement with full Monte-Carlo calculations (see § 9) when $a_v \tau_0 \gtrsim 1000$ (e.g. Neufeld 1990, Dijkstra et al. 2006).

There are other interesting aspects of Ly$\alpha$ transfer that we can discuss: the first is the mean number of scattering events that each Ly$\alpha$ photon undergoes, $N_{\rm scat}$. This number of scattering events was calculated by Adams (1972),

---

[33] Apart from a small recoil effect that can be safely ignored (Adams 1971), photons are equally likely to scatter to the red and blue sides of the resonance.

[34] Escape in a 'single excursion' can be contrasted with escape in a 'single flight': gases with lower $N_{\rm HI}$ can become optically thin to Ly$\alpha$ photons when they first scattered into the wing of the line profile. For example, gas with $N_{\rm HI} = 10^{17}$ cm$^{-2}$ has a line center optical depth $\tau_0 = 5.9 \times 10^3 (T/10^4~{\rm K})^{-1/2}$ (Eq 55). However, Figure 17 shows that the cross-section is $\gtrsim 4$ orders of magnitude smaller when $|x| \gtrsim 3$. A photon that first scattered into the wing would be free to escape from this gas without further scattering.

## Key quantities in Lyα transfer.

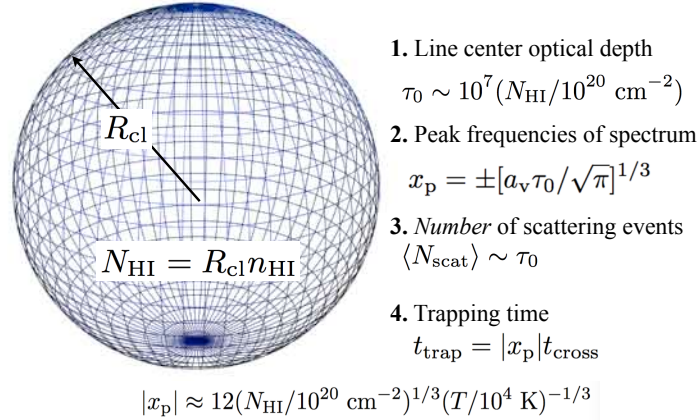

1. Line center optical depth
$$\tau_0 \sim 10^7 (N_{\rm HI}/10^{20}~{\rm cm}^{-2})$$

2. Peak frequencies of spectrum
$$x_{\rm p} = \pm[a_v \tau_0/\sqrt{\pi}]^{1/3}$$

3. *Number* of scattering events
$$\langle N_{\rm scat} \rangle \sim \tau_0$$

4. Trapping time
$$t_{\rm trap} = |x_{\rm p}|t_{\rm cross}$$

$$|x_{\rm p}| \approx 12(N_{\rm HI}/10^{20}~{\rm cm}^{-2})^{1/3}(T/10^4~{\rm K})^{-1/3}$$

FIG. 25 Lyα radiative transfer through a static, uniform sphere has several interesting features. We denote the line-center optical depth from the center to the edge of the sphere with $\tau_0$. Photons emitted in the center of the sphere diffuse in frequency space as they scatter outward, and emerge with a characteristic double peaked spectrum, for which the peaks occur at $x_{\rm p} = \pm[a_v \tau_0/\sqrt{\pi}]^{1/3}$. It takes an average of $N_{\rm scat} \sim \tau_0$ scattering events for a Lyα photon to escape, which 'traps' the photon inside the cloud for a time $t_{\rm trap} = |x_{\rm p}|t_{\rm cross}$ (where $t_{\rm cross} = R/c$, $|x_{\rm p}| \approx 12(N_{\rm HI}/10^{20}~{\rm cm}^{-2})^{1/3}(T/10^4~{\rm K})^{-1/3}$).

after observing that his numerical results implied that the Lyα spectrum of photons was flat within some frequency range centered on $x = 0$ (as in Fig 24), where the range increases with distance from the center of the sphere. Adams (1972) noted that for a flat spectrum, the scattering rate at frequency $x$ is $\propto \phi(x)$. Since partially coherent scattering does not change the frequency much (recall that $\sqrt{\langle \Delta x^2 | x_{\rm in} \rangle} = 1$, see Eq 77), we expect that if we pick a random photon after a scattering event, then the probability that this photon lies in the range $x \pm dx/2$ equals $\phi(x)dx$ (this means that we are more likely to pick a photon close to the core, which simply reflects that these photons scatter more). Adams (1972) then noted that photons at $x$ typically scatter $x^2$ times before returning to the core, and argues that the probability that a photon *first* scattered into the frequency range $x \pm dx/2$ equals $\phi(x)dx/x^2$. The probability that a photon scatters to some frequency which allows for escape in a single excursion then equals

$$P_{\rm esc} = \int_{-\infty}^{-x_{\rm p}} \frac{\phi(x)dx}{x^2} + \int_{x_{\rm p}}^{\infty} \frac{\phi(x)dx}{x^2} = 2\int_{x_{\rm p}}^{\infty} \frac{\phi(x)dx}{x^2} \underset{\rm wing}{\approx} \frac{2a_v}{3\sqrt{\pi}x_{\rm p}^3} \underset{\rm Eq~86}{=} \frac{2\sqrt{\pi}}{3\tau_0 k^3}. \tag{88}$$

We thus expect photons to escape after $N_{\rm scat} = 1/P_{\rm esc}$ scattering events, which equals

$$\boxed{N_{\rm scat} = C\tau_0}, \tag{89}$$

where $C \equiv \frac{3k^3}{2\sqrt{\pi}} \approx 1.1$ for a slab, and $C \approx 0.6$ for a sphere. This is an important result: Lyα photons typically scatter $\sim \tau_0$ times before escaping from extremely opaque media. This differs from standard random walks where a photon would scatter $\propto \tau^2$ times before escaping. The reduced number of scattering events is due to frequency diffusion, which forces photons into the wings of the line, where they can escape more easily.

We can also estimate the time it takes for a photon to escape. We now know that photons escape in a single excursion with peak frequency $x_{\rm p} = \pm[a_v \tau_0/\sqrt{\pi}]^{1/3}$. During this excursion the photon scattered $x_{\rm p}^2$ times in the wing of the line profile. We also know it took $N_{\rm scat} = 0.6\tau_0$ scattering events on average before the excursion started. Generally, $N_{\rm scat} = 0.6\tau_0 \gg x_{\rm p}^2$, and the vast majority of scattering events occurred in the line core, where the mean free path was very short. The total distance that the photon travelled while scattering in the core is $D_{\rm core} = N_{\rm scat}\lambda_{\rm mfp} \sim \tau_0 \times \tau_0^{-1} \sim 1$, where we used that the mean free path at line center is $\tau_0^{-1}$ (in units of the radius of the sphere). The total distance that was travelled in the wing - i.e. during the excursion - equals $D_{\rm wing} \sim x_{\rm p}^2 \times [\tau_0\phi(x_{\rm p})]^{-1} = x_{\rm p}^4\sqrt{\pi}/(a_v\tau_0)$. If we now substitute the expression for $x_{\rm p}$, and add the distance travelled in the core and in the wing we get

$$D = D_{\rm core} + D_{\rm wing} \approx 1 + \left(\frac{a_v\tau_0}{\sqrt{\pi}}\right)^{1/3} = 1 + |x_{\rm p}| \approx |x_{\rm p}|, \tag{90}$$

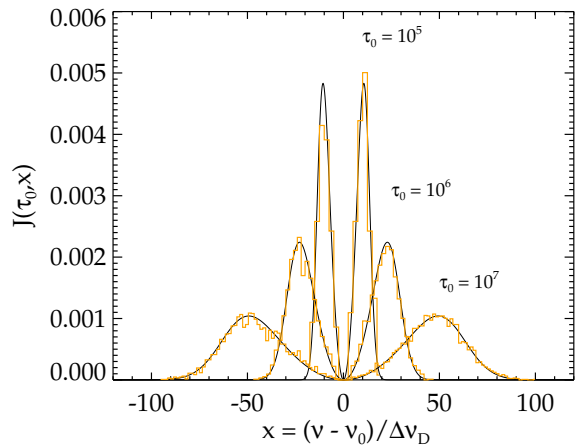

FIG. 26 Lyα spectra emerging from a uniform spherical, static gas cloud surrounding a central Lyα source which emits photons at line centre $x = 0$. The total line-center optical depth, $\tau_0$ increases from $\tau_0 = 10^5$ (*narrow histogram*) to $\tau_0 = 10^7$ (*broad histogram*). The *solid lines* represent analytic solutions (*Credit: from Figure* **A2** *of Orsi et al. 2012, 'Can galactic outflows explain the properties of Lyα emitters?', MNRAS, 425, 87O*).

where we used that $|x_p| \gg 1$, i.e. that $D_{\text{wing}} \gg D_{\text{core}}$. If we express this in terms of travel time, then the vast majority of scattering events take up negligible time. The total time it takes for Lyα photons to escape - or the total time for which they are 'trapped' - thus equals

$$\boxed{t_{\text{trap}} = |x_p| t_{\text{cross}}}, \quad |x_p| \approx 12 (N_{\text{HI}}/10^{20} \text{ cm}^{-2})^{1/3} (T/10^4 \text{ K})^{-1/3}. \tag{91}$$

This is an interesting, and often overlooked result: for $\tau_0 \sim 10^7$ we now know that Lyα photons scatter 10 million times on average before escaping. Yet, they are only trapped for $\sim 15$ light crossing times, which is not long.

The last thing I will only mention is that the diffusion equation (Eq 85) can be solved analytically for the angle averaged intensity $J(x,\tau)$ for static, uniform gaseous spheres, slabs, and cubes. These solutions were presented by Harrington (1973) & Neufeld (1990), who showed (among other things) that the angle-averaged Lyα spectrum emerging from a semi-infinite 'slab' has the following analytic solution:

$$J(x) = \frac{\sqrt{6}}{24\sqrt{\pi}a\tau_0} \left( \frac{x^2}{1 + \cosh\left[\sqrt{\frac{\pi^3}{54}} \frac{|x^3|}{a\tau_0}\right]} \right), \tag{92}$$

when the photons were emitted in the center of the slab. Here, $\tau_0$ denotes the line center optical depth from the center to the edge of the slab (which extends infinitely in other directions). Similar solutions have been derived for spheres (Dijkstra et al., 2006) and cubes (Tasitsiomi, 2006b). The *solid lines* shown in Figure 26 shows analytic solutions for $J(x)$ emerging from *spheres* for $\tau_0 = 10^5, 10^6$ and $\tau_0 = 10^7$. *Histograms* show the spectra emerging from Monte-Carlo simulations of the Lyα radiative transfer process (see § 9), which should provide us with the most accurate solution. Figure 26 illustrates that analytic solutions derived by Harrington (1973), Neufeld (1990), and Dijkstra et al. (2006) approach the true solution remarkably well. The peak flux density for the flux emerging from a slab [sphere] occurs at $x_{\text{max}} = \pm 1.1 (a_v \tau_0)^{1/3} \sim x_p$ [$x_{\text{max}} = \pm 0.9 (a_v \tau_0)^{1/3}$], where $x_p$ is the characteristic frequency estimated above (see Eq 86).

## 8.2. Lyα Transfer through Uniform, Expanding and Contracting Gas Clouds

Our previous analysis focussed on static gas clouds. Once we allow the clouds to contract or expand, no analytic solutions are known (except when $T = 0$, see below). We can qualitatively describe what happens when the gas clouds are not static.

Consider an expanding sphere: the predicted spectral line shape must also depends on the outflow velocity profile $v_{\text{out}}(r)$. Qualitatively, photons are less likely to escape on the blue side (higher energy) than photons on the red side of the line resonance because they appear closer to resonance in the frame of the outflowing gas.

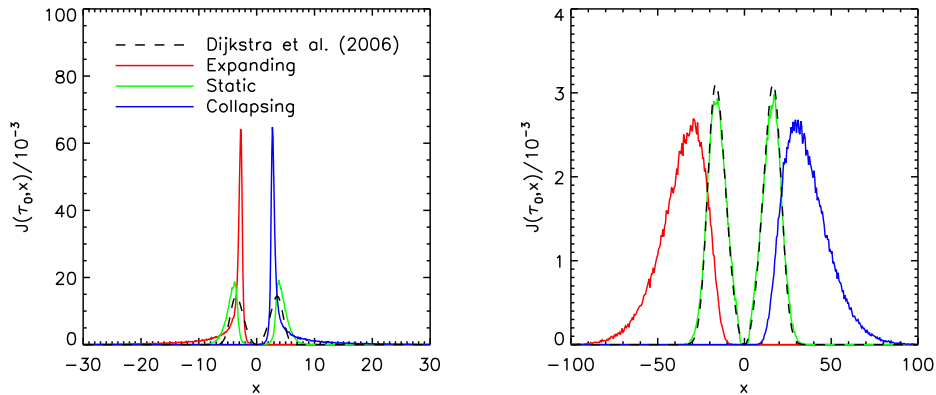

FIG. 27 This Figure illustrates the impact of bulk motion of optically thick gas to the emerging Lyα spectrum of Lyα: The *green lines* show the spectrum emerging from a static sphere (as in Fig 26). In the *left/right panel* the HI column density from the centre to the edge of the sphere is $N_{HI} = 2 \times 10^{18}/2 \times 10^{20}$ cm$^{-2}$. The *red/blue lines* show the spectra emerging from an expanding/a contracting cloud. Expansion/contraction gives rise to an overall redshift/blueshift of the Lyα spectral line (*Credit: from Figure 7 of Laursen et al. 2009b ©AAS. Reproduced with permission*).

Moreover, as the Lyα photons are diffusing outward through an expanding medium, they loose energy because the do 'work' on the outflowing gas (Zheng & Miralda-Escudé, 2002). Both these effects combined enhance the red peak, and suppress the blue peak, as illustrated in Figure 27 (taken from Laursen et al. 2009b). In detail, how much the red peak is enhanced, and the blue peak is suppressed (and shifted in frequency directions) depends on the outflow velocity and the HI column density of gas[35]. Not unexpectedly, the same arguments outlined above can be applied to a collapsing sphere: here we expect the blue peak to be enhanced and the red peak to be suppressed (e.g. Zheng & Miralda-Escudé 2002, Dijkstra et al. 2006). It is therefore thought that the Lyα line shape carries information on the gas kinematics through which it is scattering. As we discuss in § 10.1, the shape and shift of the Lyα spectral line profile has been used to infer properties of the medium through which they are scattering.

There exists one analytic solution to radiative transfer equation through an expanding medium: Loeb & Rybicki (1999) derived analytic expressions for the radial dependence of the angle-averaged intensity $J(\nu, r)$ of Lyα radiation as a function of distance $r$ from a source embedded within a neutral intergalactic medium undergoing Hubble expansion, (i.e. $v_{out}(r) = H(z)r$, where $H(z)$ is the Hubble parameter at redshift $z$). Note that this solution was obtained assuming completely coherent scattering (i.e. $x_{out} = x_{in}$ which corresponds to the special case of $T = 0$), and that the photons frequencies change during flight as a result of Hubble expansion. Formally it describes a somewhat different scattering process than what we discussed before (see Dijkstra et al., 2006, for a more detailed discussion). However, Dijkstra & Loeb (2008a) have compared this analytic solution to that of a Monte-Carlo code that uses the proper frequency redistribution functions and find good agreement between the Monte-Carlo and analytic solutions.

Finally, it is worth pointing out that in expanding/contracting media, the number of scattering events $N_{scat}$ and the total trapping time both decrease. The main reason for this is that in the presence of bulk motions in the gas, it becomes easier to scatter Lyα photons into the wings of the line profiles where they can escape more easily. The reduction in trapping time has been quantified by Bonilha et al. (1979), and in shell models (which will be discussed in § 10.1 and Fig 38) by Dijkstra & Loeb (2008a, see their Fig 6).

### 8.3. Lyα Transfer through Dusty, Uniform & Multiphase Media

The interstellar medium (ISM) contains dust. A key difference between a dusty and dust-free medium is that in the presence of dust, Lyα photons can be destroyed during the scattering process when the albedo (also known as the reflection coefficient) of the dust grains (see § 7.4) $A_d < 1$. Dust therefore causes the 'escape fraction' ($f_{esc}^{\alpha}$), which denotes the fraction Lyα photons that escape from the dusty medium, to fall below unity, i.e. $f_{esc}^{\alpha} < 1$. For a medium with a uniform distribution of gas and dust, the probability that a Lyα photon is destroyed by a dust grain increases with the distance travelled by the Lyα photon. We learned before that as photons diffuse spatially, they also diffuse in frequency. This implies that *dust affects the emerging Lyα spectral line profile in a highly frequency-dependent way, even though the absorption cross-section for this process is practically a constant over the same range of frequencies.* Specifically, if we compare Lyα spectra emerging from two scattering media which differ only in their dust content, then we find that the impact of the dust is largest at large frequency/velocity off-sets from line center (see Fig 28).

---

[35] Max Gronke has developed an online tool which allows users to vary column density, outflow/inflow velocity of the scattering medium, and directly see the impact on the emerging Lyα spectrum: see `http://bit.ly/man-alpha`.

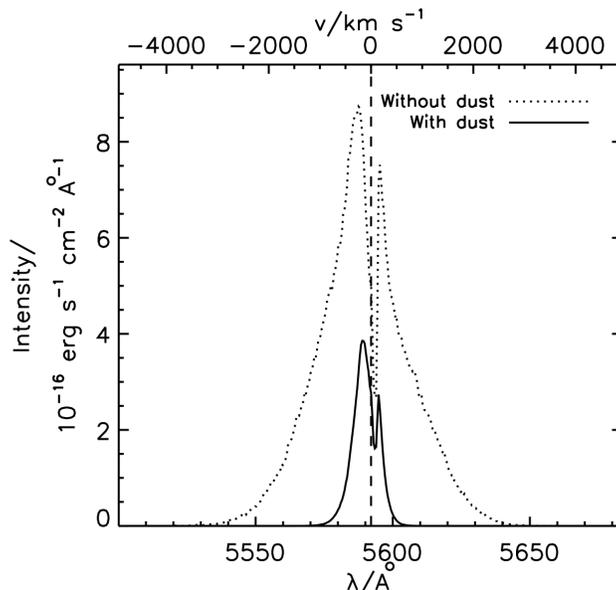

FIG. 28 This figure illustrates that dust can suppress the Lyα spectral line profile in a highly frequency dependent way, even though the dust absorption cross-section barely varies over the same frequency range (*Credit: from Figure 9 of Laursen et al. 2009b ⓒAAS. Reproduced with permission*). The main reason for this is that dust limits the distance Lyα photons can travel before they are destroyed, which limits how much they can diffuse in frequency space. Dust therefore has the largest impact at frequencies furthest from line center.

Dust also destroys UV-continuum photons, but because Lyα photons scatter and diffuse spatially through the dusty medium, the impact of dust on Lyα and UV-continuum is generally different. This can affect the 'strength' (i.e. the equivalent width) of the Lyα line compared to the underlying continuum emission. In a uniform mixture of HI gas and dust, Lyα photons have to traverse a larger distance before escaping, which increases the probability to be destroyed by dust. In these cases we expect dust to reduce the EW of the Lyα line. The ISM is not smooth and uniform however, which can drastically affect Lyα radiative transfer. The interstellar medium is generally thought to consist of the 'cold neutral medium' (CNM), the 'warm neutral/ionized medium' (WNM/WIM), and the 'hot ionized medium' (HIM, see e.g. the classical paper by McKee & Ostriker 1977). In reality, the cold gas is not in 'clumps' but rather in a complex network of filaments and sheets. Lyα transfer calculations through realistic ISM models have only just begun, partly because modeling the multiphase nature of the ISM with simulations is a difficult task which requires extremely high spatial resolution. There is substantially more work on Lyα transfer through 'clumpy' media that consists of cold clumps containing neutral hydrogen gas and dust, embedded within a (hot) ionized, dust free medium (Neufeld 1991, Hansen & Oh 2006, Laursen et al. 2013, Gronke et al. 2016), and which represent simplified descriptions of the multi-phase ISM.

Clumpy media facilitate Lyα escape from dusty interstellar media, and can help explain the detection of Lyα emission from dusty galaxies such as submm galaxies (e.g. Chapman et al. 2005), and (U)LIRGs (e.g Nilsson & Møller 2009, Martin et al. 2015). In a clumpy medium dust can even *increase* the EW of the Lyα line, i.e. preferentially destroy (non-ionizing) UV continuum photons over Lyα photons: Lyα photons can propagate freely through the 'interclump' medium. Once a Lyα photon enters a neutral clump, its mean free path can be substantially smaller than the clump size. Because scattering is partially coherent, and the frequency of the photon in the clump frame changes only by an r.m.s amount of $\sqrt{\langle \Delta x^2 | x_{in} \rangle} = 1$ (see Eq 77), the mean free path remains roughly the same. Since the photon penetrated the clump by a distance corresponding to $\tau \sim 1$, the photon is able to escape after each scattering event with a significant probability. A Lyα photon that penetrates a clump is therefore likely to escape after $\sim 5$ scattering events (Hansen & Oh 2006), rather than penetrate deeper into the clump via a much larger number of scattering events. The Lyα photons effectively scatter off the clump surface (this is illustrated in Fig 29), thus avoiding exposure to dust grains. In contrast, UV continuum photons will penetrate the dusty clumps unobscured by hydrogen and are exposed to the full dust opacity. Clumpy, dusty media may therefore preferentially let Lyα photons escape over UV-continuum.

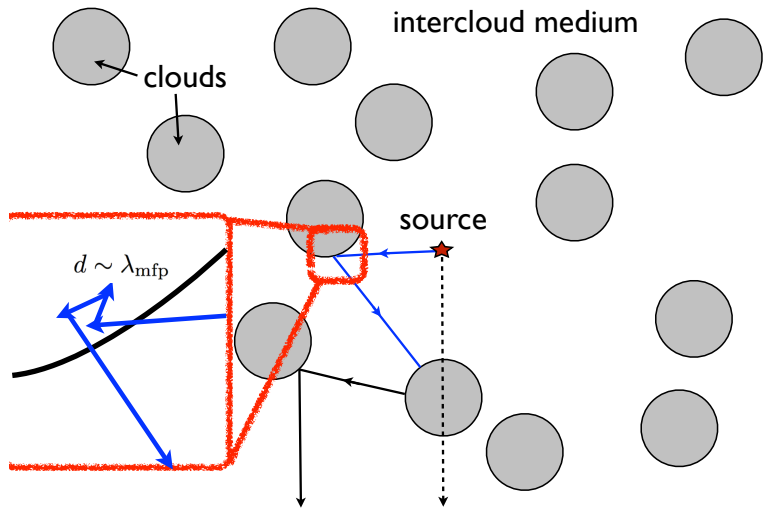

FIG. 29 Schematic illustration of how a multiphase medium may favour the escape of Lyα line photons over UV-continuum photons: *Solid/dashed lines* show trajectories of Lyα/UV-continuum photons through clumpy medium. If dust is confined to the cold clumps, then Lyα may more easily escape than the UV-continuum (*Credit: from Figure 1 of Neufeld 1991 ©AAS. Reproduced with permission.*).

Laursen et al. (2013) and Duval et al. (2014) have recently shown however that - while clumpy media facilitate Lyα escape - EW boosting only occurs under physically unrealistic conditions in which the clumps are very dusty, have a large covering factor, have very low velocity dispersion and outflow/inflow velocities, and in which the density contrast between clumps and interclump medium is maximized. While a multiphase (or clumpy) medium definitely facilitates the escape of Lyα photons from dusty media, EW boosting therefore appears uncommon. We can understand this result as follows: the preferential destruction of UV-continuum photons over Lyα requires at least a significant fraction of Lyα photons to avoid seeing the dust by scattering off the surface of the clumps. How deep the Lyα photons actually penetrate, depends on their mean-free path, which depends on their frequency in the frame of the clumps. If clumps are moving fast, then it is easy for Lyα photons to be Doppler boosted into the wing of the line profile (in the clump frame), and they would not scatter exclusively on the clump surfaces. Finally, we note that the conclusions of Laursen et al. (2013) and Duval et al (2014) apply to the EW-boost averaged over *all* photons emerging from the dusty medium. Gronke & Dijkstra (2014) have investigated that for a given model, there can be directional variations in the predicted EW, with large EW boosts occurring in a small fraction of sightlines in directions where the UV-continuum photon escape fraction was suppressed, thus partially restoring the possibility of EW boosting by a multiphase ISM.

## 9. MONTE-CARLO LYα RADIATIVE TRANSFER

Analytic solutions to the radiative transfer equation (Eq 58) only exist for a few idealised cases. A modern approach to solve this equation is via *Monte-Carlo* methods, which refer to a 'broad class of computational algorithms [...] which change processes described by certain differential equations into an equivalent form interpretable as a succession of random operations' (S. Ulam, see `https://en.wikipedia.org/wiki/Monte_Carlo_method`)[36].

---

[36] The term 'Monte-Carlo' was coined by Ulam & Metropolis as a code-name for their classified work on nuclear weapons (radiation shielding, and the distance that neutrons would likely travel through various materials). 'Monte-Carlo' was the name of the casino where Ulam's uncle had a (also classified) gambling addiction. I was told most of this story over lunch by M. Baes. For questions, please contact him.

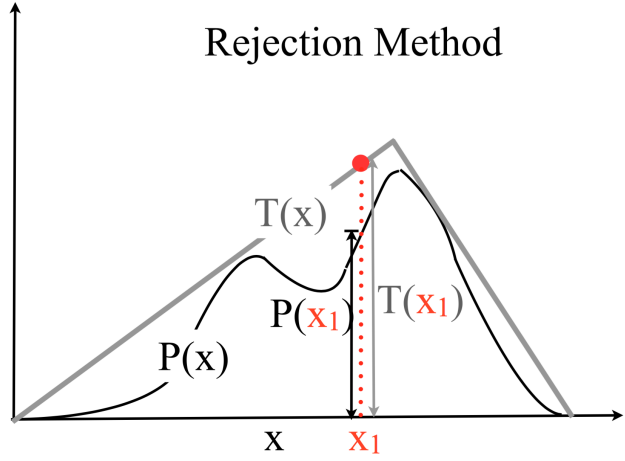

FIG. 30 The *rejection method* provides a simple way to randomly draw variables $x$ from arbitrary probability distributions $P(x)$ (see text).

In Ly$\alpha$ Monte-Carlo radiative transfer, we represent the integro-differential equation (Eq 58) by a succession of random scattering events until Ly$\alpha$ photons escape (Lee & Ahn 1998, Loeb & Rybicki 1999, Ahn et al. 2001, Zheng & Miralda-Escudé 2002, Cantalupo et al. 2005, Dijkstra et al. 2006, Verhamme et al. 2006, Tasitsiomi 2006, Semelin et al. 2007, Laursen et al. 2009a, Pierleoni et al. 2009, Faucher-Giguère et al. 2010, Kollmeier et al. 2010, Zheng et al. 2010, Barnes et al. 2011, Forero-Romero et al. 2011, Orsi et al. 2012, Yajima et al. 2012, Behrens & Niemeyer 2013, Gronke & Dijkstra 2014, Lake et al. 2015). Details on how the Monte-Carlo approach works can be found in many papers (see e.g. the papers mentioned above, and Chapters 6-8 of Laursen, 2010, for an extensive description). I will first provide a brief description of drawing random variables, which is central to the Monte-Carlo method. Then I will describe Monte-Carlo Ly$\alpha$ radiative transfer.

### 9.1. General Comments on Monte-Carlo Methods

Central in Monte-Carlo methods is generating random numbers from probability distributions. If we denote a probability distribution of variable $x$ with $P(x)$, the $P(x)dx$ denotes the probability that $x$ lies in the range $x \pm dx/2$, and we must have $\int_{-\infty}^{\infty} dx P(x) = 1$. The cumulative probability is $C(x) \equiv P(< x) = \int_{-\infty}^{x} P(x')dx'$, where clearly $C(x \to -\infty) = 0$ and $C(x \to \infty) = 1$. We can draw a random $x$ from the distribution $P(x)$ by randomly generating a number ($R$) between 0 and 1. We then transform $\mathcal{R}$ into $x$ by inverting its cumulative probability distribution $C(x)$, i.e.

$$\mathcal{R} \equiv C(x) \to \mathcal{R} \equiv \int_{-\infty}^{x} P(x')dx'. \tag{93}$$

This implies that we need to ($i$) be able to integrate $P(x)$, and ($ii$) be able to invert the integral equation. Generally, there is no analytic way of inverting Eq 93. One way of randomly drawing $x$ from $P(x)$ is provided by the *rejection method*. This method consists of picking another $T(x)$ which lies above $P(x)$ everywhere, and which we can integrate and invert analytically. We are completely free to pick $T(x)$ however we want. Because $T(x)$ lies above $P(x)$ everywhere, $\int T(x)dx = A$, where $A > 1$. We first generate $x_1$ from $T(x)$ by generating a random number $\mathcal{R}_1$ between 0 and 1 and inverting

$$\mathcal{R}_1 \equiv \frac{1}{A} \int_{-\infty}^{x_1} T(x')dx'. \tag{94}$$

Once we have $x_1$ we then generate another random number $\mathcal{R}_\in$ between 0 and 1, and accept $x_1$ as our random pick when

$$\mathcal{R}_2 \leq \frac{P(x_1)}{T(x_1)}. \tag{95}$$

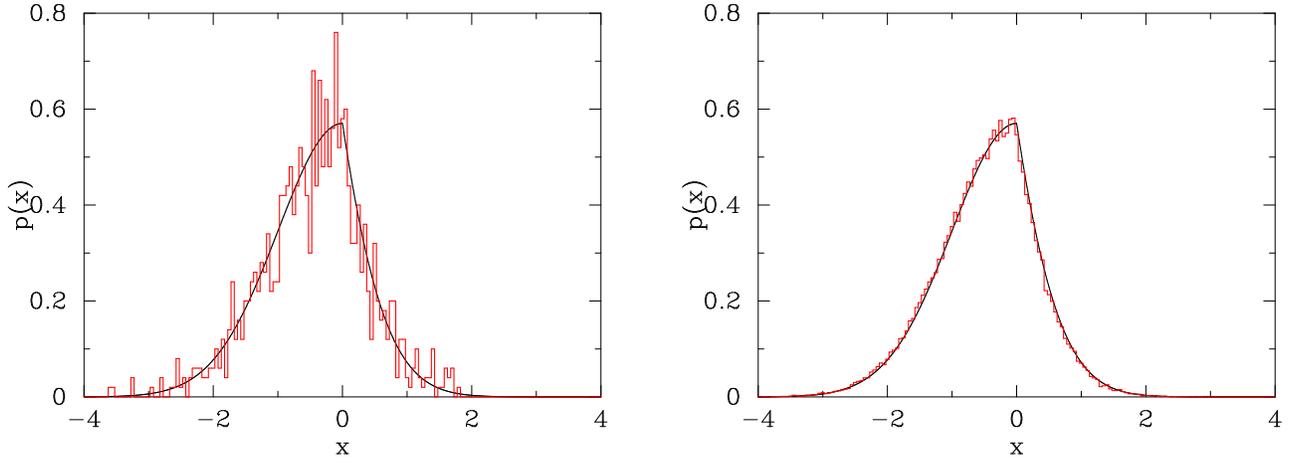

FIG. 31 *Red histograms* show the distributions of $10^3$ (*left panel*) and $10^5$ (*right panel*) randomly drawn values for $x$ from its PDF $P(x)$ by using the rejection method. In this case we randomly truncated a Gaussian PDF with $\sigma = 1$ by suppressing the PDF by an arbitrary function $f(x)$ only for $x > 0$ (note that $f(x) < 1$). This PDF is shown as the *black solid line*.

If we repeat this procedure a large number of times, and make a histogram for $x$, we can see that this traces $P(x)$ perfectly (this is illustrated with an example in Fig 31). Eq 95 shows that we ideally want $T(x)$ to lie close to $P(x)$ in order not to have to reject the majority of trials (the better the choice for $T(x)$, the smaller the rejected fraction).

### 9.2. Ly$\alpha$ Monte-Carlo Radiative Transfer

Here, we briefly outline the basic procedure that describes the Monte-Carlo method applied to Ly$\alpha$ photons. For each photon in the Monte-Carlo simulation:

1. We first randomly draw a position, $\mathbf{r}$, from which the photon is emitted from the emissivity profile[37] $j_\nu(\mathbf{r})$ (see Eq 58). We the assign a random frequency $x$, which is drawn from the Voigt function $\phi(x)$, and a random propagation direction $\mathbf{k}$.

2. We randomly draw the optical depth $\tau$ the photon propagates into from the distribution $P(\tau) = \exp(-\tau)$.

3. We convert $\tau$ into a physical distance $s$ by (generally numerically) inverting the line integral $\tau = \int_0^s d\lambda\, n_{\rm HI}(\mathbf{r}')\sigma_\alpha(x'[\mathbf{r}'])$, where $\mathbf{r}' = \mathbf{r} + \lambda\mathbf{k}$ and $x' = x - \mathbf{v}(\mathbf{r}') \cdot \mathbf{k}/(v_{\rm th})$. Here, $\mathbf{v}(\mathbf{r}')$ denotes the 3D *bulk* velocity vector of the gas at position $\mathbf{r}'$. Note that $x'$ is the dimensionless frequency of the photon in the 'local' frame of the gas at $\mathbf{r}'$.

4. Once we have selected the scattering location, we need to draw the *thermal* velocity components of the atom that is scattering the photon (we only need the thermal velocity components, as we work in the local gas frame). As in § 7.3, we decompose the thermal velocity of the atom into a direction parallel to that of the incoming photon, $v_{||}$ (or its dimensionless analogue $u$, see Eq 66), and a 2D-velocity vector perpendicular to $\mathbf{k}$, namely $\mathbf{v}_\perp$. We discussed in § 7.3 what the functional form of the conditional probability $P(u|x)$ (see discussion below Eq 68), and apply the rejection method to draw $u$ from this functional form (see the Appendix Zheng & Miralda-Escudé 2002 for a functional form of $T(u|x)$). The 2 components of $\mathbf{v}_\perp$ can be drawn from a Maxwell-Boltzmann distribution (see e.g. Dijkstra et al. 2006).

5. Once we have determined the velocity vector of the atom that is scattering the photon, we draw an outgoing direction of the photon after scattering, $\mathbf{k}_{\rm out}$, from the phase-function, $P(\mu)$ (see Eq 41 and § 6.1). We will show below that this procedure of generating the atom's velocity components and random new directions generates the proper frequency redistribution functions, as well as their angular dependence.

---

[37] For arbitrary gas distributions, the emissivity profile is a 3D-field. We can still apply the rejection method. One way to do this is to discretize the 3D field $j_\nu(x, y, z) \rightarrow j_\nu(i, j, k)$, where we have $N_x$, $N_y$, and $N_z$ of cells into these three directions. We can map this 3D-array onto a long 1D array $j_\nu(m)$, where $m = 0, 1, ..., N_x \times N_y \times N_z$, and apply the rejection method to this array.

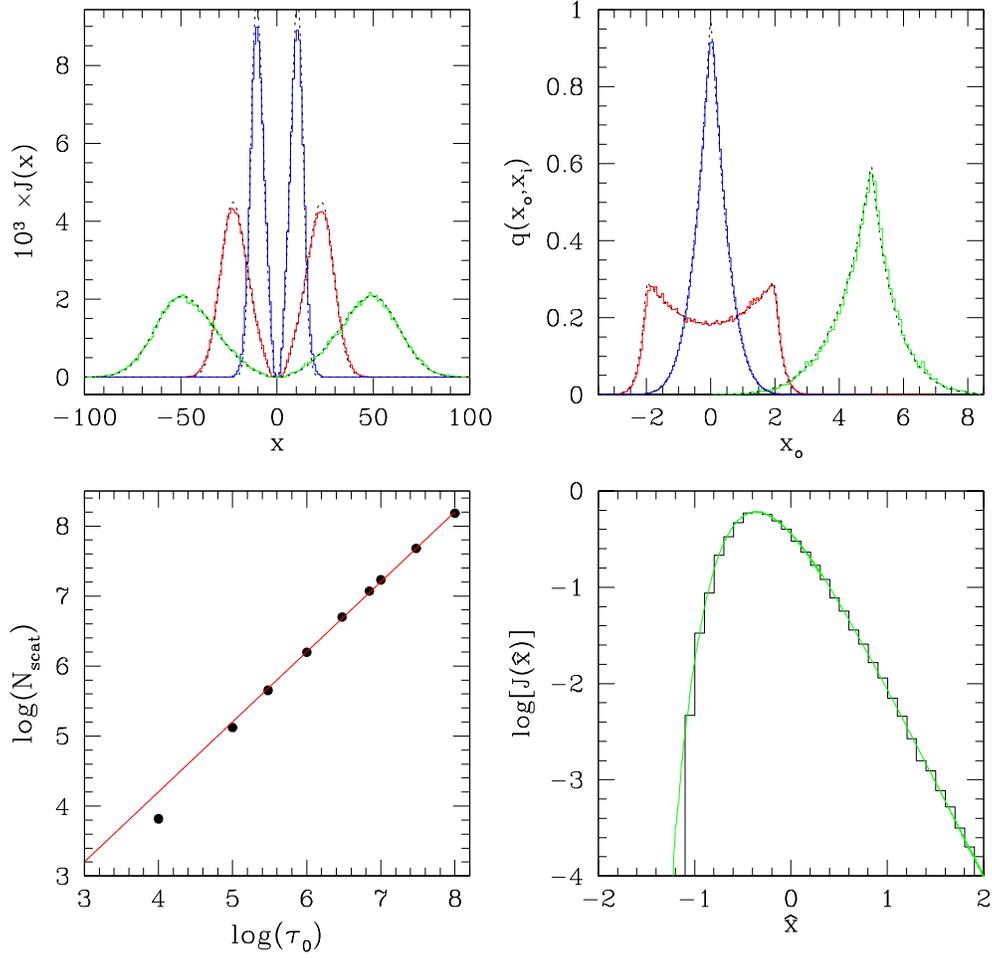

FIG. 32 This Figure shows an example set of tests that Monte-Carlo codes must be able pass (*Credit: from Figure 1 of Dijkstra et al. 2006 ©AAS. Reproduced with permission.*). In the *top left panel* show Monte-Carlo calculations of the Lyα spectra emerging from a uniform spherical gas cloud, in which Lyα photons are injected in the center of the sphere, at the line center (i.e. $x = 0$). The total line center optical depth, $\tau_0$ from the center to the edge is $\tau_0 = 10^5$ (*blue*), $\tau_0 = 10^6$ (*red*) and $\tau_0 = 10^7$ (*green*). Overplotted as the *black dotted lines* are the analytic solutions. The agreement is perfect at high optical depth ($a_v \tau \gtrsim 10^3$). *Upper right panel:* The *colored histograms* show the frequency redistribution functions, $R(x_{\rm out}, x_{\rm in})$, for $x_{\rm in} = 0$ (*blue*), $x_{\rm in} = 2$ (*red*) and $x_{\rm in} = 5$ (*green*), as generated by the Monte-Carlo simulation. The *solid lines* are the analytic solutions given by Eq 73. *Lower left panel:* The total number of scattering events that a Lyα photon experiences before it escapes from a slab of optical thickness $2\tau_0$ according to a Monte-Carlo simulation (*circles*). Overplotted as the *red–solid line* is the theoretical prediction by Harrington (1973). *Lower right panel:*. The spectrum emerging from an infinitely large object that undergoes Hubble expansion. The histogram is the output from our code, while the *green–solid line* is the (slightly modified) solution obtained by Rybicki & Loeb (1999) using their Monte Carlo algorithm (see Dijkstra et al. 2006 for more details).

6. Unless the photon escapes, we replace the photon propagation direction & frequency and go back to 1). Once the photon escapes we record information we are interested in such as the location of last scattering, the frequency of the photon, the thermal velocity components of the atom that last scattered the photon, the number of scattering events the photon underwent, the total distance it travelled through the gas, etc.

It is important to test Lyα Monte-Carlo codes in as many ways as possible. Figure 32 shows a minimum set of tests Monte-Carlo codes must be able to reproduce. These comparisons with analytic solutions test different aspects of the code.

The *histograms* in the *top left panel* show Monte-Carlo realizations of the Lyα spectra emerging from a uniform spherical gas cloud, in which Ly$\alpha$ photons are injected in the center of the sphere, at the line center (i.e. $x = 0$). The total line center optical depth, $\tau_0$ from the center to the edge is $\tau_0 = 10^5$ (*blue*), $\tau_0 = 10^6$ (*red*) and $\tau_0 = 10^7$ (*green*). Overplotted as the *black dotted lines* are the corresponding analytic solutions (see Eq 92, but modified for a sphere, see Dijkstra et al. 2006a). The agreement is perfect at high optical depth ($a_v\tau \gtrsim 10^3$). At lower optical depth, $a_v\tau_0 \lesssim 10^3$, the analytic solutions are not expected to be accurate any more (see § 8.1). Because the analytic solutions were obtained under the assumption that scattering occured in the wing, the agreement between analytic and Monte-Carlo techniques at high $\tau_0$ only confirms that the Monte-Carlo procedure accurately describes scattering in the wing of the line profile. This comparison does not test core scattering, which make up the vast majority of scattering events (see § 8.1). The fact that Ly$\alpha$ spectra emerging from optically (extremely) thick media is insensitive to core scattering implies that we need additional tests to test core scattering. However, it also implies we can skip these core-scattering events, which account for the vast majority of all scattering events. That is, Monte-Carlo simulations can be 'accelerated' by skipping core scattering events. We discuss how we can do this in more detail in § 9.4.

In the *upper right panel* the colored histograms show Monte-Carlo realization of the frequency redistribution functions, $R(x_{\rm out}, x_{\rm in})$ (see Eq 73), for $x_{\rm in} = 0$ (*blue*), $x_{\rm in} = 2$ (*red*) and $x_{\rm in} = 5$ (*green*). The solid lines are the analytic solutions given by Eq 73 (here for dipole scattering). This comparison tests individual Ly$\alpha$ core scattering events, and thus complements the test we described above.

In the *lower left panel* the *circles* show the total number of scattering events that a Ly$\alpha$ photon experiences before it escapes from a slab of optical thickness $2\tau_0$ in a Monte-Carlo simulation. Overplotted as the *red–solid line* is the theoretical prediction that $N_{\rm scat} = C\tau_0$, with $C = 1.1$ (see Eq 89, Adams 1972, Harrington 1973, Neufeld 1990). The break down at low $\tau_0$ corresponds to the range of $\tau_0$ where analytic solutions are expected to fail. This test provides another way to test core scattering events, as $N_{\rm scat}$ is set by the probability that a Ly$\alpha$ photon is first scattered sufficiently far into the wing of the line such that it can escape in a single 'excursion'. This test also shows how accurate the analytic prediction is, despite the fact that the derivation presented in § 8.1 (following Adams 1972) did not feel like it should be this accurate. This plot also underlines how computationally expensive Ly$\alpha$ transfer can be if we simulate each scattering event.

The *lower right panel* shows Ly$\alpha$ spectrum emerging from a Ly$\alpha$ point source surrounded by an infinitely large sphere that undergoes Hubble expansion. The *black histogram* shows a Monte-Carlo realization. The *green line* represents a 'pseudo-analytic' solution of Loeb & Rybicki (1999): as we mentioned earlier in § 8.2, Loeb & Rybicki (1999) provided a fully analytic solution for the angle averaged intensity $J(r, x)$ where $r$ denotes the distance from the galaxy. Unfortunately, their analytic solution does *not* apply to emerging spectrum because as $r \to \infty$, the Ly$\alpha$ photons have redshifted far enough into the wing that the IGM is optically thin to the Ly$\alpha$ photons, in which case the Ly$\alpha$ radiative transfer problem cannot be reduced to a diffusion problem anymore. Though not shown here, Dijkstra & Loeb (2008) compared the analytic *integrated* $U(r) \propto \int J(r, x)$ to that extracted from Monte-Carlo simulations, and found excellent agreement. The *green line* shown here represents the spectrum[38] obtained from a simplified Monte-Carlo simulation (see Loeb & Rybicki 1999 for details), and agrees well with the full Monte-Carlo code. This test provides us with a way to test the code when bulk motions in the gas are present.

### 9.3. Extracting Observables from Ly$\alpha$ Monte-Carlo Simulations in 3D Simulations

In Monte-Carlo radiative transfer calculations applied to arbitrary 3D gas distributions, extracting observables requires (a bit) more work than recording the location of last scattering, the photon's frequency etc. This is because formally we are interested only in a tiny subset of Ly$\alpha$ photons that escape, *and* end up in the mirror of our telescope. We denote the direction from the location of last scattering towards the telescope with $\mathbf{k_t}$. The mirror of our telescope only subtends a solid angle $d\Omega_{\rm telescope} = \frac{dA_{\rm telescope}}{d_A^2(z)}$, where $dA_{\rm telescope}$ denotes the area of the mirror, and $d_A(z)$ denotes the angular diameter distance to redshift $z$. The probability that a Ly$\alpha$ photon in our Monte-Carlo simulation escapes from the scattering medium into this tiny solid angle is negligible.

---

[38] The frequency $\hat{x}$ here shows another dimensionless frequency variable that was used by Loeb & Rybicki (1999), and relates to 'standard' dimensionless frequency $x$ as $\hat{x} \sim -3.8 \times 10^3 \ x$ for $T = 10$ K (see Dijkstra et al. 2006).

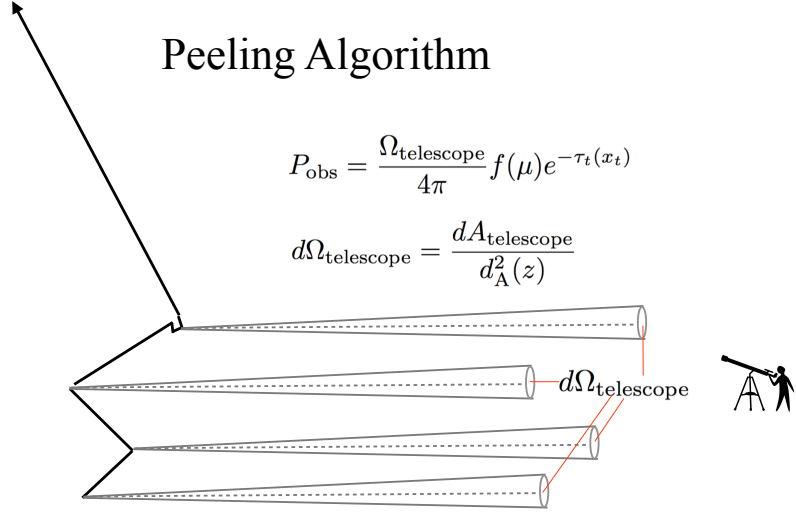

**FIG. 33** A schematic representation of the *peeling algorithm*. This algorithm allows us to efficiently and accurately extract Lyα observables (spectra, surface brightness profiles, etc) from arbitrary 3D gas distributions. This algorithm overcomes the problem that in Monte-Carlo simulations, the probability that a Lyα photon escapes - *and* into the direction of the telescope such that it lands on the mirror - is practically zero due to the infinitesimally small angular scale of the telescope mirror when viewed from the Lyα source. The peeling algorithm treats each scattering event as a point source with a (scattering induced anisotropic) luminosity $L_\alpha/N_{\rm phot}$, where $L_\alpha$ is the total luminosity of the Lyα source and $N_{\rm phot}$ denotes the number of photons used in the Monte-Carlo run to simulate the source.

One way to get around this problem is by relaxing the restriction that photons must escape into the solid angle $d\Omega_{\rm telescope}$ centered on the direction $\mathbf{k_t}$, by increasing $d\Omega_{\rm telescope}$ to a larger solid angle. This approximation corresponds to averaging over all viewing directions within some angle $\Delta\alpha$ from the real viewing direction. If observable properties of Lyα do not depend strongly on viewing direction, then this approximation is accurate. However, this method implies we are not using information from the vast majority of photons ($\sim \Delta\alpha^2/4\pi$) that we used in the Monte-Carlo simulation. To circumvent this problem in a more efficient (and accurate) way is provided by the so-called 'peeling algorithm'. This algorithm treats each scattering event in the simulation domain as a point source with luminosity $L_\alpha/N_{\rm phot}$, where $L_\alpha$ is the total Lyα luminosity of the source and $N_{\rm phot}$ denotes the total number of Lyα photons used in the Monte-Carlo simulation to represent this source (see e.g. Yusef-Zadeh et al. 1984, Zheng & Miralda-Escudé 2002, Tasitsiomi 2006). The total flux $S$ we expect to get from each point source is

$$S = \frac{L_\alpha}{4\pi d_{\rm L}^2(z) N_{\rm phot}} \times \frac{P(\mu_{\rm t})}{2} e^{-\tau_t(x_t)}, \tag{96}$$

where $d_{\rm L}(z)$ denotes the luminosity distance to redshift $z$. Furthermore, $\mu_{\rm t} \equiv \mathbf{k_{\rm in}} \cdot \mathbf{k_t}$, in which $\mathbf{k_{\rm in}}$ (like before) denotes the propagation direction of the photon prior to scattering. The presence of the scattering phase-function $P(\mu)$ reflects that the 'point source' is not emitting isotropically, but that the emission in the direction $\mu_{\rm t}$ is enhanced by a factor of $P(\mu_{\rm t})/2$ in direction $\mathbf{k_t}$ (the factor of '2' reflects that $\int d\mu\, P(\mu) = 2$, see Eq 42). The factor $e^{-\tau_t(x_t)}$ denotes the escape fraction in direction $\mathbf{k_t}$, where $x_t$ is the frequency of the photon it *would* have had if it truly had scattered in direction $\mathbf{k_t}$. This frequency $x_t$ can be obtained from Eq 66. Eq 96 should be applied for each scattering event: for each scattering event there is a tiny/infinitesimal probablity - $\sim \frac{d\Omega_{\rm telescope}}{4\pi} e^{-\tau_t(x_t)}$ - that the Lyα scatters into the telescope-mirror. We formally have to reduce the weight of the photon by this probability after each scattering event (we are 'peeling' off the weight of this photon), though in practice this can be ignored because of the tiny probability that a photon scattered into the telescope mirror.

An example of an image generated with the peeling algorithm is shown in Figure 34. Here, a Lyα source is at the origin of a cartesian coordinate system. Each of the 3 coordinate axes has 2 spheres of HI gas at identical distances from the origin. There are no hydrogen atoms outside the sphere, and the Lyα scattering should only occur inside the 6 spheres. Resulting images (taken from Dijkstra & Kramer 2012) from 6 viewing directions are shown in the *left panel* of Figure 34. The 'darkness' of a pixel represents its Lyα surface brightness. The average of these 6 images is shown in the box. The *right panel* shows a close-up view of the image associated with the sphere on the $+y$-axis. The side of the sphere facing the Lyα source (on the bottom at $y = 0$) is brightest. The *red line* in the *inset* shows an analytic calculation of the expected surface brightness, under the assumption that the sphere as a whole is optically thin (only this assumption allows for analytic solutions). This image shows that the Peeling algorithm gives rise to the sharp features in the surface brightness profiles that should exist. Note that the alternative method we briefly mentioned above, which averages over all viewing directions within some angle $\Delta\alpha$ from $\mathbf{k}_t$, would introduce some blurring to these images.

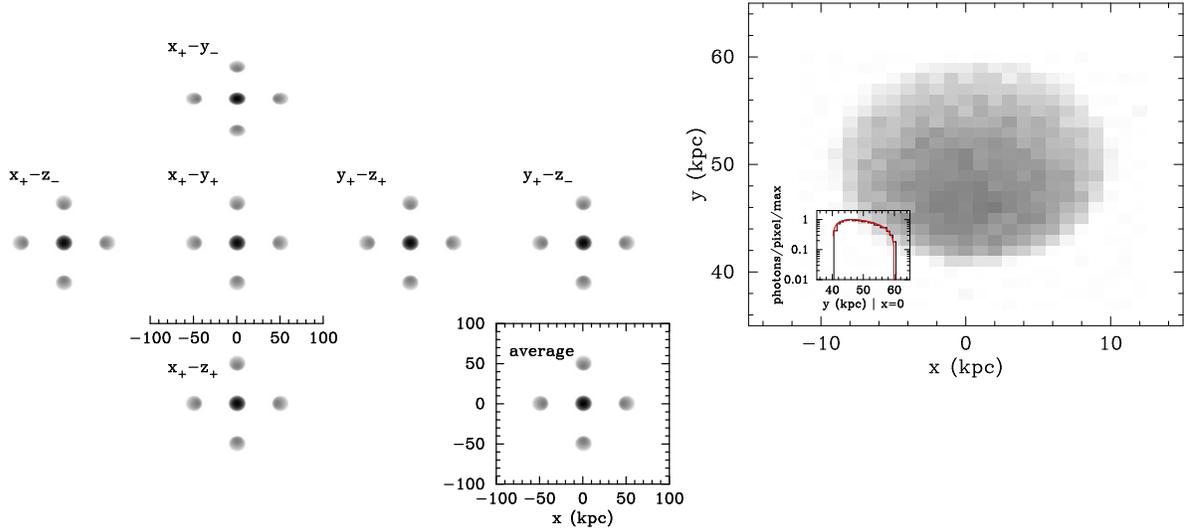

FIG. 34 Examples of Lyα images generated with the *peeling algorithm* (see text for a more detailed description, *Credit: from Figure* **A3** *of Dijkstra & Kramer. 2012, 'Line transfer through clumpy, large-scale outflows: Lyα absorption and haloes around star-forming galaxies', MNRAS, 424, 1672D*).

### 9.4. Accelerating Lyα Monte-Carlo Simulations

As we mentioned above, we do not care about the vast majority of core scattering events for Lyα transfer through extremely optically thick media ($a_v\tau_0 \gtrsim 10^3$). We can accelerate Monte-Carlo simulations by 'forcing' photons into the wing of the line profile. Ordinarily, the physical mechanism that puts a Lyα photon from the core into the wing of the line profile is an encounter with a fast moving atom. We can force the scattering atom to have a large velocity when generating its velocity components. A simple way to do this is by forcing the velocity vector of the atom perpendicular to $\mathbf{k}_{in}$, $\mathbf{v}_\perp$, to be large (Ahn et al. 2002, Dijkstra et al. 2006). We know from § 7.3 that $\mathbf{v}_\perp$ follows a 2D Maxwell-Boltzmann distribution $g(\mathbf{v}_\perp)d^2\mathbf{v}_\perp \propto v_\perp \exp(-v_\perp^2)$, where $v_\perp \equiv |\mathbf{v}_\perp|$ is the magnitude of $\mathbf{v}_\perp$. In dimensionless units $u_\perp \equiv v_\perp/v_{th}$, and we have $g(u_\perp)du_\perp = 2\pi u_\perp \exp(-u_\perp^2)/\pi$ (see the discussion under Eq 68). We can force $u_\perp$ to be large by drawing it from a truncated Maxwell-Boltzmann distribution, which states that

$$p(u_\perp)du_\perp = \begin{cases} 0 & |u_\perp| < x_{crit} \\ \mathcal{N}u_\perp \exp(-u_\perp^2) & |u_\perp| > x_{crit}. \end{cases}, \tag{97}$$

where $\mathcal{N}$ ensures that the truncated distribution function for $u_\perp$ is normalized. Furthermore, $x_{crit}$ is a parameter that determines how far into the wing we force the Lyα photons. This parameter therefore sets how much Lyα transfer is accelerated. Clearly, one has to be careful when choosing $x_{crit}$: forcing photons too far into the wing may cause them to escape at frequencies where frequency diffusion would otherwise never take them. Various authors have experimented with choosing $x_{crit}$ based on the local HI-column density of a cell in a simulation, etc (Tasitsiomi 2006a,b, Laursen et al. 2009a, Smith et al. 2015).

## 10. LYα TRANSFER IN THE UNIVERSE

Previous sections discussed the basics of the theory describing Lyα transfer through optically thick media. The goal of this section is to discuss what we know about Lyα transfer in the real Universe. We decompose this problem into several scales: (*i*) Lyα photons have to escape from the interstellar medium (ISM) of galaxies into the circum galactic/intergalactic medium (CGM/IGM). We discuss this in § 10.1. We then go to large scales, and describe the subsequent radiative transfer through the CGM/IGM at lower redshift (§ 10.2) and at higher redshift when reionization is still ongoing (§ 10.4).

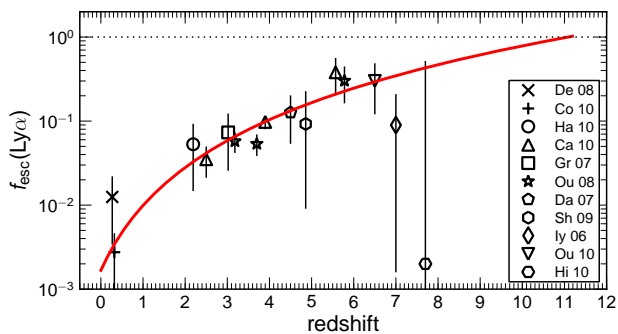

FIG. 35 Observational constraints on the redshift-dependence of the volume averaged 'effective' escape fraction, $f_{esc}^{eff}$, which contains constraints on the true escape fraction $f_{esc}^{\alpha}$ (*Credit: from Figure 1 of Hayes et al. 2011 ©AAS. Reproduced with permission*).

## 10.1. Interstellar Radiative Transfer

[39]Understanding interstellar Lyα radiative transfer requires us to understand gaseous flows in a multiphase ISM, which lies at the heart of understanding star and galaxy formation. Modelling the neutral component of interstellar medium is an extremely challenging task, as it requires resolving the multiphase structure of interstellar medium, and how it is affected by feedback from star-formation (via supernova explosions, radiation pressure, cosmic ray pressure, etc). Instead of taking an 'ab-initio' approach to understanding Lyα transfer, it is illuminating to use a 'top-down' approach in which we try to constrain the broad impact of the ISM on the Lyα radiation field from observations (for this also see the lecture notes by M. Ouchi and M. Hayes for more extended discussions of the observations).

We first focus on observational constraints on the *escape fraction* of Lyα photons, $f_{esc}^{\alpha}$. To estimate $f_{esc}^{\alpha}$ we would need to compare the observed Lyα luminosity to the *intrinsic* Lyα luminosity. The intrinsic Lyα luminosity corresponds to the Lyα luminosity that is actually produced. The best way to estimate the intrinsic Lyα luminosity is from some other non-resonant nebular emission line such as Hα. The observed Hα luminosity can be converted into an intrinsic Hα luminosity once nebular reddening is known (from joint measurements of e.g. the Hα and Hβ lines see lecture notes by M. Hayes). Once the intrinsic Hα luminosity is known, then we can compute the intrinsic Lyα luminosity assuming case-B (or case-A) recombination. This procedure indicates that $f_{esc}^{\alpha} \sim 1-2\%$ at $z \sim 0.3$ (Deharveng et al., 2008) and $z \sim 5\%$ at $z \sim 2$ (Hayes et al., 2010).

We do not have access to Balmer series lines at higher redshifts until JWST flies. Instead, it is common to estimate the intrinsic Lyα luminosity from the inferred star formation rate of galaxies (and apply Eq 17 or Eq 18). These star formation rates can be inferred from the dust corrected (non-ionizing) UV-continuum flux density and/or from the IR flux density (e.g. Kennicutt, 1998). These analyses have revealed that $f_{esc}^{\alpha}$ is anti-correlated with the dust-content[40] of galaxies (Atek et al. 2009, Kornei et al. 2010, Hayes et al. 2011). This correlation may explain why $f_{esc}^{\alpha}$ increases with redshift from $f_{esc}^{\alpha} \sim 1-3\%$ at $z \sim 0$ (Deharveng et al., 2008; Wold et al., 2014) to about $f_{esc}^{\alpha} \sim 30-50\%$ at $z \sim 6$ (Hayes et al. 2011, Blanc et al. 2011, Dijkstra & Jeeson-Daniel 2013, also see Fig 35), as the overall average dust content of galaxies decreases towards higher redshifts (e.g. Bouwens et al., 2012; Finkelstein et al., 2012). It is worth cautioning here that observations are not directly constraining $f_{esc}^{\alpha}$: Lyα photons that escape from galaxies can scatter frequently in the IGM (or circum-galactic medium) before reaching earth in a low surface brightness glow that cannot be detected yet (see § 10.2). These photons would effectively be removed from observations, even though they did escape.

---

[39] This discussion represents an extended version of the discussion presented in the review by Dijkstra (2014).
[40] There is also little observational evidence for EW-boosting by a multiphase medium (e.g. Finkelstein et al., 2008; Scarlata et al., 2009).

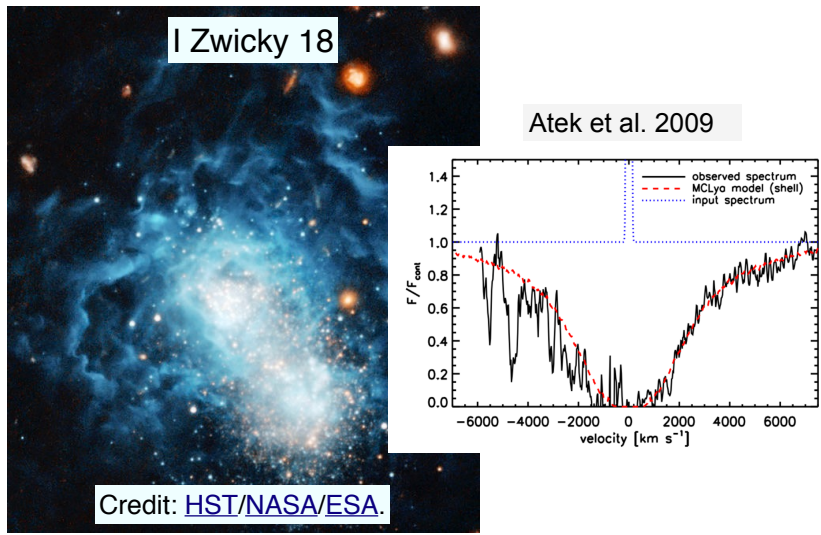

FIG. 36 1Zwicky18. A nearby, metal poor, blue young star forming galaxy (*Image Credit: HST/NASA/ESA, and A. Aloisi*). While this galaxy is expected to be dust-poor, no Lyα is detected in emission (*Credit: Atek et al, A&A, 502, page 791-801, 2009, reproduced with permission ©ESO*). It is thought that this is because there are no (or little) outflows present in this galaxy, which could have facilitated the escape of Lyα. More enriched nearby star forming galaxies that do show Lyα in emission, show evidence for outflows (e.g. Kunth et al. 1998, Atek et al. 2008).

The dependence of $f_{esc}^{\alpha}$ on dust content of galaxies is an intuitive result, as it is practically the only component of the ISM that is capable of destroying Lyα. However, there is more to Lyα escape. This is probably best illustrated by nearby starburst galaxy 1Zwicky18 (shown in Fig 36). This is a metal poor, extremely blue galaxy, and it has even been argued to host Population III stars (i.e. stars that formed our primordial gas). We would expect this galaxy to have a high $f_{esc}^{\alpha}$. However, the spectrum of 1Zwicky18 (also shown in Fig 36) shows strong Lyα absorption. In contrast, the more enriched ($Z \sim 0.1 - 0.3 Z_{\odot}$) nearby galaxy ESO 350 does show strong Lyα emission (see Fig 1 of Kunth et al. 1998). The main difference between the two galaxies is that ESO 350 shows evidence for the presence of outflowing gas. Kunth et al. (1998) observed that for a sample of 8 nearby starburst galaxies, 4 galaxies that showed evidence for outflows showed Lyα in emission, while no Lyα emission was detected for the 4 galaxies that showed no evidence for outflows (irrespective of the gas metallicity of the galaxies). These observations indicate that gas kinematics is a key parameter that regulates Lyα escape (Kunth et al. 1998, Atek et al. 2008, Wofford et al. 2013, Rivera-Thorsen et al. 2015). This result is easy to understand qualitatively: in the absence of outflows, the Lyα sources are embedded within a static optically thick scattering medium. The traversed distance of Lyα photons is enhanced compared to that of (non-ionizing) UV continuum photons (see § 8.1), which makes them more 'vulnerable' to destruction by dust. In contrast, in the presence of outflows Lyα photons can be efficiently scattered into the wing of the line profile, where they can escape easily.

The role of outflows is apparent at all redshifts. Simultaneous observations of Lyα and other non-resonant nebular emission lines indicate that Lyα lines typically are redshifted with respect to these other lines by $\Delta v_{Ly\alpha}$. This redshift is more prominent for drop-out (Lyman break) galaxies, in which the average $\Delta v_{Ly\alpha} \sim 460$ km s$^{-1}$ in LBGs (Steidel et al. 2010, Kulas et al. 2012), which is larger than the shift observed in LAEs, where the average $\Delta v \sim 200$ km s$^{-1}$ (McLinden et al. 2011, Chonis et al. 2013, Hashimoto et al. 2013, Erb et al. 2014, McLinden et al. 2014, Song et al 2014, Trainor et al. 2015, Prescott et al. 2015)[41]. These observations indicate that outflows generally affect Lyα radiation while it is escaping from galaxies. This is not surprising:

---

[41] The different $\Delta v$ in LBGs and LAEs likely relates to the different physical properties of both samples of galaxies. Shibuya et al. (2014) argue that LAEs may contain smaller $N_{HI}$ which facilitates Lyα escape, and results in a smaller shift (also see Song et al., 2014).

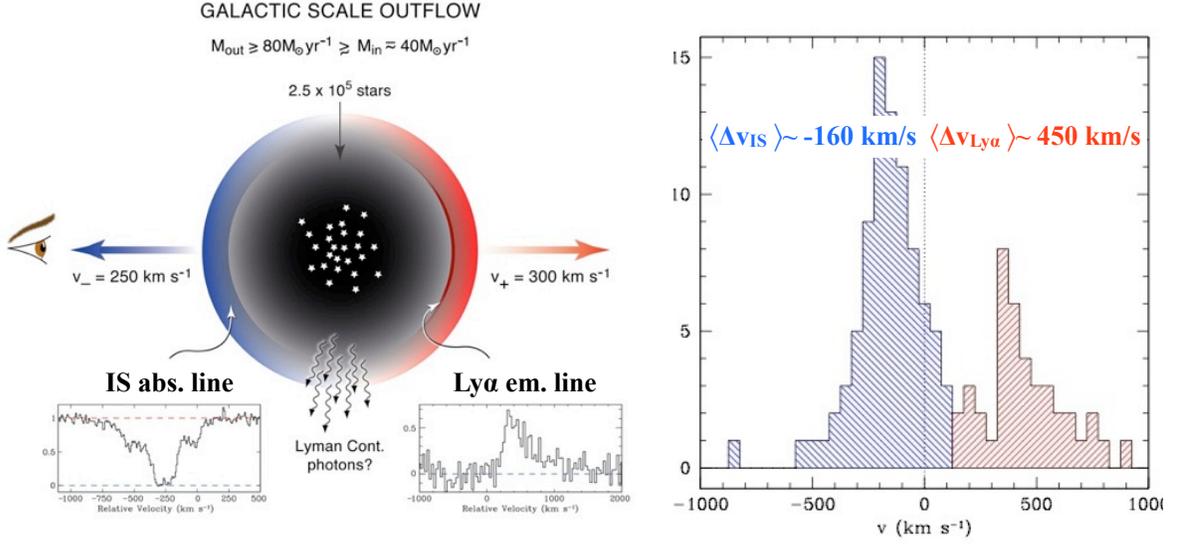

FIG. 37 The Figure shows (some) observational evidence for the ubiquitous existence of cold gas in outflows in star-forming galaxies, and that this cold gas affects the Lyα transport: the *right panel* shows the vast majority of low-ionization interstellar (IS) absorption lines are blueshifted relative to the systemic velocity of the galaxy, which is indicative of outflows (as illustrated in the *left panel*. Moreover, the *right panel* illustrates that the Lyα emission line is typically redshifted by an amount that is ∼ 2 − 3 times larger than typical blueshift of the IS lines *in the same galaxies*. These observations are consistent with a scenario in which Lyα photons scatter back to the observer from the far-side of the nebular region (indicated schematically in the *left panel*). *Credit: figure as a whole corresponds to Figure 12 of Dijkstra 2014, Lyman Alpha Emitting Galaxies as a Probe of Reionization, PASA, 31, 40D.*

outflows are detected ubiquitously in absorption in other low-ionization transitions (e.g. Steidel et al., 2010). Moreover, the Lyα photons appear to interact with the outflow, as the Lyα line is redshifted by an amount that is correlated with the outflow velocity inferred from low-ionization absorption lines (e.g. Steidel et al. 2010, Shibuya et al. 2014). The presence of winds and their impact on Lyα photons is illustrated schematically in Figure 37.

As modelling the outflowing component in interstellar medium is an extremely challenging task (as we mentioned in the beginning of this section), simplified representations, such as the popular 'shell model', have been invoked. In the shell model the outflow is represented by a spherical shell with a thickness that is $0.1\times$ its inner/outer radius. Figure 38 summarizes the different ingredients of the shell model. The two parameters that characterize the Lyα sources are (i) its equivalent width (EW) which measures the 'strength' of the source compared to the underlying continuum, (ii) its full width at half maximum (FWHM) which denotes the width of the spectral line prior to scattering. This width may reflect motions in the Lyα emitting gas. The main properties that characterise the shell are its (i) HI-column density, $N_{HI}$, (ii) outflow velocity, $v_{sh}$, (iii) 'b-parameter' $b^2 \equiv v_{turb}^2 + v_{th}^2$. Here, $v_{th} = \sqrt{2k_B T/m_p}$ (which we encountered before), and $v_{turb}$ denotes its turbulent velocity dispersion; (iv) its dust content (e.g. Ahn et al., 2003; Verhamme et al., 2006, 2008).

The shell-model can reproduce observed Lyα spectral line shapes remarkably well (e.g. Verhamme et al. 2008, Schaerer & Verhamme 2008, Dessauges-Zavadsky et al. 2010, Vanzella et al 2010, Hashimoto et al.2015, Yang et al. 2016), though not always (see e.g. Barnes & Haehnelt 2010, Kulas et al. 2012, Chonis et al. 2013, Forero-Romero et al. in prep.). One example of a good shell model fit to an observed spectrum is shown in Figure 39. Here, the galaxy is a $z \sim 0.3$ green pea galaxy (fit taken from Yang et al. 2016 using fitting algorithm from Gronke et al. 2015 and data from Henry et al. 2015). The 'triangle' diagram (produced using a modified version of `triangle.py`, Foreman-Mackey et al. 2013) shows constraints on the 6 shell model parameters, and their correlations. 'Typical' HI column densities in shell models are $N_{HI} = 10^{18} - 10^{21}$ cm$^{-2}$ and $v_{sh} \sim$ a few tens to a few hundreds km s$^{-1}$. For a limited range of column densities, the Lyα spectrum peaks at $\sim 2v_{sh}$. This peak consists of photons that scatter 'back' to the observer on the far side of the Lyα source, and are then Doppler boosted to twice the outflow velocity[42], where they are sufficiently far in the wing of the absorption cross section to escape from the medium (the cross section at $\Delta v = 200$ km s$^{-1}$ is only $\sigma_\alpha \sim$ a few times $10^{-20}$ cm$^2$, see Eq 55 and Fig 17).

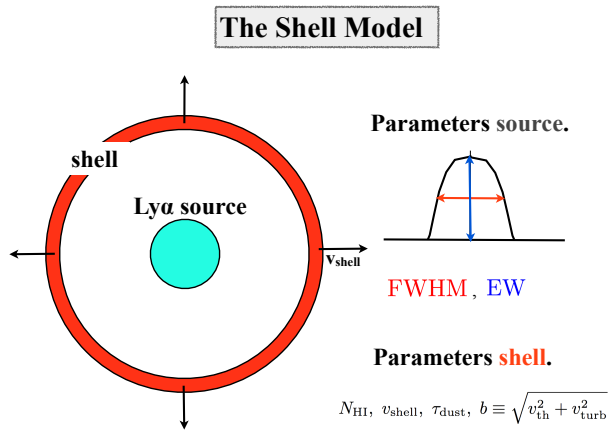

FIG. 38 The 'shell model' is a simplified representation of the Lyα transfer process on interstellar scales. The shell model contains six parameters, and generally reproduces observed Lyα spectral line profiles remarkably well (see Fig 39).

In spite of its success, there are two issues with the shell-models: (i) gas in the shells has a single outflow velocity and a small superimposed velocity dispersion, while observations of low-ionization absorption lines indicate that outflows typically cover a much wider range of velocities (e.g. Kulas et al. 2012, Henry et al. 2015); and (ii) observations of low-ionization absorption lines also suggest that outflows - while ubiquitous - do not completely surround UV-continuum emitting regions of galaxies. Observations by Jones et al. (2013) show that the maximum low-ionization covering fraction is 100% in only 2 out of 8 of their $z > 2$ galaxies (also see Heckman et al. 2011, who find evidence for a low covering factor of optically thick, neutral gas in a small fraction of lower redshift Lyman Break Analogues). There is thus some observational evidence that there exist sight lines that contain no detectable low-ionization (i.e. cold) gas, which may reflect the complex structure associated with outflows which cannot be captured with spherical shells. Two caveats are that (a) the inferred covering factors are measured as a function of velocity (and can depend on spectral resolution, see e.g. Prochaska, 2006, but Jones et al. 2013 discuss why this is likely not an issue in their analysis). Gas at different velocities can cover different parts of the source, and the outflowing gas may still fully cover the UV emitting source. This velocity-dependent covering is nevertheless not captured by the shell-model; (b) the low-ionization metal absorption lines only probe enriched cold (outflowing) gas. Especially in younger galaxies it may be possible that there is additional cold (outflowing) gas that is not probed by metal absorption lines.

Shibuya et al. (2014) have shown that Lyα line emission is stronger in galaxies in which the covering factor of low-ionization material is smaller (see their Fig 10, also see Trainor et al. 2015). Similarly, Jones et al. (2012) found the average absorption line strength in low-ionization species to decrease with redshift, which again coincides with an overall increase in Lyα flux from these galaxies (Stark et al., 2010). Besides dust, the covering factor of HI gas therefore plays an additional important role in the escape of Lyα photons. These cavities may correspond to regions that have been cleared of gas and dust by feedback processes (see Nestor et al., 2011, 2013, who describe a simple 'blow-out' model).

---

[42] This argument implicitly assumes that the scattering is partially coherent (see § 7.3): photons experience a Doppler boost $\Delta\nu/\nu \sim -v_{out}/c$ when they enter the shell, and an identical Doppler boost $\Delta\nu/\nu \sim -v_{out}/c$ when they exit the shell in opposite direction (as is the case for 'back scattered' radiation). In the case of partially coherent scattering, the frequency of the photon changes only little in the frame of the gas (because $\sqrt{\langle \Delta x_{in}^2 | x_{in} \rangle} = 1$), and the total Doppler boost equals the sum of the two Doppler boosts imparted upon entry and exit from the shell.

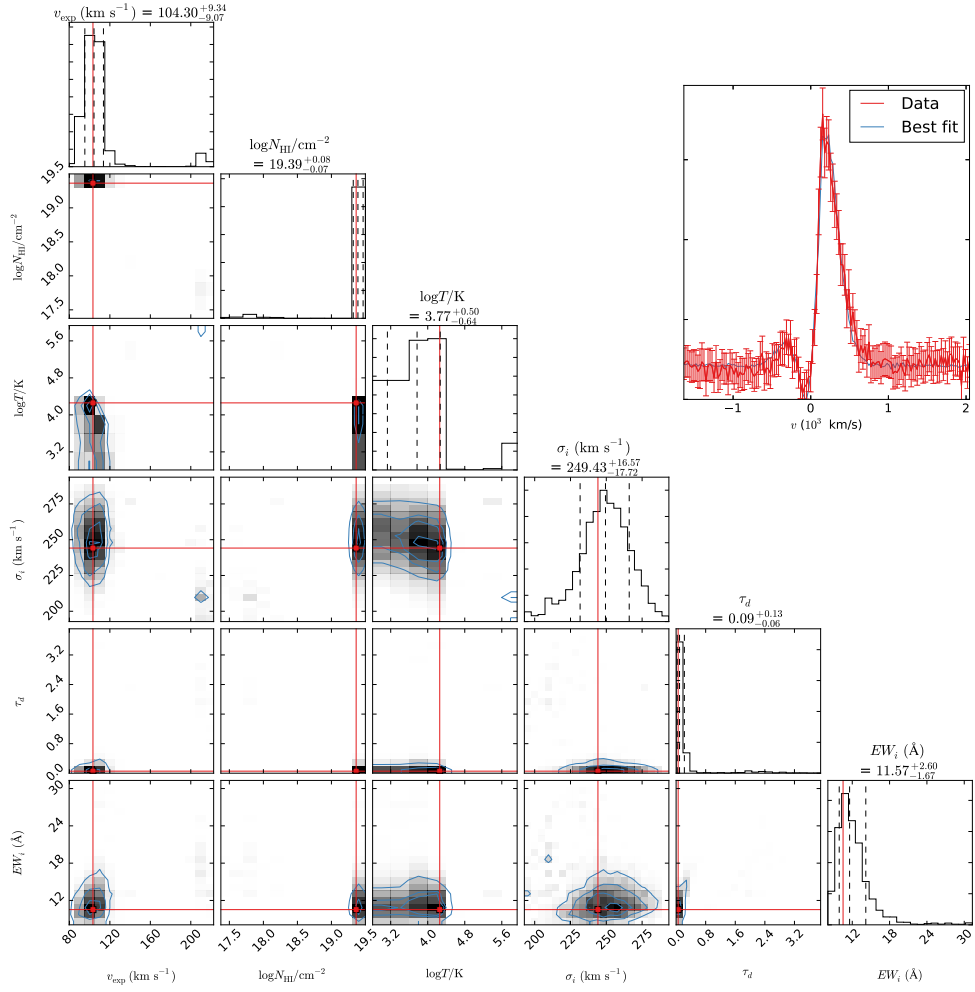

FIG. 39 This figure shows an example of an observed Lyα spectral line shape of a $z \sim 0.3$ green pea galaxy (Henry et al. 2015), which can be reproduced well by the shell model (see Yang et al. 2016, using fitting algorithm from Gronke et al. 2015). The triangle diagram shows the constraints on the shell model parameters, and degeneracies that exist between them (*Image Credit: Max Gronke*).

In short, dusty outflows appear to have an important impact on the interstellar Lyα radiative process, and give rise to redshifted Lyα lines. Low HI-column density holes further facilitate the escape of Lyα photons from the ISM, and can alter the emerging spectrum such that Lyα photons can emerge closer to the galaxies' systemic velocities (Behrens et al. 2014, Verhamme et al. 2015, Gronke & Dijkstra 2016, Dijkstra et al. 2016b, also see Zheng & Wallace 2014).

## 10.2. Transfer in the ionised IGM/CGM.

HI gas that exists outside of the galaxy can further scatter Lyα that escaped from the ISM. If this scattering occurs sufficiently close to the galaxy, then this radiation can be detected as a low surface brightness glow (e.g. Zheng et al., 2010, 2011). As we showed previously in Figure 13, observations indicate that spatially extended Lyα halos appear to exist generally around star-forming galaxies (see e.g. Fynbo et al. 1999, Hayashino et al. 2004, Hayes et al. 2005, 2007, Rauch et al. 2008, Östlin et al. 2009, Steidel et al. 2011, Matsuda et al. 2012, Hayes et al. 2013, Momose et al. 2014, Guaita et al. 2015, Wisotzki et al., 2016, Xue et al. 2017). Understanding what fraction of the Lyα flux in these halos consists of scattered Lyα radiation that escaped from the ISM, and what fraction was produced in-situ (as recombination, cooling, and/or fluorescence, see § 5) is still an open question[43], which we can address with integral field spectographs such as MUSE and the Keck Cosmic Web Imager. Polarization measurements (see § 11) should also be able to distinguish between in-situ production and scattering (Dijkstra & Loeb 2008b, Hayes et al. 2011b, Beck et al. 2016), although these differences can be subtle (Trebitsch et al., 2016).

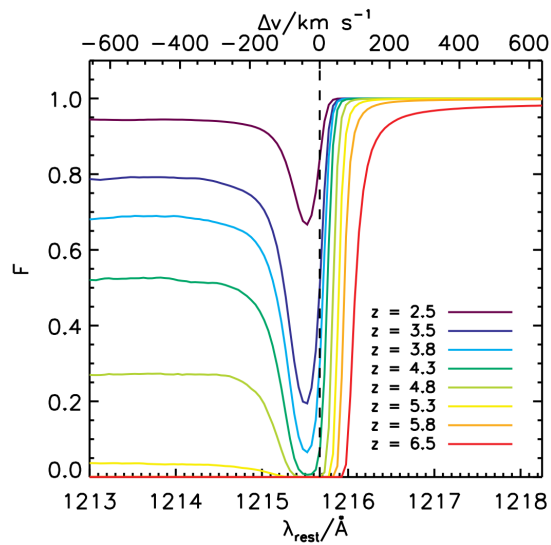

FIG. 40 The average fraction of photons that are transmitted though the IGM to the observer as a function of (restframe) wavelength. Overdense gas in close proximity to the galaxy - this gas can be referred to as 'circum-galactic gas' - enhances the opacity in the forest at velocities close to systemic (i.e $v_{sys} = 0$). Inflowing circum-galactic gas gives rise to a large IGM opacity even at a range of velocities redward of the Ly$\alpha$ resonance. Each line represents a different redshift. At wavelengths well on the blue side of the line, we recover the mean transmission measured from the Ly$\alpha$ forest. Overdense gas at close proximity to the galaxy increases the IGM opacity close to the Ly$\alpha$ resonance (and causes a dip in the transmission curve, *Credit: from Figure 2 of Laursen et al. 2011 ©AAS. Reproduced with permission*).

Radiation that scatters 'sufficiently' far out is too faint to be detected with direct observations, and is effectively removed from observations. This applies especially to photons that emerge on the blue side of the line resonance (i.e. $\nu > \nu_\alpha$ and/or $x > 0$), as these will redshift into the Ly$\alpha$ resonance due to the Hubble expansion of the Universe. Once these photons are at resonance there is a finite probability to be scattered. These photons are clearly not destroyed, but they are removed from the intensity of the radiation pointed at us. For an observer on Earth, these photons are effectively destroyed. Quantitatively, we can take 'sufficiently far' to mean that $r_{IGM} \gtrsim 1.5 r_{vir}$, where $r_{vir}$ denotes the virial radius of halos hosting dark matter halos (Laursen et al., 2011), and $r_{IGM}$ denotes the radius beyond which scattered radiation is effectively removed from obsevations. Clearly, $r_{IGM}$ depends on the adopted surface brightness sensitivity and the total Ly$\alpha$ luminosity of the source, and especially with sensitive MUSE observations, it will be important to verify what '$r_{IGM}$' is. At $r > r_{IGM}$ 'intergalactic' radiative transfer consists of simply suppressing the emerging Ly$\alpha$ flux at frequency $\nu_{em}$ by a factor of $\mathcal{T}_{IGM}(\nu_{em}) \equiv e^{-\tau_{IGM}(\nu_{em})}$, where $\tau_{IGM}(\nu)$ equals

$$\tau_{IGM}(\nu_{em}) = \int_{r_{IGM}}^{\infty} ds \, n_{HI}(s) \sigma_\alpha(\nu[s, \nu_{em}]). \tag{98}$$

As photons propagate a proper differential distance $ds$, the cosmological expansion of the Universe redshifts the photons by an amount $d\nu = -ds H(z)\nu/c$. Photons that were initially emitted at $\nu_{em} > \nu_\alpha$ will thus redshift into the line resonance. Because $\sigma_\alpha(\nu)$ is peaked sharply around $\nu_\alpha$ (see Fig 17), we can approximate this integral by taking $n_{HI}(s)$ and $c/\nu \approx \lambda_\alpha$ outside of the integral. We make an additional approximation and assume that $n_{HI}(s)$ corresponds to $\bar{n}_{HI}(z)$, where $\bar{n}_{HI}(z) = \Omega_b h^2 (1 - Y_{He})(1+z)^3/m_p$ denotes the mean number density of hydrogen atoms in the Universe at redshift $z$. If we evaluate this expression at $\nu_{em} > \nu_\alpha$, i.e. at frequencies blueward of the Ly$\alpha$ resonance, then we obtain the famous Gunn-Peterson optical depth (Gunn & Peterson, 1965):

$$\tau_{IGM}(\nu_{em} > \nu_\alpha) \equiv \tau_{GP} = \frac{\bar{n}_{HI}(z)\lambda_\alpha}{H(z)} \int_0^\infty d\nu \, \sigma_\alpha(\nu) = \frac{\bar{n}_{HI}(z)\lambda_\alpha}{H(z)} f_\alpha \frac{\pi e^2}{m_e c} \approx 7.0 \times 10^5 \left(\frac{1+z}{10}\right)^{3/2}, \tag{99}$$

where we used that $\int d\nu \, \sigma(\nu) = f_\alpha \frac{\pi e^2}{m_e c}$ (e.g. Rybicki & Lightman, 1979, p 102). The redshift dependence of $\tau_{IGM}$ reflects that $n_{HI}(z) \propto (1+z)^3$ and that at $z \gg 1$ $H(z) \propto (1+z)^{3/2}$. Eq 99 indicates that if the IGM were 100% neutral, it would be extremely opaque to photons emitted blue-ward of the Ly$\alpha$ resonance. Observations of quasar absorption line spectra indicate that the IGM transmits an *average* fraction $F \sim 85\%$ [$F \sim 40\%$] of Ly$\alpha$ photons at $z = 2$ [$z = 4$] (see lecture notes by X. Prochaska) which imply 'effective' optical depths of $\tau_{eff} \equiv -\ln[F] \sim 0.15$ [$\tau_{eff} \sim 0.9$] (e.g. Faucher-Giguère et al., 2008). The measured values $\tau_{eff} \ll \tau_{GP}$ which is (of course) because the Universe was highly ionized at these redshifts. A common approach to model the impact of the IGM is to reduce the Ly$\alpha$ flux on the blue side of the Ly$\alpha$ resonance by this observed (average) amount, while transmitting all flux on the red side.

_______

[43] Modeling the distribution of cold, neutral gas in the CGM is also challenging as there is increasing observational support that the CGM of galaxies contains cold, dense clumps (of unknown origin) on scales that are not resolved with current cosmological simulations (Cantalupo et al., 2014; Hennawi et al., 2015).

The values of $F$ and $\tau_{eff}$ mentioned above are averaged over a range of frequencies. In detail, density fluctuations in the IGM give rise to enhanced absorption in overdense regions which is observed as the Ly$\alpha$ forest. It is important to stress that galaxies populate overdense regions of the Universe in which: (*i*) the gas density was likely higher than average (see e.g. Fig 2 of Barkana, 2004), (*ii*) peculiar motions of gas attracted by the gravitational potential of dark matter halos change the relation between $ds$ and $d\nu$, (*iii*) the local ionising background was likely elevated. We thus clearly expect the impact of the IGM[44] at frequencies close to the Ly$\alpha$ emission line to differ from the mean transmission in the Ly$\alpha$ forest: Figure 40 shows the transmitted fraction of Ly$\alpha$ photons averaged over a large number of sight lines to galaxies in a cosmological hydrodynamical simulation (Laursen et al. 2011). This Figure shows that infall of over dense gas (and/or retarded Hubble flows) around dark matter halos hosting Ly$\alpha$ emitting galaxies can give rise to an increased opacity of the IGM around the Ly$\alpha$ resonance, and even extending somewhat into the red side of the Ly$\alpha$ line (Santos 2004, Dijkstra et al. 2007, Iliev et al. 2008, Laursen et al. 2011, Dayal et al. 2011).

Because these models predict that the IGM can strongly affect frequencies close to the Ly$\alpha$ resonance, the overall impact of the IGM *depends strongly on the Ly$\alpha$ spectral line shape* as it emerges from the galaxy (also see Haiman, 2002; Santos, 2004). This is illustrated by the *lower three panels* in Figure 41. For Gaussian and/or generally symmetric emission lines centered on the galaxies' systemic velocities, the IGM can transmit as little as[45] $\mathcal{T}_{IGM} = 10 - 30\%$ even for a fully ionized IGM (e.g. Dijkstra et al. 2007, Zheng et al. 2010, Dayal et al. 2011, Laursen et al. 2011). However, when scattering through outflows shifts the line sufficiently away from line centre, the overall impact of the IGM can be reduced tremendously (e.g. Haiman 2002, Santos 2004, Dijkstra et al. 2011, Garel et al. 2012). *That is, not only do we care about how much Ly$\alpha$ escapes from the dusty ISM, we must care as much about how the emerging photons escape in terms of the line profile.* The fraction $\mathcal{T}_{IGM}$ (also indicated in Fig 41) denotes the 'IGM transmission' and denotes the total fraction of the Ly$\alpha$ radiation emitted by a galaxy that is transmitted by the IGM. The IGM transmission $\mathcal{T}_{IGM}$ is given by the integral over the frequency-dependent transmission, $e^{-\tau_{IGM}(\nu)}$. This frequency-dependence can be expressed as a function of the dimensionless frequency variable $x$, or as a function of the velocity offset $\Delta v \equiv -x v_{th}$ as:

$$\mathcal{T}_{IGM} = \int_{-\infty}^{\infty} d\Delta v \, J_\alpha(\Delta v) \exp[-\tau_{IGM}(z_g, \Delta v)] \tag{100}$$

where $J_\alpha(\Delta v)$ denotes the line profile of Ly$\alpha$ photons as they escape from the galaxy at $z_g$. The IGM opacity discussed above originates in mildly over dense ($\delta = 1 - 20$, see Dijkstra et al. 2007), highly ionized gas. Another source of opacity is provided by Lyman-limit systems (LLSs) and Dampled Ly$\alpha$ absorbers (DLAs). The precise impact of these systems on Ly$\alpha$ radiation has only been studied recently (Bolton & Haehnelt 2013, Mesinger et al. 2015, Choudhury et al. 2015, Kakiichi et al. 2016), and depends most strongly on how they cluster around Ly$\alpha$ emitting galaxies.

---

[44] Early studies defined the IGM to be all gas at $r > 1 - 1.5$ virial radii, which would correspond to the 'circum-galactic' medium by more recent terminology. Regardless of what we call this gas, scattering of Ly$\alpha$ photons would remove photons from a spectrum of a galaxy, and redistribute these photons over faint, spatially extended Ly$\alpha$ halos.

[45] It is worth noting that these models predict that the IGM can reduce the observed Ly$\alpha$ line by as much as $\sim 30\%$ between $z = 5.7$ and $z = 6.5$ (Laursen et al., 2011). Observations of Ly$\alpha$ halos around star forming galaxies provide hints that scattering in this CGM may be more prevalent at $z = 6.5$ than at $z = 5.7$, although the statistical significance of this claim is weak (Momose et al., 2014).

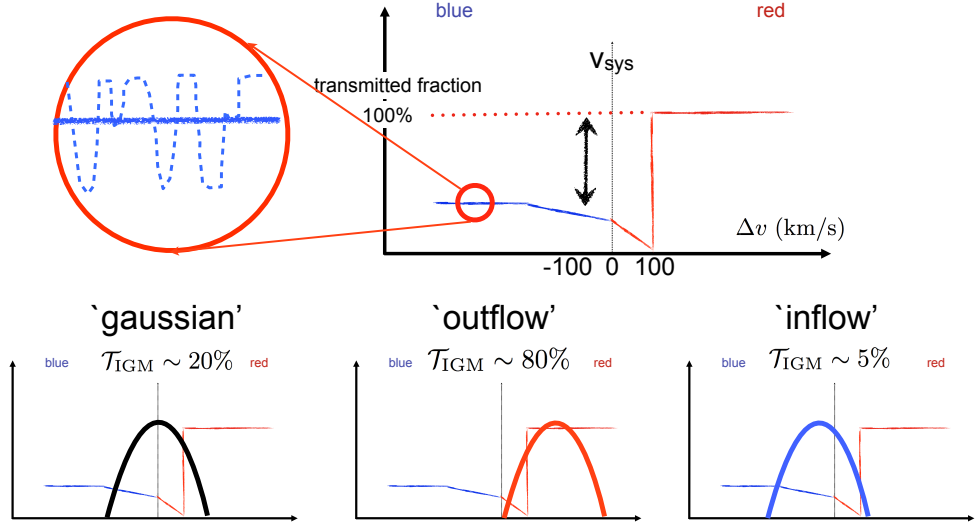

FIG. 41 Schematic representation of the impact of a residual amount of intergalactic HI on the fraction of photons that is directly transmitted to the observer, $\mathcal{T}_{\rm IGM}$ for a set of different line profiles. The *top panel* shows that the Ly$\alpha$ forest (shown in the *inset*) suppresses flux on the blue side of the Ly$\alpha$ line. The *lower left figure* shows that for lines centered on $v_{\rm sys} = 0$ (here symmetric around $v_{\rm sys} = 0$ for simplicity), the IGM cuts off a significant fraction of the blue half of the line, and some fraction of the red half of the line. For lines that are redshifted [blueshifted] w.r.t $v_{\rm sys}$, larger [smaller] fraction of emitted Ly$\alpha$ photons falls outside of the range of velocities affected by the IGM. The line profiles set at the interstellar level thus plays a key role in the subsequent intergalactic radiative transfer process. *Credit: from 'Understanding the Epoch of Cosmic Reionization: Challenges and Progress', Vol 423, Fig 3 of Chapter 'Constraining Reionization with Ly$\alpha$ Emitting Galaxies' by Mark Dijkstra, 2016, page 145-161, With permission of Springer.*

### 10.3. Intermezzo: Reionization

Reionization refers to the transformation of the intergalactic medium from fully neutral to fully ionized. For reviews on the Epoch of Reionization (EoR) we refer the reader to e.g. Barkana & Loeb (2001), Furlanetto et al. (2006), Morales & Wyithe (2010), and the recent book by Mesinger (2016). The EoR is characterized by the existence of patches of diffuse neutral intergalactic gas, which provide an enormous source of opacity to Ly$\alpha$ photons: the Gunn-Peterson optical depth is $\tau_{\rm GP} \sim 10^6$ (see Eq 99) in a fully neutral medium. It is therefore natural to expect that detecting Ly$\alpha$ emitting galaxies from the EoR is hopeless. Fortunately, this is not the case, as we discuss in § 10.4.

Reionization was likely not a homogeneous process in which the ionization state of intergalactic hydrogen changed everywhere in the Universe at the same time, and at the same rate. Instead, it was likely a temporally extended inhomogeneous process. The first sources of ionizing radiation were highly biased tracers of the underlying mass distribution. As a result, these galaxies were clustered on scales of tens of comoving Mpc (cMpc, Barkana & Loeb 2004). The strong clustering of these first galaxies in overdense regions of the Universe caused these regions to be reionized first (e.g. Fig 1 of Wyithe & Loeb, 2007), which thus created fluctuations in the ionization field over similarly large scales. As a result a proper description of the reionization process requires simulations that are at least 100 cMpc on the side (e.g. Trac & Gnedin, 2011).

Ideally, one would like to simulate reionization by performing full radiative transfer calculations of ionising photons on cosmological hydrodynamical simulations. A number of groups have developed codes that can perform these calculations in 3D (e.g. Gnedin 2000, Sokasian et al. 2001, Ciardi et al. 2003, Mellema et al. 2006, Trac & Cen 2007, Pawlik & Schaye 2008, Finlator et al. 2009, So et al. 2014, Gnedin 2016). These calculations are computationally challenging as one likes to simultaneously capture the large scale distribution of HII bubbles, while resolving the photon sinks (such as Lyman Limit systems) and the lowest mass halos ($M \sim 10^8$ M$_\odot$) which can contribute to the ionising photon budget (see e.g. Trac & Gnedin, 2011). Modeling reionization contains many poorly known parameters related to galaxy formation, the ionising emissivity of star-forming galaxies, their spectra etc. Alternative, faster 'semi-numeric' algorithms have been developed which allow for a more efficient exploration of the full parameter space (e.g. Mesinger & Furlanetto 2007, Mesinger et al. 2011, Majumdar et al. 2014, Sobacchi & Mesinger 2014). These semi-numeric algorithms utilize excursion-set theory to determine if a cell inside a simulation is ionized or not (Furlanetto et al., 2004). Detailed comparisons between full radiation transfer simulations and semi-numeric simulations show both methods produce very similar ionization fields (Zahn et al., 2011).

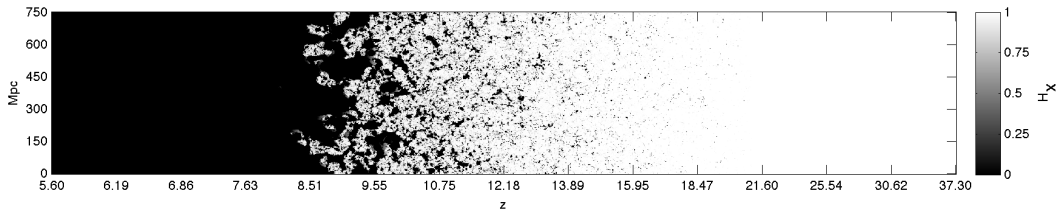

FIG. 42 The predicted redshift evolution of the ionization state of the IGM in a realistic reionization model (*Credit: Figure kindly provided by Andrei Mesinger, published previously as Fig 14 in Dijkstra 2014, Lyman Alpha Emitting Galaxies as a Probe of Reionization, PASA, 31, 40D*). The white/black represents fully neutral/ionized intergalactic gas. This Figure demonstrates the inhomogeneous nature of the reionization process which took place over an extended range of redshifts: at $z > 16$ the first ionized regions formed around the most massive galaxies in the Universe (at that time). During the final stages of reionization - here at $z \sim 9$ the IGM contains ionized bubbles several tens of cMpc across.

The picture of reionization that has emerged from analytical consideration and large-scale simulations is one in which the early stages of reionization are characterized by the presence of HII bubbles centered on overdense regions of the Universe, completely separated from each other by a neutral IGM (Furlanetto et al., 2004; Iliev et al., 2006; McQuinn et al., 2007). The ionized bubbles grew in time, driven by a steadily growing number of star-forming galaxies residing inside them. The final stages of reionization are characterized by the presence of large bubbles, whose individual sizes exceeded tens of cMpc (e.g. Zahn et al. 2011, Majumdar et al. 2014). Ultimately these bubbles overlapped (percolated), which completed the reionization process. The predicted redshift evolution of the ionization state of the IGM in a realistic reionization model is shown in Figure 42. This Figure illustrates the inhomogeneous, temporally extended nature of the reionization process.

## 10.4. Intergalactic Lyα Radiative Transfer during Reionization

There are indications that we are seeing Lyα emission from galaxies in the reionization epoch: there is increasing observational support for the claim that Lyα emission experiences extra opacity at $z > 6$ compared to the expectation based on extrapolating observations from lower redshift, and which is illustrated in Figure 43. The *left panel* shows the drop in the 'Lyα fraction' in the drop-out (Lyman Break) galaxy population. The 'Lyα fraction' denotes the fraction of continuum selected galaxies which have a 'strong' Lyα emission line (Stark et al. 2010). The 'strength' of the Lyα emission line is quantified by some arbitrary equivalent width threshold. The Lyα fraction rises out to $z \sim 6$ (this trend continues down to lower redshift), and then suddenly drops at $z > 6$ (Pentericci et al. 2011, Schenker et al. 2012, Ono et al. 2012, Caruana et al. 2012, Treu et al. 2013, Pentericci et al. 2014, Caruana et al. 2014). The *right panel* shows the sudden evolution in the Lyα luminosity function of Lyα selected galaxies (LAEs). The Lyα luminosity function does not evolve much between $z \sim 3$ and $z \sim 6$ (Ouchi et al. 2008, also see Hu et al. 1998), but then suddenly drops at $z \gtrsim 6$ (Kashikawa et al. 2006, Ouchi et al. 2010, Kashikawa et al. 2011, Faisst et al. 2014, Santos et al. 2016, Zheng et al. 2017, Ota et al. 2017), with this evolution being most apparent at $z \gtrsim 6.5$ (see e.g. Ota et al. 2010, Matthee et al. 2015). The observed redshift-evolution of the Lyα luminosity function of LAEs and Lyα fractions have been shown to be quantitatively consistent (see Dijkstra & Wyithe 2012, Gronke et al. 2015).

During reionization we denote the opacity of the IGM in the *ionized bubbles* at velocity off-set $\Delta v$ and redshift $z$ with $\tau_{\rm HII}(z, \Delta v)$. This allows us to more explicitly distinguish this source of intergalactic opacity from the 'damping wing' opacity, $\tau_D(z, \Delta v)$, which is the opacity due to the diffuse neutral IGM and which is only relevant during reionization. In other words, $\tau_{HII}$ refers to the Lyα opacity in neutral gas that survived in the ionized bubbles. This neutral gas can reside in dense self-shielding clouds[46], or as residual neutral hydrogen that survived in the ionized bubbles[47]. Eq 100 therefore changes to

$$\mathcal{T}_{IGM} = \int_{-\infty}^{\infty} d\Delta v \, J_\alpha(\Delta v) \exp[-\tau_{IGM}(z_g, \Delta v)], \quad \tau_{IGM}(z_g, \Delta v) = \underbrace{\tau_D(z_g, \Delta v)}_{\text{reionization only}} + \tau_{HII}(z_g, \Delta v). \quad (101)$$

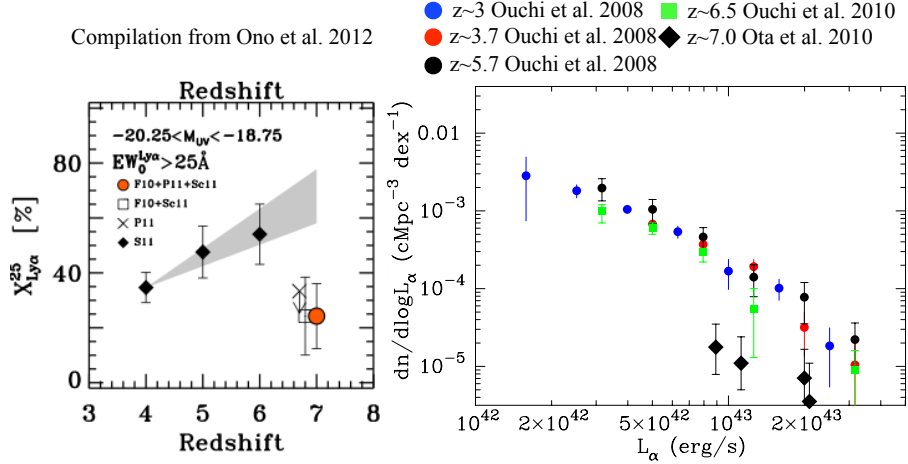

FIG. 43 There are two independent observational indications that the Lyα flux from galaxies at $z > 6$ is suppressed compared to extrapolations from lower redshifts. The *left panel* shows the drop in the 'Lyα fraction' in the drop-out (Lyman Break) galaxy population, the *right panel* shows the sudden evolution in the Lyα luminosity function of Lyα selected galaxies (LAEs). This evolution has been shown to be quantitatively consistent (see Dijkstra & Wyithe 2012, Gronke et al. 2015). *Credit: from 'Understanding the Epoch of Cosmic Reionization: Challenges and Progress', Vol 423, Fig 1 of Chapter 'Constraining Reionization with Lyα Emitting Galaxies' by Mark Dijkstra, 2016, page 145-161, With permission of Springer.*

Decomposing $\tau_{IGM}(z_g, \Delta v)$ into $\tau_D(z_g, \Delta v)$ and $\tau_{HII}(z_g, \Delta v)$ is helpful as they depend differently on $\Delta v$: the *left panel* of Figure 44 shows the IGM transmission, $e^{-\tau_D(z_g, \Delta v)}$, as a function of velocity off-set ($\Delta v$) for the diffuse neutral IGM for $x_{HI} = 0.8$. This Figure shows clearly that the neutral IGM affects a range of frequencies that extends much further to the red-side of the Lyα resonance than the ionized IGM (compare with Fig 40). This large opacity $\tau_D(z_g, \Delta v)$ far redward of the Lyα resonance is due to the damping wing of the absorption cross-section (and not due to gas motions at these large velocities), which is why we refer to it as the 'damping wing optical depth'. The *right panel* shows that there is a contribution to the damping wing optical depth from neutral, self-shielding clouds (with $N_{HI} \gtrsim 10^{17}$ cm$^{-2}$, also see the high-redshift curves in Fig 40) which can theoretically mimic the impact of a diffuse neutral IGM (Bolton & Haehnelt 2013), though this requires a large number density of these clouds (see Bolton & Haehnelt 2013, Mesinger et al. 2015, Choudhury et al. 2015, Kakiichi et al. 2016).

Star-forming galaxies that are luminous enough to be detected with existing telescopes likely populated dark matter halos with masses in excess of $M \gtrsim 10^{10}$ M$_\odot$ (see e.g. Sobacchi & Mesinger 2015). These halos preferentially reside in over dense regions of the Universe, which were reionized earliest. It is therefore likely that (Ly$\alpha$ emitting) galaxies preferentially resided inside these large HII bubbles. This has an immediate implication for the visibility of the Ly$\alpha$ line. Ly$\alpha$ photons emitted by galaxies located inside these HII regions can propagate (to the extent that is permitted by the ionized IGM) - and therefore redshift away from line resonance - through the ionized IGM before encountering the neutral IGM. Because of the strong frequency-dependence of the Ly$\alpha$ absorption cross section, these photons are less likely to be scattered out of the line of sight inside the neutral IGM. A non-negligible fraction of Ly$\alpha$ photons may be transmitted directly to the observer, which is illustrated schematically in Figure (45). *Inhomogeneous reionization thus enhances the prospect for detecting Ly$\alpha$ emission from galaxies inside HII bubbles* (see Dijkstra, 2014, for a review, and an extensive list of references). It also implies that the impact of diffuse neutral intergalactic gas on the visibility of Ly$\alpha$ flux from galaxy is more subtle than expected in models in which reionization proceeds homogeneously, and that the observed reduction in Ly$\alpha$ flux from galaxies at $z > 6$ requires a significant volume filling fraction of neutral gas (which is indeed the case, as we discuss below). This Figure also illustrates that Ly$\alpha$ photons emitted by galaxies that lie outside of large HII bubbles, scatter repeatedly in the IGM. These photons diffuse outward, and are visible only as faint extended Ly$\alpha$ halos (Loeb & Rybicki 1999, Kobayashi et al. 2006, Jeeson-Daniel et al. 2012).

---

[46] To make matters more confusing: self-shielding absorbers inside the ionized bubbles with sufficiently large HI column densities can be optically thick in the Lyα damping wing, and can give rise to damping wing absorption as well. This damping wing absorption is included in $\tau_{HII}(z, \Delta v)$.

[47] Just as the Lyα forest at lower redshifts - where hydrogen reionization was complete - contains neutral hydrogen gas with different densities, ionization states and column densities.

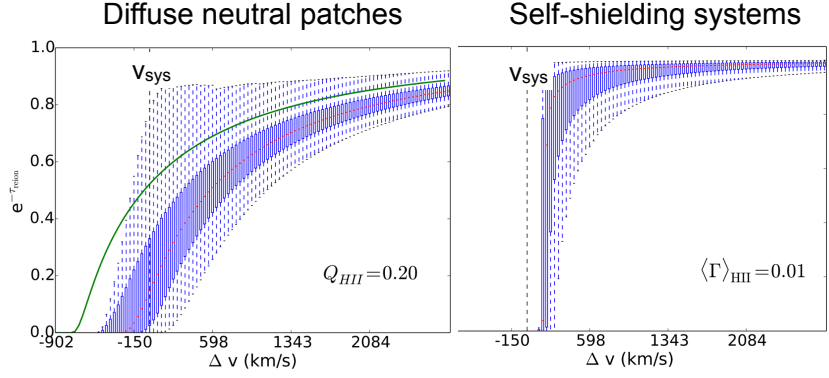

FIG. 44 Neutral gas in the intergalactic medium can give rise to a large 'damping' wing opacity $\tau_D(z_g, \Delta v)$ that extends far to the red side of the Ly$\alpha$ resonance, i.e. $\Delta v \gg 1$. The *left panel* shows the IGM transmission, $e^{-\tau_D(z_g, \Delta v)}$, as a function of velocity off-set ($\Delta v$) for the diffuse neutral IGM for $x_{HI} = 0.8$. The *right panel* shows the IGM transmission for a (large) number of self-shielding clouds (see text). To obtain the *total* IGM transmission, one should multiply the transmission curves shown here and in Fig 40. *Credit: from 'Understanding the Epoch of Cosmic Reionization: Challenges and Progress', Vol 423, Fig 4 of Chapter 'Constraining Reionization with Ly$\alpha$ Emitting Galaxies' by Mark Dijkstra, 2016, page 145-161, With permission of Springer.* Figures adapted from *Figures 2 and 4 of Mesinger et al. 2015, 'Can the intergalactic medium cause a rapid drop in Ly$\alpha$ emission at $z > 6$?', MNRAS, 446, 566.*

We can quantify the impact of neutral intergalactic gas on the Ly$\alpha$ flux from galaxies following our analysis in § 10.2. We denote the optical depth in the neutral intergalactic patches with $\tau_D$. We first consider the simplest case in which a Ly$\alpha$ photon encounters one fully neutral patch which spans the line-of-sight coordinate from $s_b$ ('b' stands for beginning) to $s_e$ ('e' stands for end):

$$\tau_D(\nu) = \int_{s_b}^{s_e} ds \; n_{HI}(s)\sigma_\alpha(\nu[s]) = \frac{n_{HI}(s)\lambda_\alpha}{H(z)} \int_{\nu_b(\nu)}^{\nu_e(\nu)} d\nu' \sigma_\alpha(\nu'). \tag{102}$$

where we followed the analysis of § 10.2, and changed to frequency variables, and assumed that $n_{HI}(s)$ is constant across this neutral patch. We eliminate $n_{HI}(s)$ by using the expression for the Gunn-Peterson optical depth in Eq 99, and recast Eq 102 as

$$\tau_D(\nu) = \tau_{GP} \frac{\int_{\nu_b(\nu)}^{\nu_e(\nu)} d\nu' \sigma_\alpha(\nu')}{\int_0^\infty d\nu' \sigma_\alpha(\nu')} = \tau_{GP} \frac{\int_{x_b(\nu)}^{x_e(\nu)} dx' \phi(x')}{\int_0^\infty dx' \phi(x')}. \tag{103}$$

The denominator can be viewed as a normalisation constant, and we can rewrite Eq 103 as

$$\tau_D(\nu) = \frac{\tau_{GP}}{\sqrt{\pi}} \int_{x_b(\nu)}^{x_e(\nu)} dx' \; \phi(x'), \tag{104}$$

where the factor of $\sqrt{\pi}$ enters because of our adopted normalisation for the Voigt profile $\phi(x)$.

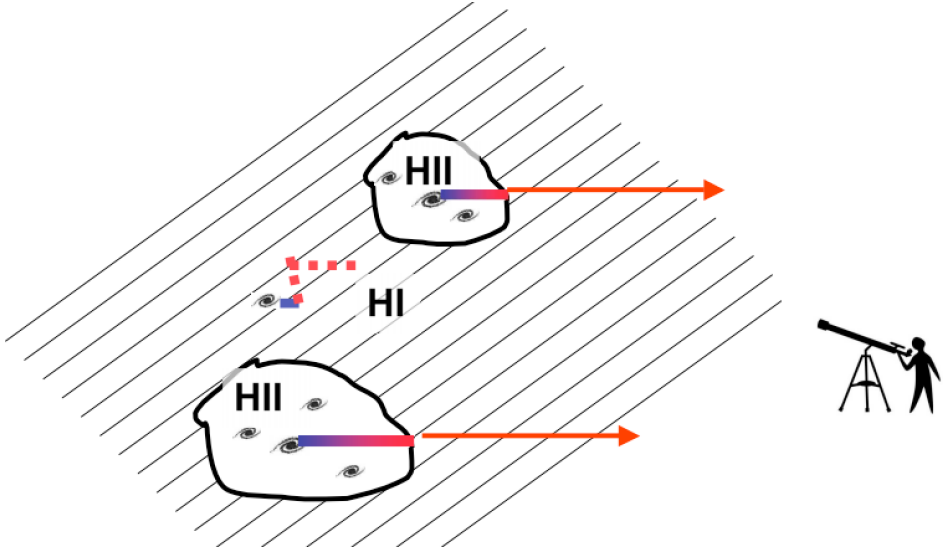

FIG. 45 This Figure schematically shows why inhomogeneous reionization boosts the visibility of Lyα emitting galaxies. During the mid and late stages of reionization star-forming - and hence Lyα emitting - galaxies typically reside in large HII bubbles. Lyα photons emitted inside these HII bubbles can propagate - redshift away from line resonance - through the ionized IGM before encountering the neutral IGM. The resulting reduced opacity of the neutral IGM (Eq 108) to Lyα photons enhances the prospect for detecting Lyα emission from those galaxies inside HII bubbles. *Credit: from Figure 15 of Dijkstra 2014, Lyman Alpha Emitting Galaxies as a Probe of Reionization, PASA, 31, 40D.*

During the EoR a Lyα photon emitted by a galaxy will generally propagate through regions that are alternating between (partially) neutral and highly ionized. The more general case should therefore contain the sum of the optical depth in separate neutral patches:

$$\tau_{\mathrm{D}}(\nu) = \frac{1}{\sqrt{\pi}} \sum_i \tau_{\mathrm{GP,i}} \, x_{\mathrm{HI,i}} \int_{x_{b,i}(\nu)}^{x_{e,i}(\nu)} dx \, \phi(x'), \tag{105}$$

where we have placed $\tau_{\mathrm{GP}}$ within the sum, because $\tau_{\mathrm{GP}}$ depends on redshift as $\tau_{\mathrm{GP}} \propto (1+z_i)^{3/2}$, and therefore differs slightly for each patch 'i' at redshift $z_i$, which has a neutral hydrogen fraction $x_{\mathrm{HI,i}}$.

More specifically, the total optical depth of the neutral IGM to Lyα photons emitted by a galaxy at redshift $z_{\mathrm{g}}$ with some velocity off-set $\Delta v$ is given by Eq 105 with $x_{b,i} = \frac{-1}{v_{\mathrm{th,i}}}[\Delta v + H(z_{\mathrm{g}})R_{b,i}/(1+z_{\mathrm{g}})]$, in which $R_{b,i}$ denotes the comoving distance to the beginning of patch 'i' ($x_{e,i}$ is defined similarly). Eq 105 must generally be evaluated numerically. However, one can find intuitive approximations: for example, if we assume that $(i)$ $x_{\mathrm{HI,i}} = 1$ for all 'i' (i.e. all patches are fully neutral), $(ii)$ $z_i \sim z_{\mathrm{g}}$, and $(iii)$ that Lyα photons have redshifted away from resonance by the time they encounter this first neutral patch[48], then

$$\tau_{\mathrm{D}}(z_{\mathrm{g}}, \Delta v) = \frac{\tau_{\mathrm{GP}}(z_{\mathrm{g}})}{\sqrt{\pi}} \sum_i \left( \frac{a_{v,\mathrm{i}}}{\sqrt{\pi} x_{e,i}} - \frac{a_{v,\mathrm{i}}}{\sqrt{\pi} x_{b,i}} \right) \equiv \frac{\tau_{\mathrm{GP}}(z_{\mathrm{g}})}{\sqrt{\pi}} \bar{x}_{\mathrm{D}} \left( \frac{a_v}{\sqrt{\pi} x_e} - \frac{a_v}{\sqrt{\pi} x_{b,1}} \right), \tag{106}$$

where $x_{e,i} = x_{e,i}(\Delta v)$ and $x_{b,i} = x_{b,i}(\Delta v)$. It is useful to explicitly highlight the sign-convention here: photons that emerge redward of the Lyα resonance have $\Delta v > 0$, which corresponds to a negative $x$. Cosmological expansion redshifts photons further, which decreases $x$ further. The $a_v/[\sqrt{\pi} x_{b,i}]$ is therefore less negative, and $\tau_{\mathrm{D}}$ is thus smaller. In the last term, we defined the 'patch-averaged' neutral fraction, $\bar{x}_D$, which is related to the volume filling factor of neutral hydrogen $\langle x_{HI} \rangle$ in a non-trivial way (see Mesinger & Furlanetto 2008).

---

[48] If a photon enters the first neutral patch on the blue side of the line resonance, then the total opacity of the IGM depends on whether the photon redshifted into resonance inside or outside of a neutral patch. If the photon redshifted into resonance inside patch 'i', then $\tau_{\mathrm{D}}(z_{\mathrm{g}}, \Delta v) = \tau_{\mathrm{GP}}(z) x_{\mathrm{HI,i}}$. If on the other hand the photon redshifted into resonance in an ionized bubble, then we must compute the optical depth in the ionized patch, $\tau_{\mathrm{HII}}(z, \Delta v = 0)$, plus the opacity due to subsequent neutral patches. Given that the ionized IGM at $z = 6.5$ was opaque enough to completely suppress Lyα flux on the blue-side of the line, the same likely occurs inside ionized HII bubbles during reionization because of $(i)$ the higher intergalactic gas density, and $(ii)$ the shorter mean free path of ionizing photons and therefore likely reduced ionizing background that permeates ionised HII bubbles at higher redshifts.

Following Mesinger & Furlanetto (2008), we can ignore the term $a_v/[\sqrt{\pi} x_e]$ and write

$$\tau_D(z_g, \Delta v) \approx \frac{\tau_{GP}(z_g)}{\pi} \bar{x}_D \frac{a_v}{|x_{b,1}|} = \frac{\tau_{GP}(z_g)}{\pi} \bar{x}_D \frac{A_\alpha c}{4\pi \nu_\alpha} \frac{1}{\Delta v_{b,1}}, \tag{107}$$

where $x_e$ denotes the frequency that photon has redshifted to when it exits from the last neutral patch, while $x_{b,1}$ denotes the photon's frequency when it encounters the first neutral patch. Because typically $|x_e| \gg |x_{b,1}|$ we can drop the term that includes $x_e$. We further substituted the definition of the Voigt parameter $a_v = A_\alpha/(4\pi\Delta\nu_\alpha)$, to rewrite $x_{b,1}$ as a velocity off-set from line resonance when a photon first enters a neutral patch, $\Delta v_{b,1} = \Delta v + H(z_g)R_{b,i}/(1+z_g)$.

Substituting numbers gives (Miralda-Escudé 1998, Dijkstra & Wyithe 2010)

$$\tau_D(z_g, \Delta v) \approx 2.3\bar{x}_D \left( \frac{\Delta v_{b,1}}{600 \text{ km s}^{-1}} \right)^{-1} \left( \frac{1+z_g}{10} \right)^{3/2}. \tag{108}$$

This equation shows that the opacity of the IGM drops dramatically once photons enter the first patch of neutral IGM with a redshift. This redshift can arise partly at the interstellar level, and partly at the intergalactic level: scattering off outflowing material[49] at the interstellar level can efficiently redshift Ly$\alpha$ photons by a few hundred km/s (see § 10.1). Because Ly$\alpha$ photons can undergo a larger cosmological subsequent redshift inside larger HII bubbles, Ly$\alpha$ emitting galaxies inside larger HII bubbles may be more easily detected. Eq 108 shows that setting $\tau_D = 1$ for $\bar{x}_D = x_{HI} = 1.0$ requires $\Delta v = 1380$ km s$^{-1}$. This cosmological redshift reduces the damping wing optical depth of the neutral IGM to $\tau_D < 1$ for HII bubbles with radii $R \gtrsim \Delta v/H(z) \sim 1$ Mpc (proper), *independent of z* (because at a fixed $R$ the corresponding cosmological redshift $\Delta v \propto H(z) \propto (1+z)^{3/2}$, Miralda-Escudé 1998). The presence of large HII bubbles during inhomogeneous reionization may have drastic implications for the prospects of detecting Ly$\alpha$ emission from the epoch.

Current models indicate that if the observed reduction in Ly$\alpha$ flux from galaxies at $z > 6$ is indeed due to reionization - which is plausible (see Greig & Mesinger 2017 for recent constraints on the reionization history from a suite of observations) - then this requires a volume filling factor of diffuse neutral gas which exceeds $\langle x_{HI} \rangle \gtrsim 40\%$, which implies that reionization is still ongoing at $z \sim 6 - 7$ (see Dijkstra 2014 for a review). This constraint is still uncertain due to the limited number of Ly$\alpha$ galaxies at $z \sim 6$ and $z \sim 7$, but this situation is expected to change, especially with large surveys for high-z Ly$\alpha$ emitters to be conducted with Hyper Suprime-Cam. These surveys will enable us to measure the variation of IGM opacity on the sky at fixed redshift, and constrain the reionization morphology (see Jensen et al. 2013, 2014, Sobacchi & Mesinger 2015).

## 11. MISCELLANEOUS TOPICS I: POLARIZATION

Ly$\alpha$ radiative transfer involves scattering. Scattered radiation can be polarized. The polarization of electromagnetic radiation measures whether there is a preferred orientation of its electric and magnetic components. Consider the example that we discussed in § 6.1 of a free electron that scatters incoming radiation (shown schematically in Fig 14). If the incoming radiation field were unpolarized, then its electric vector is distributed randomly throughout the plane perpendicular to its propagation direction prior to scattering (denoted with $\mathbf{k}_{in}$).

In § 6.1 we discussed how the intensity of scattered radiation $I \propto \sin^2 \Psi$, where $\cos \Psi \equiv \mathbf{k}_{out} \cdot \mathbf{e}_E$ in which $\mathbf{e}_E$ denotes the normalized direction of the electric vector (see Fig 15). Similarly, we can say that the amplitude of the electric-field scales as $E \propto \sin \Psi$ (note at $I \propto |E|^2$), i.e. we project the electric vector onto the plane perpendicular to $\mathbf{k}_{out}$ (see the *left panel* of Figure 46). This same argument can be applied to demonstrate that a free electron can transform unpolarized into a polarized radiation if there is a 'quadrupole anisotropy' in the incoming intensity: the *right panel* of Figure 46 shows a free electron with incident radiation from the left and from the top. If the incident radiation is unpolarized, then the electric field vector points in arbitrary direction in the plane perpendicular to the propagation direction. Consider scattering by 90°. If we apply the projection argument, then for radiation incident from the left we only 'see' the component of the E-field that points upward (shown in *blue*). Similarly, for radiation coming in from the top we only see the E-field that lies horizontally. The polarization of the scattered radiation vanishes if the *blue* and *red* components are identical, which - for unpolarized radiation - requires that the total intensity of radiation coming in from the top must be identical to the that coming in from the left. For this reason, electron scattering can polarize the Cosmic Microwave Background if the intensity varies on angular scales of 90°. If fluctuation exist on these scales, then the CMB is said to have a non-zero quadrupole moment. Similarly, if there were a point source irradiating the electron from the top, then we would also expect only the *red* E-vector to be transmitted.

---

[49] Scattering through an extremely opaque *static* medium gives rise to a spectrally broadened double-peaked Ly$\alpha$ spectrum (see Fig 26). Of course, photons in the red peak start with a redshift as well, which boosts their visibility especially for large $N_{HI}$ (see Fig 2 in Dijkstra & Wyithe, 2010, also see Haiman 2002).

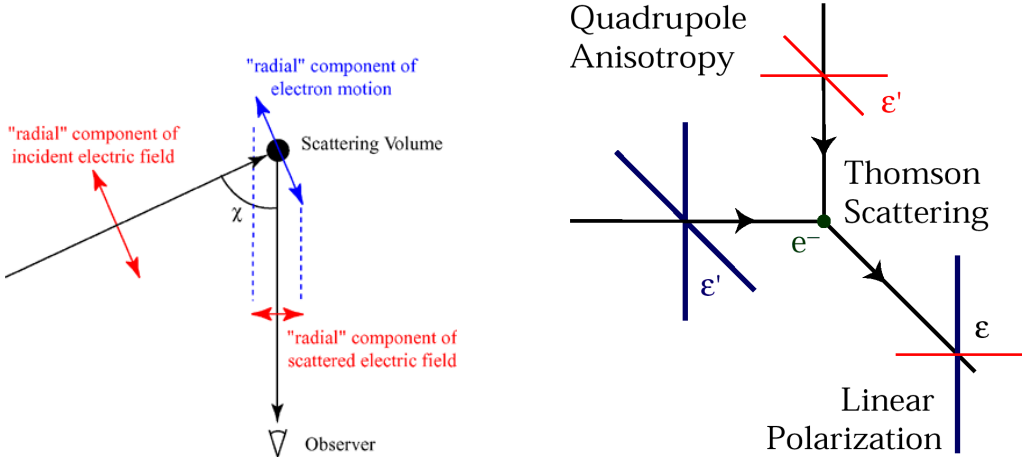

FIG. 46 This Figure illustrates how scattering by a free electron can polarize radiation. Both panels illustrate how scattering by a free electron transmits only the electric vector of the incident radiation, projected onto the outgoing radiation direction (also see Fig 15).

The first detections of polarization in spatially extended Ly$\alpha$ sources have been reported (Hayes et al. 2011b, Humphrey et al. 2013, Beck et al. 2016). The *left panel* in Figure 47 shows Ly$\alpha$ polarization vectors overlaid on a Ly$\alpha$ surface brightness map of 'LAB1' (Ly$\alpha$ Blob 1, see Hayes et al. 2011b). The *lines* here denote the linear polarization (see a more detailed discussion of this quantity below), which denotes the preferred orientation of the electric vector of the observed Ly$\alpha$ radiation. The longer the lines, the more polarized the radiation. Figure 47 shows how the polarization vectors appear to form concentric circles around spots of high Ly$\alpha$ surface brightness. This is consistent with a picture in which most Ly$\alpha$ was emitted in the spots with high surface brightness, and then scattered towards the observer at larger distance from these sites. This naturally gives rise to the observed polarization pattern (also see the *right panel* of Fig 47 for an artistic illustration of this, from Bower 2011).

The previous discussion can be condensed into a compact equation if we decompose the intensity of the radiation into a component parallel and perpendicular to the scattering plane, which is spanned by the propagation directions $\mathbf{k}_{\mathrm{in}}$ and $\mathbf{k}_{\mathrm{out}}$ (see Fig 48). We write $I \equiv I_{||} + I_{\perp}$, with $I_{||} \equiv |\mathbf{e}_{||}|^2 I$, in which $\mathbf{e}_{||}$ denotes the component of $\mathbf{e}_{\mathrm{E}}$ in the scattering plane. Similarly, we have $I_{\perp} \equiv |\mathbf{e}_{\perp}|^2 I$. We define the *scattering matrix*, $R$, as

$$
\begin{pmatrix} I'_{||} \\ I'_{\perp} \end{pmatrix} = \begin{pmatrix} S_1 & S_2 \\ S_3 & S_4 \end{pmatrix} \equiv R \begin{pmatrix} I_{||} \\ I_{\perp} \end{pmatrix},
\tag{109}
$$

where the total outgoing intensity is given by $I' = I'_{||} + I'_{\perp}$. The scattering matrix quantifies the angular redistribution of both components of a scattered electromagnetic wave. For comparison, the phase-function quantified the angular redistribution of the intensity total $I'$ only. For the case of a free electron, the scattering matrix $R_{\mathrm{Ray}}$ is given by

$$
R_{\mathrm{dip}} = \frac{3}{2} \begin{pmatrix} \cos^2 \theta & 0 \\ 0 & 1 \end{pmatrix}.
\tag{110}
$$

This expression indicates that for scattering by an angle $\theta$, $\mathbf{e}'_{||} = \cos\theta \mathbf{e}_{||}$ while $\mathbf{e}'_{\perp} = \mathbf{e}_{\perp}$ (see Fig 48). The total outgoing intensity $I' \equiv I'_{||} + I'_{\perp} = \frac{3}{2}(\cos^2 \theta I_{||} + I_{\perp})$, which for unpolarized incoming radiation ($I_{||} = I_{\perp} = 0.5I$)

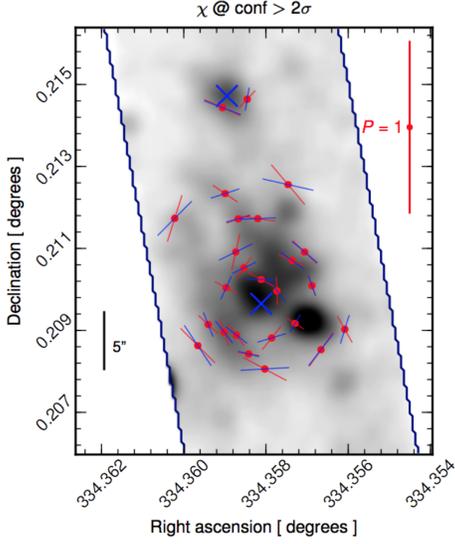

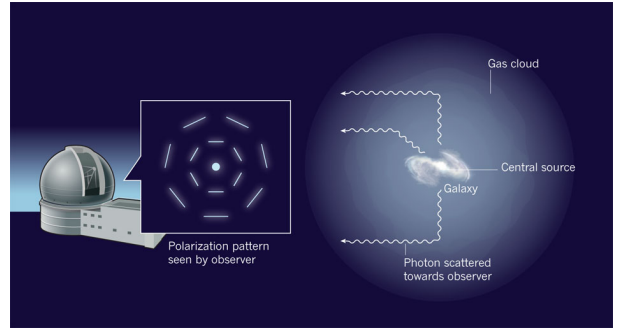

FIG. 47 *Left Panel*: Lines indicate the magnitude and orientation of the linear polarization of Lyα radiation in Lyα blob 1, overlaid on the observed Lyα surface brightness (in *gray scale*, (*Reprinted by permission from Macmillan Publishers Ltd: Hayes et al, 2011b, Nature 476, 304H, copyright*). Lyα polarization vectors form concentric circles around the most luminous Lyα 'spots' in the map. This is consistent with a picture in which Lyα emission is produced in the locations of high surface brightness, and where lower surface brightness regions correspond to Lyα that was scattered back into the line-of-sight at larger distance from the Lyα source. This process is illustrated visually in the *right panel* (*Reprinted by permission from Macmillan Publishers Ltd: Bower, 2011, Nature 476, 288B, copyright*).

reduces to $I' = \frac{3}{4}(1 + \cos^2 \theta)I$. The ratio $I'/I$ corresponds to the phase-function for Rayleigh scattering encountered in § 6.1 (see Eq 42). Furthermore, the *linear polarization* of the radiation is defined as

$$P \equiv \frac{I_\perp - I_{||}}{I_{||} + I_\perp}, \tag{111}$$

and for the scattered intensity of unpolarized radiation we find

$$P = \frac{1 - \cos^2 \theta}{1 + \cos^2 \theta}. \tag{112}$$

Note how this reflects our previous discussion: unpolarized radiation that is scattered by 90° becomes 100% linearly polarized (also see the *right panel* of Fig 46). For comparison, the scattering matrix for an 'isotropic' scatterer is

$$R_{\text{iso}} = \frac{1}{2} \begin{pmatrix} 1 & 1 \\ 1 & 1 \end{pmatrix}. \tag{113}$$

That is, the outgoing intensity $I'_{||} = I'_\perp = 0.5$ and has no directional dependence. Furthermore, isotropic scattering produces no polarization. An 'isotropic scatterer' scatters photons without caring about the properties of the incoming photon. We introduce this concept because isotropic scattering plays a role in Lyα scattering.

An electron is bound to a hydrogen atom is confined to orbits defined by quantum physics of the atom (see Fig 3 & § 3.2). That is, unlike the case of the free electron discussed above, an electron that is bound to a H nucleus is not completely free to oscillate along the polarization vector of the incoming photon, but is bound to orbits set by quantum mechanics. However, depending on which (sub) quantum state of the $2p$ state (in § 11.1 below we discuss how the $2p$ state splits up into several substates) the hydrogen atom is in, the electron may have some memory of the direction and polarization of the incoming photon. In other words, the wavefunction of the electron in the $2p$ state may be aligned along the polarization vector of the incoming photon. It turns out that scattering of Lyα photons by hydrogen atoms can be described as some linear combination of dipole and isotropic scattering, which is described as 'scattering by anisotropic particles' (Chandrasekhar, 1960). The scattering matrix for anisotropic particles is given by (Chandrasekhar, 1960, Eq 250-258)

$$R = \frac{3}{2}E_1 \begin{pmatrix} \cos^2\theta & 0 \\ 0 & 1 \end{pmatrix} + \frac{1}{2}E_2 \begin{pmatrix} 1 & 1 \\ 1 & 1 \end{pmatrix}, \tag{114}$$

where $E_1 + E_2 = 1$. Precisely what $E_1$ and $E_2$ are is determined by the quantum numbers that describe the electron in the $2p$ state. The number $E_1$ gives the relative importance of dipole scattering, and is sometimes referred to as the '*polarizability*', as it effectively measures how efficiently a scatterer can polarize incoming radiation. Both the angular redistribution - or the phase function - and the polarization of scattered Ly$\alpha$ radiation can therefore be characterized entirely by this single number $E_1$, which is discussed in more detail next.

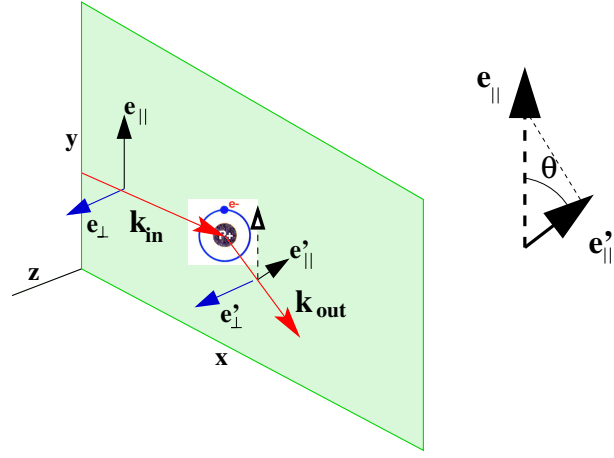

FIG. 48 $I_{||}$ ($I'_{||}$) denotes the intensity of the radiation parallel to the scattering plane before (after) scattering (spanned by $\mathbf{k}_{in}$ and $\mathbf{k}_{out}$, indicated in *green*). Furthermore, $I_{\perp}$ ($I'_{\perp}$) denotes the intensity perpendicular to this plane. The intensities relate to the polarization vectors $\mathbf{e}_{\perp}$ and $\mathbf{e}_{||}$ as $I_{||} \equiv (\mathbf{e}_{||} \cdot \mathbf{e}_{||})I$ etc. Classically, an incoming photon accelerates an electron in the direction of the polarization vector $\mathbf{e}$. The oscillating electron radiates as a classical dipole, and the angular redistribution of the outgoing intensity $I' = I'_{||} + I'_{\perp}$ scales as $I \propto 1 + \cos^2\theta$, where $\cos\theta = \mathbf{k}_{in} \cdot \mathbf{k}_{out}$. For scattering by an angle $\theta$, $\mathbf{e}'_{||} = \cos\theta\,\mathbf{e}_{||}$ while $\mathbf{e}'_{\perp} = \cos\theta$. For a classical dipole, the intensity $I'_{||}$ is therefore reduced by a factor of $\cos^2\theta$ relative to $I'_{\perp}$ (see text).

### 11.1. Quantum Effects on Ly$\alpha$ Scattering: The Polarizability of the Hydrogen Atom

In order to accurately describe how H atoms scatter Ly$\alpha$ radiation, we must consider the *fine-structure* splitting of the $2p$ level. The spin of the electron causes the $2p$ state quantum state to split into the $2p_{1/2}$ and $2p_{3/2}$ levels, which are separated by $\sim 10$ Ghz (see Fig. 49). The notation that is used here is $nL_J$, in which $\mathbf{J} = \mathbf{L} + \mathbf{s}$ denotes the total (orbital +spin) angular momentum of the electron. The $1s_{1/2} \rightarrow 2p_{1/2}$ and $1s_{1/2} \rightarrow 2p_{3/2}$ is often referred as the K-line and H-line, respectively.

It turns out that a quantum mechanical calculation yields that $E_1 = \frac{1}{2}$ for the H transition, while $E_1 = 0$ for the K transition (e.g. Hamilton 1947, Brandt & Chamberlain 1959, Lee et al. 1994, Ahn et al. 2002). When a Ly$\alpha$ scattering event goes through the K-transition, the hydrogen atom behaves like an isotropic scatterer. This is because the wavefunction of the $2p_{1/2}$ state is spherically symmetric (see White 1934), and the atom 'forgot' which direction the photon came from or which direction the electric field was pointing to. For the $2p_{3/2}$ state, the wavefunction is not spherically symmetric and contains the characteristic 'double lobes' shown in Figure 3. The hydrogen in the $2p_{3/2}$ state thus has some memory of the direction of the incoming Ly$\alpha$ photon and its electric vector, and behaves partially as a classical dipole scatterer, and partly as an isotropic scatterer.

However, in reality the situation is more complex and Stenflo (1980) showed that $E_1$ depends strongly on frequency as

$$E_1 = \frac{\left(\frac{\omega_K}{\omega_H}\right)^2(b_H^2 + d_H^2) + 2\frac{\omega_K}{\omega_H}(b_H b_K + d_H d_K)}{b_K^2 + d_K^2 + 2\left(\frac{\omega_K}{\omega_H}\right)^2(b_H^2 + d_H^2)}, \tag{115}$$

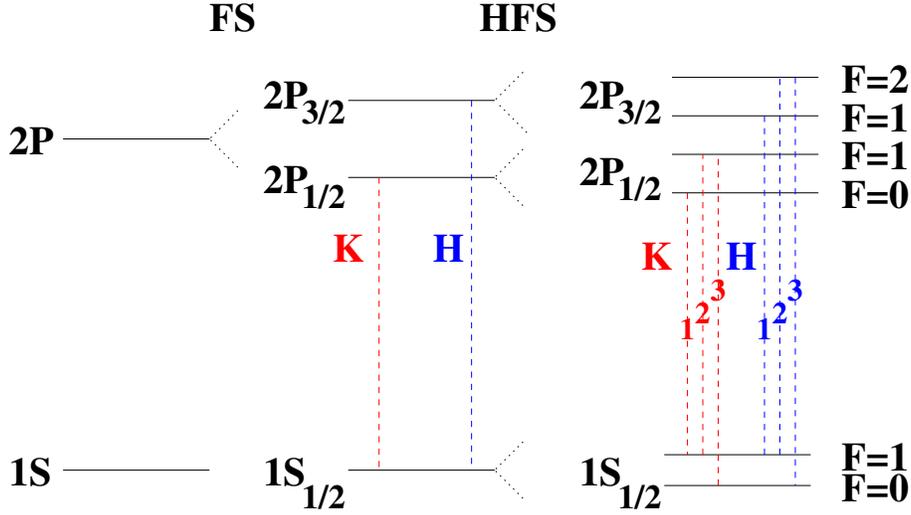

FIG. 49 Schematic diagram of the energy levels of a hydrogen atom. The notation for each level is $nL_J$, where $n$ is the principle quantum number, and $L$ denotes the orbital angular momentum number, and $J$ the total angular momentum ('total' angular momentum means orbital+spin angular momentum, i.e. $\mathbf{J} = \mathbf{L} + \mathbf{S}$). The fine structure splitting of the $2p$ level shifts the $2p_{1/2}$ and $2p_{3/2}$-level by $\Delta E/h_p \sim 11$ Ghz (e.g. Brasken & Kyrola 1998). A transition of the form $1s_{1/2} \rightarrow 2p_{1/2}$ ($1s_{1/2} \rightarrow 2p_{3/2}$) is denoted by a K (H) transition. The spin of the nucleus induces further 'hyperfine' splitting of the line. This is illustrated in the *right panel*, where each fine structure level breaks up into two lines which differ only in their quantum number F which measures the total + nuclear spin angular momentum, i.e. $\mathbf{F} = \mathbf{J} + \mathbf{I}$. Hyperfine splitting ultimately breaks up the Ly$\alpha$ transition into six individual transitions. Fine and hyperfine structure splitting plays an important role in polarizing scattered Ly$\alpha$ radiation.

where $b_{H,K} = \omega^2 - \omega_{H,K}^2$ and $d_{H,K} = \omega_{H,K}\Gamma_{H,K}$. Here, $\omega_H = 2\pi\nu_H$ ($\omega_K = 2\pi\nu_K$) denotes the resonant angular frequency of the H (K) transition, and $\Gamma_{H,K} = A_\alpha$. The frequency dependence of $E_1$ is shown in Figure 50, where we have plotted $E_1$ as a function of wavelength $\lambda$. The *black solid line* shows $E_1$ that is given by Eq 115. This plot shows that

1. $E_1 = 0$ at $\lambda = \lambda_K$ and that $E_1 = \frac{1}{2}$ at $\lambda = \lambda_H$, which agrees with earlier studies (e.g. Hamilton 1947, Brandt & Chamberlain 1959), and which reflects what we discussed above.

2. $E_1$ is negative for most wavelengths in the range $\lambda_H < \lambda < \lambda_K$. The classical analogue to this would be that when an atom absorbs a photon at this frequency, that then the electron oscillates along the propagation direction of the incoming wave, which is strange because the electron would be oscillating in a direction orthogonal to the direction of the electric vector of the electromagnetic wave. However, scattering at these frequencies is very unlikely (see § 11.2).

3. $E_1 = 1$ when a photon scatters in the wings of the line, which is arguably the most bizarre aspect of this plot. Stenflo (1980) points out that, when a Ly$\alpha$ photon scatters in the wing of the line profile, it goes simultaneously through the $2p_{1/2}$ and $2p_{3/2}$ states, and as a result, the bound electron is permitted to behave as if it were free.

The impact of 'quantum interference' on the scattering phase function is more than just interesting from a 'fundamental' viewpoint, because it makes a physical distinction between 'core' and 'wing' scattering. This distinction arises because $E_1$ affects the scattering phase function: the phase function for wing scattering $P_{\text{wing}} \propto (1 + \mu^2)$, while for core scattering the phase function can behave like a (sometimes strange, i.e. when $E_1 < 0$) superposition of isotropic and wing scattering (see Eq 117 for an example of how to compute this phase function from $E_1$).

For reference: there is hyperfine splitting in the fine structure lines that is induced by the spin of the proton which can couple to the electron spin, which induces further splitting of the line. This is illustrated in Figure 49, where each fine structure level breaks up into two lines which differ only in their quantum number F which measures the total + nuclear spin angular momentum, i.e. $\mathbf{F} = \mathbf{J} + \mathbf{I}$. The dependence of $E_1$ on frequency when this hyperfine splitting is accounted for has been calculated by Hirata (2006, his Appendix B. Note however, that $E_1$ is not computed explicitly. Instead, the angular redistribution functions are given and it is possible to extract $E_1$ from these). The formula for $E_1$ is quite lengthy, and the reader is referred to Hirata (2006) for the full expression. We have overplotted as the *red dotted line* wavelength dependence of $E_1$ when hyperfine splitting is accounted for. The general frequency dependence of $E_1$ is not affected. However, there is an overall higher level of polarization. The box in the *lower left corner* of Figure 50 shows a close-up view of $E_1$ near the resonances. Around the H-resonance, hyperfine splitting introduces new resonances (indicated by the *blue dotted lines*) which affect $E_1$ somewhat. At wavelengths longward of the H resonance, the $E_1$ is boosted slightly, which causes scattering through the K resonance to not be perfectly isotropic (another interpretation is that the hyperfine splitting breaks the perfect spherical symmetry of the $2p_{1/2}$ state). An overall boost in the polarizability as a result of hyperfine splitting has also been found by other authors (e.g. Brasken & Kyrola, 1998).

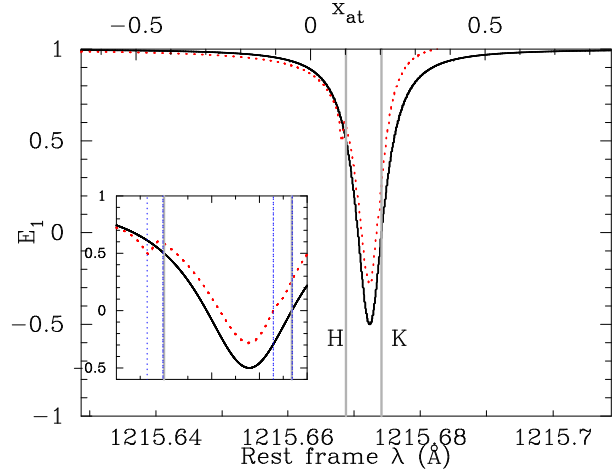

FIG. 50 The frequency dependence of the polarizability $E_1$ of a hydrogen atom is shown. The *black solid line* shows $E_1$ as a function of $\lambda$ when fine structure splitting is accounted for (Eq 115, taken from Stenflo 1980). This plot shows that $E_1 = 0$ at $\lambda = \lambda_K$ and that $E_1 = \frac{1}{2}$ at $\lambda = \lambda_H$. Interestingly, $E_1$ is negative for most wavelengths in the range $\lambda_H < \lambda < \lambda_K$. However, scattering at these frequencies is very unlikely (see § 11.2). More importantly, $E_1 = 1$ when a photon scatters in the wings of the line. That is, for photons that are scattered in the wing of the line profile, the electron in the hydrogen atom behaves like a classical dipole (i.e. as if it were a free electron!). The *red dotted line* shows the wavelength dependence of $E_1$ when hyperfine splitting is accounted for. Hyperfine splitting introduces an overall slightly higher level of polarization. The box in the *lower left corner* shows a close-up view of $E_1$ near the resonances (see text).

## 11.2. Lyα Propagation through HI: Resonant vs Wing Scattering

We highlighted the distinction between 'core' versus 'wing' scattering previously in § 11. As we already mentioned, thd polarizability can be negative $E_1 < 0$ for a range of frequencies between the H and K resonance frequencies (see Fig 50). However, scattering at these frequencies does not occur often enough to leave an observable imprint. The reason for this is that the natural width of the line for both the H and K transitions is much smaller than their separation, i.e $\gamma_{H,K} \equiv \Gamma_{H,K}/[4\pi] \sim 10^8$ Hz $\ll \nu_H - \nu_K = 1.1 \times 10^{10}$ Hz, and the absorption cross–section in the atom's rest-frame scales as, $\sigma(\nu) \propto [(\nu - \nu_{H,K})^2 + \gamma_{H,K}^2]^{-1}$ (see Eq 48). A Lyα photon is therefore much more likely to be absorbed by an atom for which the photon appears exactly at resonance, than by an atom for which the photon has a frequency corresponding to a negative $E_1$. Quantitatively, the Maxwellian probability $P$ that a photon of frequency $x$ is scattered in the frequency range $x_{at} \pm dx_{at}/2$ in the atom's rest-frame is given by

$$P(x_{at}|x)dx_{at} = \frac{a}{\pi H(a_v, x)} \frac{e^{-(x_{at}-x)^2}}{x_{at}^2 + a_v^2} dx_{at}, \tag{116}$$

where we used that $P(x_{at}|x) = P(u_{at}|x) = P(x|u_{at})P(u_{at})/P(x)$, where $u_{at} = x - x_{at}$ denotes the atom velocity in units of $v_{th}$ (also see the discussion above Eq 69). The probability $P(x_{at}|x)$ is shown in Figure 51 for[50] $x = 3.3$ (*solid line*) and $x = -5.0$ (*dashed line*). Figure 51 shows that for $x = 3.3$, photons are scattered either when they are exactly at resonance or when they appear $\sim 3$ Doppler widths away from resonance. The inset of Figure 51 zooms on the region near $x_{at} = 0.0$. For frequencies $x \lesssim 3$ there are enough atoms moving at velocities such that the Lyα photon appears at exactly at resonance in the frame of the atom. However, this is not the case any more for frequencies $|x| \gtrsim 3$. Instead, the majority of photons is scattered while it is in the wing of the absorption profile. This is illustrated by the *dashed line* which shows the case $x = -5.0$, for which resonant scattering is less likely by orders of magnitude. This discussion illustrates that the transition from 'core' to 'wing' scattering is continuous (though it occurs over a narrow range of frequencies). Photons can 'resonantly' scatter while they are in the wing with a finite probability, and vice versa. While in practise, this is not an important effect, it is worth keeping in mind.

---

[50] Note that the transition from core to wing scattering occurs at $x \sim 3$, see Fig 17.

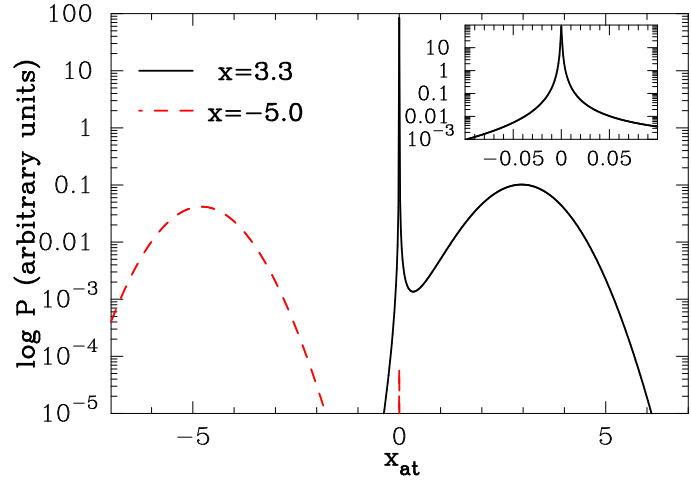

FIG. 51 The probability $P(x_{at}, x)$ (Eq. 116) that a photon of frequency $x$ is scattered by an atom such that it appears at a frequency $x_{at}$ in the frame of the atom (*Credit: from Figure A2 of Dijkstra & Loeb. 2008, 'The polarization of scattered Lyα radiation around high-redshift galaxies', MNRAS, 386, 492D*). The *solid* and *dashed* lines correspond to $x = 3.3$ and $x = -5.0$ respectively. For $x = 3.3$, photons are either scattered by atoms to which they appear *exactly* at resonance (see inset), or to which they appear $\sim 3$ Doppler widths away. For $x = -5$, resonant scattering is less important by orders of magnitude. In combination with Fig 50, this figure shows that if a photon is resonantly scattered then $E_1$ is either 0 or $\frac{1}{2}$.

### 11.3. Polarization in Monte-Carlo Radiative Transfer

Incorporating polarization in a Monte-Carlo is complicated if you want to do it correctly. A simple procedure was presented by Angel (1969) and Rybicki & Loeb (1999), which is accurate for wing-scattering only. In practise this is often sufficient, as Lyα wing photons are the ones that are most likely to escape from a scattering medium, and are thus most likely to be observed.

Rybicki & Loeb (1999) assigned 100% linear polarization to each Lyα photon used in the Monte-Carlo simulation by attaching an (normalized) electric vector $\mathbf{e}_E$ to the photon perpendicular to its propagation direction (see discussion above Eq 37). For each scattering event, the outgoing direction ($\mathbf{k}_{out}$) is drawn from the phase-function $P \propto \sin^2 \Psi$ (see the discussion above Eq 37), where $\cos \Psi = \mathbf{k}_{out} \cdot \mathbf{e}$ (see Angel 1969, Rybicki & Loeb 1999, Dijkstra & Loeb 2008b for technical details). This approach thus accounts for the polarization-dependence of the phase function described previously in § 6.1.

Consider a Lyα photon in the Monte-Carlo simulation that was propagating in a direction $\mathbf{k}_{in}$, and with an electric vector $\mathbf{e}_E$. It was then scattered towards the observer. The polarization of this photon when it is observed can be obtained as follows: we know that when a photon reaches us, its propagation direction, $\mathbf{k}_{out}$, is perpendicular to the plane of the sky. We therefore know that the polarization vector $\mathbf{e}'_E$ must lie in the plane of the sky. The linear polarization measures the difference in intensity when measured in the two orthogonal directions in the plane of the sky (denoted previously with $I_{\parallel}$ and $I_{\perp}$). We now define $\mathbf{r}$ to be the vector that connects the location of last scattering to the Lyα source, projected onto the sky. Both $\mathbf{r}$ and $\mathbf{e}'_E$ therefore lie in the plane of the sky, and we let $\chi$ denote the angle between them (i.e. $\cos \chi \equiv \mathbf{r} \cdot \mathbf{e}'_E / |\mathbf{r}|$). The photon then contributes $\cos^2 \chi$ to $I_l$ and $\sin^2 \chi$ to $I_r$. This geometry is depicted in Figure 52. This procedure was tested successfully by Rybicki & Loeb (1999) against analytic solutions obtained by Schuster (1879).

For core scattering the situation is more complicated. We know that scattering through the $H$-transition corresponds to isotropic scattering, while scattering through the $K$-transition corresponds to scattering off an anisotropic particle with $E_1 = 0.5$. From their statistical weights we can infer that scattering through the $K$-transition is twice is likely as scattering through the H-transition ($g_H = 2$, $g_K = 4$). The weighted average implies that core scattering corresponds to scattering by an anisotropic particle with $E_1 = 1/3$. Eq 114 shows that the scattering matrix and phase function take on the following form:

$$R = \begin{pmatrix} \frac{1}{2}\cos^2\theta + \frac{1}{3} & \frac{1}{3} \\ \frac{1}{3} & \frac{5}{6} \end{pmatrix} \quad \Rightarrow \quad P(\mu) = \frac{11}{12} + \frac{3}{12}\mu^2, \tag{117}$$

for unpolarized incident radiation. Formally, the frequency redistribution function for core scattering is therefore neither given by $R_A(x_{out}|x_{in})$ nor $R_B(x_{out}|x_{in})$ (Eq 73), but by some intermediate form. Given the similarity of $R_A(x_{out}|x_{in})$ and $R_B(x_{out}|x_{in})$ (see Fig 21) this difference does not matter in practice. Note that this phase function can be implemented naturally in a Monte-Carlo simulation by treating $1/3$ of all core scattering events as pure dipole scattering events, and the remaining $2/3$ as pure isotropic scattering events (Trebitsch et al., 2016).

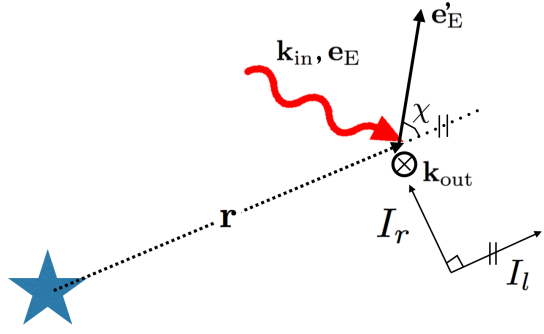

FIG. 52 This figure illustrates visually what the 'polarization' angle $\chi$ is, which can be used in Monte-Carlo calculations to compute the linear polarization of scattered Ly$\alpha$ as a function of projected distance $r \equiv |\mathbf{r}|$ from the galaxy/Ly$\alpha$ source.

The previous approach assigns electric vectors to each Ly$\alpha$ photon in the Monte-Carlo simulation, and therefore implicitly assumes that each individual Ly$\alpha$ photon is 100% linearly polarized. It would be more realistic if we could assign a fractional polarization to each photon, which would be more representative of the radiation field. Recall that the phase-functions depend on the polarization of the radiation field. An alternative way of incorporating polarization which allows fractional polarization to be assigned to individual Ly$\alpha$ photons is given by the *density-matrix formalism* described in Ahn & Lee (2015). In this formalism all polarization information is encoded in 2 parameters (the 2 parameters reflect the degrees of freedom for a mass-less spin-0 'particle' within the 'density matrix'. We will not discuss this formalism in this lecture. Both methods should converge for scattering in optically thick gas, but they have not been compared systematically yet (but see Chang et al. 2017 for recent work in this direction).

## 12. APPLICATIONS BEYOND LY$\alpha$: WOUTHUYSEN-FIELD COUPLING AND 21-CM COSMOLOGY/ASTROPHYSICS

### 12.1. The 21-cm Transition and its Spin Temperature

Detecting the redshifted 21-cm line from neutral hydrogen gas in the young Universe is one of the main challenges of observational cosmology for the next decades, and serves as the science driver for many low frequency arrays that were listed in § 3.1. The 21-cm transition links the two hyperfine levels of the ground state ($1s$) of atomic hydrogen. The energy difference arises due to coupling of the proton to the electron spin: the proton spin $S_p$ gives it a magnetic moment[51] $\mu_p = \frac{g_p e\hbar}{2m_p c} S_p$, where the proton's 'g-factor' is $g_p \sim 5.59$. This magnetic dipole generates a magnetic field which interacts with the magnetic moment of the electron ($\mu_e$) due to its spin. Classically, the energy difference between the two opposite electron spin states equals $\Delta E = 2|\mu_e||B_p|$, where $B_p$ denotes the magnetic field generated by the spinning proton. This is illustrated schematically in Figure 53. The 21-cm transition corresponds to the electron's spin 'flipping' in the magnetic field generated by the proton. The 21-cm transition is therefore often referred to as the 'spin-flip' transition. What is interesting about this classical picture is that it fails to convey that quantum mechanically, the energy difference between the two hyperfine levels of the $1s$ level is actually zero, *unless* **r**=0. In quantum mechanics, there is a finite probability that the electron finds itself inside the proton. Formally, this is what causes the different hyperfine levels of the ground state of atomic hydrogen to have different energies (special thanks to D. Spiegel for pointing this out to me).

---

[51] A spinning proton can be seen as a rotating charged sphere, which produces a magnetic field.

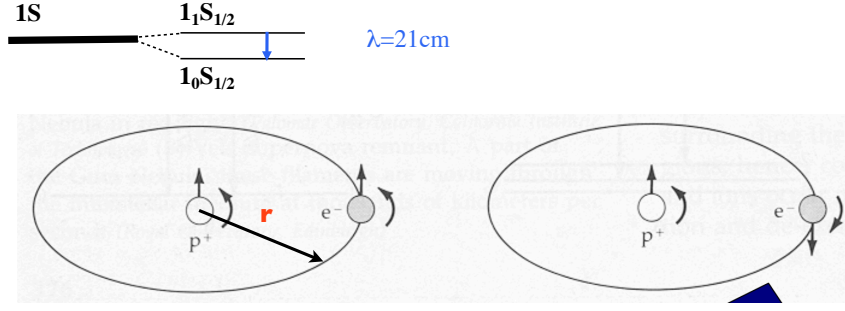

FIG. 53 This figure illustrates the classical picture of the origin of the energy difference between the two hyperfine levels of the ground state of atomic hydrogen. The proton spin generates a magnetic dipole moment which in turn generates a magnetic field. This magnetic field introduces an energy dependence to the orientation of the electron spin. In quantum mechanics however, the two states shown here have equal energy, unless the electron finds itself inside the proton (i.e unless **r** = 0).

The 21-cm transition is a highly forbidden transition with a natural life-time of $t \equiv A_{21}^{-1} \sim (2.87 \times 10^{-15}\ \mathrm{s}^{-1})^{-1} \sim 1.1 \times 10^7\ \mathrm{yr}$ (one way to interpret this long lifetime is to connect it to the low probability of the electron and proton overlapping). The 21-cm line has been observed routinely in nearby galaxies, and has allowed us to map out the distribution & kinematics of HI gas in galaxies. Observations of the kinematics of HI gas have given us galaxy rotation curves, which further confirmed the need for dark matter on galaxy scales. Because of its intrinsic faintness, it is difficult to detect HI gas in emission beyond $z \gtrsim 0.5$ until the Square Kilometer Array (SKA) becomes operational.

## 12.2. The 21-cm Brightness Temperature

It is theoretically possible to detect the 21-cm line from neutral gas during the reionization epoch and even the 'Dark Ages', which refers to the epoch prior to the formation of the first stars, black holes, etc., and when the only source of radiation was the Cosmic Microwave Background which had redshifted out of the visual band and into the infrared. The visibility of HI gas in its 21-cm line can be expressed in terms of a *differential* brightness temperature[52], $\delta T_b(\nu)$, with respect to the background CMB, and equals (e.g. Furlanetto et al. 2006, Morales & Wyithe 2010, and references therein):

$$\delta T_b(\nu) \approx 9x_{\mathrm{HI}}(1+\delta)(1+z)^{1/2}\left(1 - \frac{T_{\mathrm{CMB}}(z)}{T_S}\right)\left[\frac{H(z)(1+z)}{dv_{||}/dr}\right]\ \mathrm{mK}, \tag{118}$$

where $\delta + 1 \equiv \rho/\bar{\rho}$ denotes the overdensity of the gas cloud, $z$ the redshift of the cloud, $T_{\mathrm{CMB}}(z) = 2.73(1+z)$ K denotes the temperature of the CMB, the factor in square brackets contains the line-of-sight velocity gradient $dv_{||}/dr$. Finally, $T_S$ denotes the *spin temperature* (also known as the excitation temperature) of the 21-cm transition, which quantifies the number densities of hydrogen atoms in each of the hyperfine transitions, i.e.

---

[52] Eq 118 follows from solving the radiative transfer equation, $\frac{dI}{d\tau} = -I + \frac{A_{10}}{B_{10}}\frac{1}{(3n_0/n_1 - 1)}$, in the (appropriate) limit that the neutral IGM is optically thin in the HI 21-cm line, and that it therefore only slightly modifies the intensity $I$ of the background CMB. It is common in radio astronomy to express intensity fluctuations as temperature fluctuations by recasting intensity as a temperature in the Rayleigh-Jeans limit: $I_\nu(\mathbf{x}) \equiv \frac{2k_B T(\mathbf{x})}{\lambda}$.

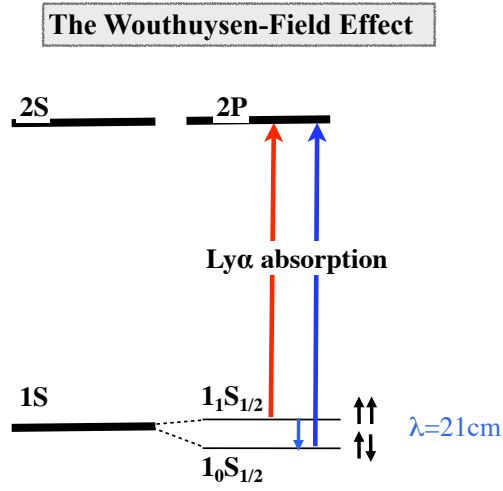

FIG. 54 Lyα photons that have a higher energy are (slightly) more likely to be absorbed from the $1_0S_{1/2}$ level because of the (again slightly) larger energy separation between this level and any $n = 2$ level. The precise spectrum of Lyα photons around the line resonance therefore affects $n_0$ (number density of hydrogen atoms in the $1_0S_{1/2}$ level) and $n_1$ (number density of hydrogen atoms in the $1_1S_{1/2}$ level). Repeated scattering of Lyα photons in turn modifies the spectrum in such as a way that Lyα scattering drives the spin temperature to the gas temperature. This is known as the Wouthuysen-Field effect.

$$\frac{n_1}{n_0} \equiv \frac{g_1}{g_0} \exp\left(\frac{-h\nu_{21\text{cm}}}{k_B T_s}\right) = 3 \exp\left(\frac{-h\nu_{21\text{cm}}}{k_B T_s}\right), \tag{119}$$

where $n_0$ $(n_1)$ denotes the number densities of hydrogen atoms in the ground (excited) level of the 21-cm transition, and where we used that $g_1 = 3$ and $g_0 = 1$. The numerical prefactor of 9 mK in Eq 118 was derived assuming that the gas was undergoing Hubble expansion. For slower expansion rates, we increase the number of hydrogen atoms within a fixed velocity (and therefore frequency) range, which enhances the brightness temperature.

Eq 118 states that when $T_{\text{CMB}}(z) < T_S$, we have $\delta T_b(\nu) > 0$, and when $T_{\text{CMB}}(z) > T_S$ we have $\delta T_b(\nu) < 0$. This means that we see HI in absorption [emission] when $T_S < T_{\text{CMB}}(z)$ $[T_S > T_{\text{CMB}}(z)]$. The spin temperature thus plays a key role in setting the 21-cm signal, and we briefly discuss what physical processes set $T_s$ below in § 12.3. Eq 118 highlights why so much effort is going into trying to detect the 21-cm transition: the 21-cm transition corresponds to a line, having a 3D map of the 21-cm line provides us with a unique way of constraining the 3D density, velocity field etc. It provides us with a powerful cosmological and astrophysical probe of the high-redshift Universe.

### 12.3. The Spin Temperature and the Wouthuysen-Field Effect

The spin temperature - i.e. how the two hyperfine levels are populated - is set by (i) collisions, which drive $T_S \rightarrow T_{\text{gas}}$, (ii) absorption by CMB photons, which drives $T_S \rightarrow T_{\text{CMB}}$ (and thus that $\delta T_b \rightarrow 0$, see Eq 118), and (iii) Lyα scattering. This is illustrated in Figure 54: absorption of Lyα photons can occur from any of the two hyperfine levels. The subsequent radiative cascade back to the ground state can leave the atom in either hyperfine level. Lyα scattering thus mixes the two hyperfine levels, which drive $T_S \rightarrow T_\alpha$, where $T_\alpha$ is known as the Lyα color temperature, which will be discussed below. Quantitatively, it has been shown that (e.g. Madau et al. 1997, Tozzi et al. 2000, Furlanetto et al. 2006)

$$\frac{1}{T_S} = \frac{T_{\text{CMB}}^{-1} + x_c T_{\text{gas}}^{-1} + x_\alpha T_\alpha^{-1}}{1 + x_c + x_\alpha}, \tag{120}$$

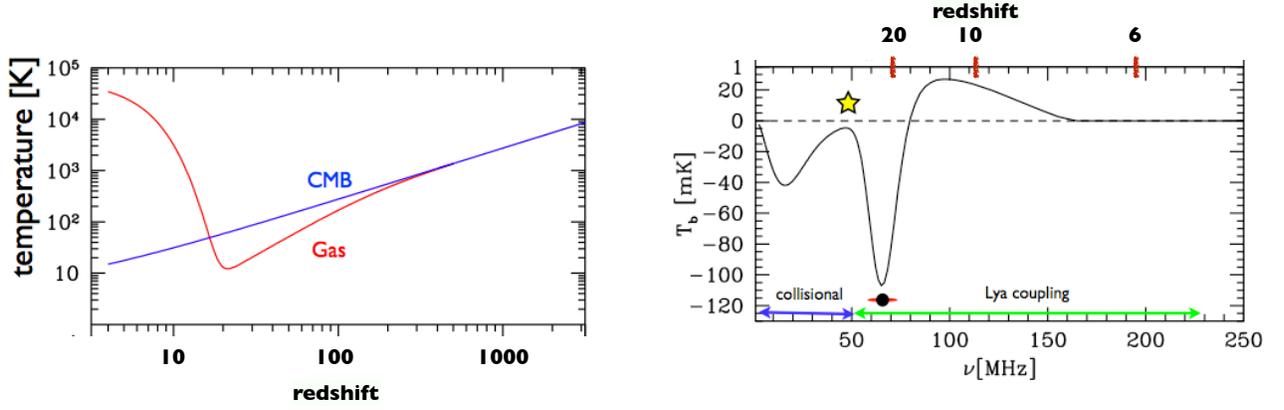

FIG. 55 *Left panel:* Globally, averaged redshift evolution of $T_{gas}$ (*red solid line*) and the $T_{CMB}$ (*blue solid line*). Right panel: The 'Global' 21-signal, represents the sky-averaged 21-cm brightness temperature $\delta T_b(\nu)$. See the main text for a description of each of these curves. The global 21-cm signal constrains when the first stars, black holes, and galaxies formed and depends on their spectra. *Credit: adapted from slides created by J. Pritchard (based on Pritchard & Loeb 2010, 2012).*

where $x_c = \frac{C_{21} T_*}{A_{21} T_{gas}}$ denotes the collisional coupling coefficient, in which $k_B T_*$ denotes the energy difference between the hyperfine levels, $C_{21}$ denotes the collisional deexcitation rate coefficient. Furthermore, $x_\alpha = \frac{4 P_\alpha}{27 A_{21} T_\alpha}$, in which $P_\alpha$ denotes the Ly$\alpha$ scattering rate[53]. The Ly$\alpha$ color temperature provides a measure of the shape of the spectrum near the Ly$\alpha$ resonance (Meiksin 2006, Dijkstra & Loeb 2008c):

$$\frac{k_B T_\alpha}{h_P} = \frac{\int J(\nu) \sigma_\alpha(\nu) d\nu}{\int \frac{\partial \sigma_\alpha}{\partial \nu} J(\nu) d\nu} = -\frac{\int J(\nu) \sigma_\alpha(\nu) d\nu}{\int \frac{\partial J}{\partial \nu} \sigma_\alpha(\nu) d\nu},$$

(121)

where $\sigma_\alpha(\nu)$ denotes the Ly$\alpha$ absorption cross-section (see Eq 55), and $J(\nu)$ denotes the angle-averaged intensity (see § 7.5). In the last step, we integrated by parts. The reason the spectral shape near Ly$\alpha$ enters is illustrated in Fig 54, which shows that higher frequency Ly$\alpha$ photons are (slightly) more likely to excite hydrogen atoms from the ground (singlet) state of the 21-cm transition. The relative number of Ly$\alpha$ photons slightly redward and blueward of the resonance therefore affects the 21-cm spin temperature (Wouthuysen, 1952). Eq 121 indicates that if $\frac{\partial J(\nu)}{\partial \nu} < 0$, then $T_\alpha > 0$. Interestingly, it has been demonstrated that repeated scattering of Ly$\alpha$ photons changes the Ly$\alpha$ spectrum around the resonance such that it drives $T_\alpha \to T_{gas}$ (Field 1959). The resulting coupling between $T_s$ and $T_{gas}$ as a result of Ly$\alpha$ scattering is known as the 'Wouthuysen-Field (WF)[54] coupling. We generally expect $T_s$ to be some weighted average of the gas and CMB temperature, where the precise weight depends on various quantities such as gas temperature, density and the WF-coupling strength.

There are two key ingredients in WF-coupling: The Ly$\alpha$ color temperature $T_\alpha$ and the Ly$\alpha$ scattering rate. Theoretically, Ly$\alpha$ scattering rates are boosted in close proximity to star forming galaxies because of the locally enhanced Ly$\alpha$ background (Chuzhoy & Zheng 2007). However, this local boost in the WF-coupling strength can be further affected by the assumed spectrum of photons emerging from the galaxy (galaxies themselves are optically thick to Ly$\alpha$ and higher Lyman series radiation), which has not been explored yet.

## 12.4. The Global 21-cm Signal

Eq 120 implies that the spin temperature $T_s$ is a weighted average of the gas and CMB temperature. The *left panel* of Figure 55 shows the universally (or globally) averaged temperatures of the gas (*red solid line*) and the CMB (*blue*

---

[53] Note that this equation uses that for each Ly$\alpha$ scattering event, the probability that it induces a scattering event is $P_{flip} = \frac{4}{27}$ (see e.g. Meiksin 2006, Dijkstra & Loeb 2008c). This probability reduces by many orders of magnitude for wing scattering as $P_{flip} \propto x^{-2}$ (see Hirata 2006, Dijkstra & Loeb 2008c). In practise wing scattering contributes little to the overall scattering rate, but it is good to keep this in mind.

[54] See Furlanetto et al. 2006 for an explanation of how to best pronounce 'Wouthuysen'. *Hint:* it helps if you hold your breath 12 seconds before trying.

*solid line*), and the corresponding 'global' 21-cm signature in the *Right panel*. We discuss these in more detail below:

- Adiabatic expansion of the Universe causes $T_{CMB} \propto (1+z)$ at all redshifts. At $z \gtrsim 100$, the gas temperature remains coupled to the CMB temperature because of the interaction between CMB photons and the small fraction of free electrons that exist because recombination is never 'complete' (i.e. there is a residual fraction of protons that never capture an electron to form a hydrogen atom). When $T_{CMB} = T_{gas}$, we must have $T_s = T_{CMB}$, and therefore that $\delta T_b(\nu) = 0$ mK, which corresponds to the high-$z$ limit in the *right panel*.

- At $z \lesssim 100$, electron scattering can no longer couple the CMB and gas temperatures, and the (non-relativistic) gas adiabatically cools faster than the CMB as $T_{gas} \propto (1+z)^2$. Because $T_{gas} < T_{CMB}$, we must have that $T_s < T_{CMB}$ and we see the 21-cm line in absorption. When $T_{gas}$ first decouples from $T_{CMB}$ the gas densities are high enough for collisions to keep $T_s$ locked to $T_{gas}$. However, at $z \sim 70$ ($\nu \sim 20$ MHz) collisions can no longer couple $T_s$ to $T_{gas}$, and $T_s$ crawls back to $T_{CMB}$, which reduces $\delta T_b(\nu)$ (at $\nu \sim 20-50$ MHz, i.e. $z \sim 70-30$).

- The first stars, galaxies, and accreting black holes emitted UV photons in the energy range $E = 10.2 - 13.6$ eV. These photons can travel freely through the neutral IGM, until they redshift into one of the Lyman series resonances, at which point a radiative cascade can produce Ly$\alpha$. The formation of the first stars thus generates a Ly$\alpha$ background, which initiates the WF-coupling, which pushes $T_s$ back down to $T_{gas}$. The onset of Ly$\alpha$ scattering - and thus the WF coupling - causes $\delta T_b(\nu)$ to drop sharply at $\nu \gtrsim 50$ Mhz ($z \lesssim 30$).

- At some point the radiation of stars, galaxies, and black holes starts heating the gas. Especially X-rays produced by accreting black holes can easily penetrate deep into the cold, neutral IGM and contain a lot of energy which can be converted into heat after they are absorbed. The *left panel* thus has $T_{gas}$ increase at $z \sim 20$, which corresponds to onset of X-ray heating. In the *right panel* this onset occurs a bit earlier. This difference reflects that the redshift of all the features (minima and maxima) in $\delta T_b(\nu)$ are model dependent, and not well-known (more on this below). With the onset of X-ray heating (combined with increasingly efficient WF coupling to the build-up of the Ly$\alpha$ background) drives $\delta T_b(\nu)$ up until it becomes positive when $T_{gas} > T_{CMB}$.

- Finally, $\delta T_b(\nu)$ reaches yet another maximum, which reflects that neutral, X-ray heated gas is reionized away by the ionizing UV-photons emitted by star forming galaxies and quasars. When reionization is complete, there is no diffuse intergalactic neutral hydrogen left, and $\delta T_b(\nu) \rightarrow 0$.

In detail, the onset and redshift evolution of Ly$\alpha$ coupling, X-ray heating, and reionization depend on the redshift evolution of the number densities of galaxies, and their spectral characteristics. All these are uncertain, and it is not possible to make robust predictions for the precise shape of the global 21-cm signature. Instead, one of the main challenges for observational cosmology is to measure the global 21-cm signal, and from this constrain the abundances and characteristic of first generations of galaxies in our Universe. Detecting the global 21-cm is challenging, but especially the deep absorption trough that is expected to exist just prior to the onset of X-ray heating at $\nu \sim 70$ MHz is something that may be detectable because of its characteristic spectral shape. It is interesting that the presence of this absorption feature relies on the presence of a Ly$\alpha$ background, which must be strong enough to enable WF-coupling.

**Acknowledgments** I thank Ivy Wong for providing a figure for a section which (unfortunately) was cut as a whole, Daniel Mortlock & Chris Hirata for providing me with tabulated values of their calculations which I used to create Fig 17, Jonathan Pritchard for permission to use one of his slides for these notes, Andrei Mesinger for providing Fig 42, and other colleagues for their permission to re-use Figures from their papers. I thank the astronomy department at UCSB for their kind hospitality when I was working on preparing these lecture notes. I thank the organizers of the school: Anne Verhamme, Pierre North, Hakim Atek, Sebastiano Cantalupo, Myriam Burgener Frick, to Matt Hayes, X. Prochaska, and Masami Ouchi for their inspiring lectures. I thank Pierre North for carefully reading these notes, and for finding and correcting countless typos. Finally, special thanks to the students for their excellent attendance, and for their interest & enthusiastic participation.